\documentclass{article}
\pdfoutput=1

\usepackage{arxiv}
\usepackage[utf8]{inputenc}
\usepackage[T1]{fontenc}    

\usepackage[sort&compress,numbers]{natbib}

\usepackage{graphicx}
\usepackage{subfig}       
\graphicspath{{figures/}} 

\usepackage{mathtools}  
\usepackage{psfrag}       
\usepackage{amsmath,bm,amsfonts,mathrsfs}   
\usepackage{shortcuts}    
\usepackage{color}

\usepackage{hyperref}       
\usepackage{url}            

\usepackage{algorithm,algorithmic}

\usepackage{soul} 

\title{Machine learning materials physics: Multi-resolution  neural networks learn the free energy and nonlinear elastic response of evolving microstructures}

\author{Xiaoxuan Zhang$^1$,~Krishna Garikipati$^{1,2,3}$ \thanks{Corresponding author. E-mail address: krishna@umich.edu \hfill \today} \\[3mm]
  $^1$Department of Mechanical Engineering, University of Michigan, United States \\
  $^2$Department of Mathematics, University of Michigan, United States \\
  $^3$Michigan Institute for Computational Discovery \& Engineering, University of Michigan, United States \\
}

\begin{document}

\maketitle

\begin{abstract}
  Many important multi-component crystalline solids undergo mechanochemical spinodal decomposition: a phase transformation in which the compositional redistribution is coupled with structural changes of the crystal, resulting in dynamically evolving microstructures. The ability to rapidly compute the macroscopic behavior based on these detailed microstructures is of paramount importance for accelerating material discovery and design. Here, our focus is on the macroscopic, nonlinear elastic response of materials harboring microstructure. Because of the diversity of microstructural patterns that can form, there is interest in taking a purely computational approach to predicting their macroscopic response. However, the evaluation of macroscopic, nonlinear elastic properties purely based on direct numerical simulations (DNS) is computationally very expensive, and hence impractical for material design when a large number of microstructures need to be tested. A further complexity of a hierarchical nature arises if the elastic free energy and its variation with strain is a small-scale fluctuation on the dominant trajectory of the total free energy driven by microstructural dynamics.
To address these challenges, we present a data-driven approach, which combines advanced neural network (NN) models with DNS to predict the homogenized, macroscopic, mechanical free energy and stress fields arising in a family of multi-component crystalline solids that develop microstructure. The microstructures are numerically generated by solving a coupled, mechanochemical spinodal decomposition problem governed by nonlinear strain gradient elasticity and the Cahn-Hilliard phase field equation. 
The hierarchical structure of the free energy's evolution induces a multi-resolution character to the machine learning paradigm:
We construct knowledge-based neural networks (KBNNs) with either pre-trained fully connected deep neural networks (DNNs), or pre-trained convolutional neural networks (CNNs) that describe the dominant characteristic of the data to fully represent the hierarchically evolving free energy.
We demonstrate multi-resolution learning of the materials physics; specifically of the nonlinear elastic response for both fixed and evolving microstructures.
\end{abstract}


\section{Introduction}

Mechanochemical spinodal decomposition refers to a continuous phase transformation mechanism due to an onset of instability with respect to the composition and/or a structural order parameter. It occurs in materials systems with a free-energy density that is non-convex in strain-composition space. Wide regimes of the state space lie far from thermodynamic equilibrium, and the resulting first-order dynamics manifests in evolving microstructures that are distinguishable by strain and composition variables \cite{Garikipati2016Rudraraju-NPJ}.
Mechanochemical spinodal decomposition exists in many important multi-component crystalline solids, such as cubic yttria-stabilized zirconia, the lithium-ion battery electrode material Li$_x$Mn$_2$O$_4$, transition metal hydrides and certain two-dimensional materials such as TaS.
In such material systems, as the first-order dynamics is driven by fluxes determined by the local free energy density, the material microstructure, controlled by strain and composition variables, undergoes changes. The macroscopic behaviors and properties are inherently related to the evolving microstructures.  Progress has been made in understanding the detailed dynamics and in modeling the resulting microstructures \cite{Garikipati2016Rudraraju-NPJ,Garikipati2016Sagiyama-Unconditionally}. However, in order to optimize the properties of existing materials and to design new materials, it also is essential to rapidly predict the material's macroscopic response based on the detailed microstructure.

Macroscopic material responses/properties can be measured from well-designed experiments or predicted from physics-based direct numerical simulations (DNS).
Numerical methods to upscale the nonlinear macroscopic behavior of a heterogeneous microstructure are commonly categorized as computational homogenization methods. They necessitate the solution of expensive boundary value problems (BVPs) on representative volume elements (RVEs) that encompass the targeted material microstructures \cite{Geers2010homogenization-trends-challenges,saeb+steinmann+javili16}.
It is impractical, if not impossible, to evaluate macroscopic material properties based on either experimental measurements or DNS when a large number of microstructures need to be tested.

Machine learning has emerged as a powerful approach among data-driven methods, and has been applied to study a wide range of problems in materials physics, such as material screening \cite{Meredig2014-screen-materials,Wolverton2016Ward-screen-material-property,Ramprasad2017-material-informatics-review}, constitutive modeling \cite{Hashash2004-NN-constitutive,Chinesta2019Ibanez-hybrid-constitutive-modeling,Sun+Wang2019-game-constitutive}, scale bridging \cite{Brockherde2017DFT-MD,Garikipati2019Teichert-ML-bridge}, and system identification \cite{Brunton2016Kutz-system-id,Garikipati2019Wang-System-Identification}. 
Interested readers are directed to Refs \cite{Bock2019Kalidindi-ML-CM-review,Dimiduk2018review-ML-on-material-process-structure} for more data-driven examples in the field of materials physics. 
Computational homogenization is yet another successful application of machine learning, where attempts to predict effective material properties \cite{Kalidindi+Cecen2018-CNN-Structure-property,Li2019Zhuang-effective-ME-DNN,Agrawal2018Yang-Composites-S-P-deep-learning,Kondo2017CNN-ionic-conductivity,Rong2019CNN-thermal-conductivity-composites} and non-linear material response \cite{Hambli2011multiscale-bone-with-NN,Bessa2017-data-driven-framework-elasticity-inelasticity,Garikipati2019Sagiyama-ML-Martensitic,Jones2019Frankel-Oligocrytal-behavior-CNN,Sun2018Wang-homogenization,Yvonnet2015Le-RVE-elasticity,Yvonnet2018Lu-NN-RVE-graphene} based on both experimentally and numerically generated data have been made by exploring different data-driven techniques. 
For example, convolutional neural networks (CNNs), which take images of microstructures as inputs, have been used to construct microstructure-property linkages \cite{Kalidindi+Cecen2018-CNN-Structure-property} and to predict macroscopic properties, such as effective ionic conductivity in ceramics \cite{Kondo2017CNN-ionic-conductivity}, effective mechanical properties in composites \cite{Agrawal2018Yang-Composites-S-P-deep-learning} and shale \cite{Li2019Zhuang-effective-ME-DNN}, effective thermal conductivity in composites \cite{Rong2019CNN-thermal-conductivity-composites}, and many others.
Artificial neural networks (ANNs)/deep neural networks (DNNs), which are trained to construct complex nonlinear relationships between predefined features (e.g.\ strain components/volume fraction) and some quantities of interest (e.g.\ averaged stress responses/averaged elastic modulus ), have been coupled with finite element simulations to accelerate multiscale homogenization for bone remodeling \cite{Hambli2011multiscale-bone-with-NN}, nonlinear elastic composites \cite{Yvonnet2015Le-RVE-elasticity}, graphene/polymer nanocomposites with nonlinear anisotropic electrical response \cite{Yvonnet2018Lu-NN-RVE-graphene}, geological materials with multi-porosity \cite{Sun2018Wang-homogenization}, oligocrystals with plastic response \cite{Jones2019Frankel-Oligocrytal-behavior-CNN}, and many others.
Data-driven computational homogenization has demonstrated the potential to drastically reduce computational time in traditional multilevel calculations, making possible the inclusion of detailed microstructural information in this setting  \cite{Yvonnet2015Le-RVE-elasticity,Yvonnet2018Lu-NN-RVE-graphene,Matous2017Review-multiscale-heter-model}.

In this work, a data-driven homogenization approach is explored to jointly predict the mechanical free energy \emph{and} homogenized stress-strain response of a family of multi-component crystalline microstructures that are numerically generated based on the computational framework laid out by Rudraraju et al. \cite{Garikipati2016Rudraraju-NPJ}. In this initial communication, we consider plane strain mechanics to leverage the reduction in feature complexity afforded by the resulting two-dimensional setting.
The physics underlying  mechanochemical spinodal decomposition delivers families of microstructures that are not at thermodynamic equilibrium. As outlined above, these microstructures evolve, driven by the free energy. There is a hierarchical nature to the free energy of this class of material phase transformations: The strain excursions imposed on a microstructure must remain ``small'' in order to prevent further evolution of the microstructure itself, or the elasticity equations drive the free energy out of local basins. The corresponding structural rearrangements could then be large enough that the microstructure itself changes. This would violate the notion of homogenization since the base pattern over which the effective properties are being sought is itself changed.
Consequently, the fluctuations in elastic free energy themselves must remain small, implying that they should be induced by small strains.
Thus, the free energy of each microstructure has a multi-resolution structure with a dominant trajectory from phase transformations that drive evolution of the microstructure, and small-scale fluctuations from strains that explore the effective elastic response of a given microstructure.
The dominant trajectory strongly depends on the microstructural information, such as the volume fraction, the location and orientation of each crystalline phase, and the interfaces.
  Similar multi-resolution data structure also exists in the field of medical diagnosis, where fluctuation of signals is used to detect diseases \cite{Ubeyli2005combined-nn-medical-diagnostic,Hayashi2002combined-nn-medical-diagnosis,Guler2005combined-nn-ECG}. 

  We use knowledge-based neural networks (KBNNs) \cite{Towell1990KBNN,Towell1994KBNN,Garikipati2019Teichert-ML-SurrogateOpt} to represent the response in the multi-resolution information structure. KBNN is a hybrid learning system, in which domain specific knowledge are translated into a NN via the form of hierarchically structured rules \cite{Towell1990KBNN,Towell1994KBNN}, or the form of pre-trained NNs \cite{Garikipati2019Teichert-ML-SurrogateOpt}.  
  Though the knowledge, whose completeness and correctness are not required, provides only approximately correct explanation of a problem, neural learning techniques can be used to train the KBNN to gain improved understanding of the problem and achieve superior performance than randomly initialized NNs \cite{Towell1990KBNN}.
  KBNN is one form of so-called transfer learning that is defined as: ``the ability of a system to recognize and apply knowledge and skills learned in previous tasks to novel tasks'' \cite{Pan2010TransferLearning}.
  In this work, the KBNNs are specifically built upon pre-trained DNNs or CNNs, which describe the dominant part of the free energy, to learn the small-scale fluctuations of free energy and predict homogenized stresses.
It is important to mention that although the term DNN refers to a large family of neural network structures, it will specifically refer to deep neural networks with fully connected layers in this work.
Our studies demonstrate that multi-resolution neural networks using both DNN-based and CNN-based KBNN models can accurately learn the macroscopic mechanical behavior of a single microstructure. 
Furthermore, CNN-enhanced KBNN models are capable of learning the macroscopic mechanical behavior of many microstructures from different DNSs.
Such KBNN models for multi-resolution learning and testing can be used to rapidly screen materials based on their microstructures for applications such as additive manufacturing, polymer blending, or materials synthesis.    

The rest of the paper is organized as follows.
In Section~\ref{sec:spinodal-framework}, we summarize the mathematical description of mechanochemical spinodal decomposition, and the computational framework that is used to generate different microstructures in materials undergoing this class of phase transformations.
The neural network (NN) model structures used in this work are presented in Section~\ref{sec:NN}.
Section~\ref{sec:data} covers the procedures of data generation, feature selection, and hyperparameter searches.
The detailed simulation results are presented in Section~\ref{sec:num-example}.
Concluding remarks and perspectives are offered in Section~\ref{sec:conclusion}.

\section{Mechanochemical spinodal decomposition} \label{sec:spinodal-framework}
We first summarize the mathematical framework of mechanochemical spinodal decomposition. Interested readers are directed to Ref. \cite{Garikipati2016Rudraraju-NPJ} for details.

\subsection{Free energy density function}
\begin{figure}[t!]
  \centering
  \includegraphics[width=1.0\linewidth]{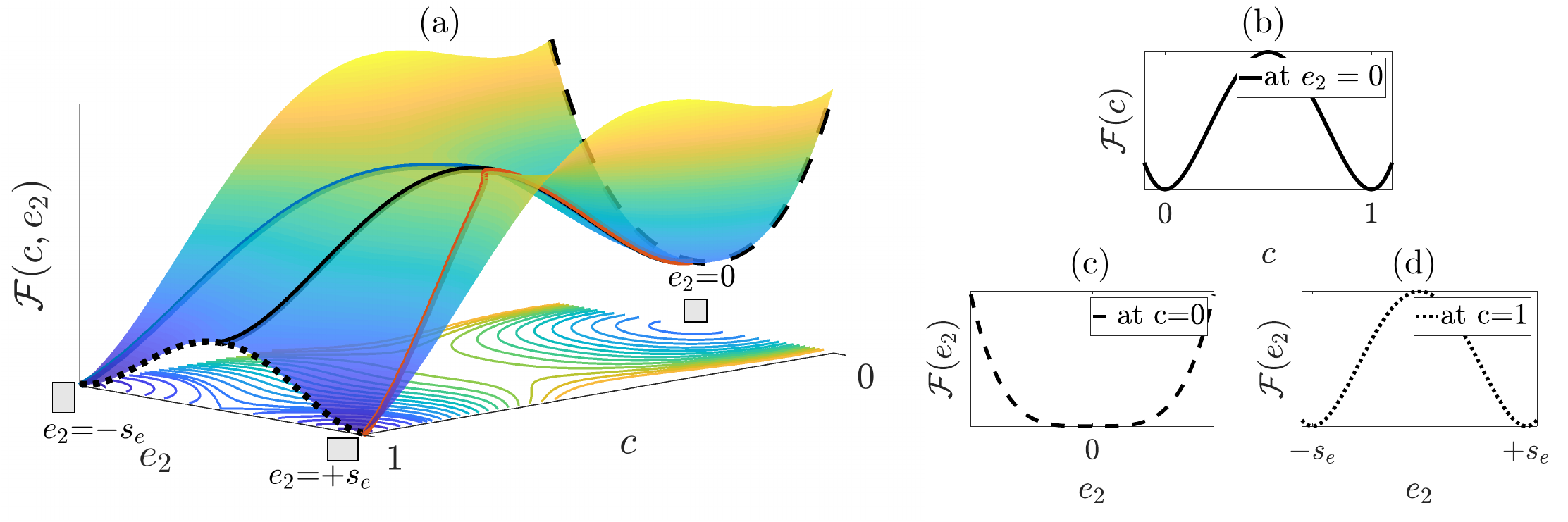}
  \caption{Illustration of (a) the homogeneous free energy density over the reduced strain-composition space in the low temperature phase. 
  The red and blue lines indicate two energetic paths, each of which has local minima. The minima of the red path correspond to the square or positive rectangular phases. The minima of the blue path correspond to the square or negative rectangular phases.
    (b) The chemical part of $\scrF$ has a double-well shape with respect to $c$, indicating a composition triggered phase transformation. This is traced out by the black path in (a). 
    (c) The mechanical part of $\scrF$ has a convex shape at $c=0$, indicating a stable square phase, and (d) a double-well shape at $c=1$, indicating a deformation triggered phase transformation into the rectangular phases.
    The corresponding locations of the free energy paths in (b,c,d) are shown in (a).
}
  \label{fig:free-energy}
\end{figure}

  In this work, we focus on mechanochemical spinodal decomposition in plane strain. We work within the effectively two-dimensional setting that results, and  defer consideration of the full complexity of three-dimensional microstructures to a later communication. Mechanochemical spinodal decomposition gives rise to coupled diffusional/martensitic phase transformations. In the two-dimensional setting of plane strain, the phenomenon manifests as a solid that has a single square phase at high temperature and undergoes a square-to-rectangle structural transformation at low temperature. These structures can be understood as resulting from the restriction to vanishing out-of-plane strains in the well-known three-dimensional cubic-to-tetragonal transformation.
  The square lattice is the high symmetry phase that serves as the reference state for strain measurement. 
  Here, the Green-Lagrange strain tensor $\BE$ is used,\footnote{Recall that $\BE = \frac{1}{2}(\BF^\text{T}\BF-\boldsymbol{1})$, where the deformation gradient is $\BF = \boldsymbol{1}+\partial\Bu/\partial\BX$, and $\Bu$ is the displacement vector.} with its relevant in-plane components denoted as $E_{11}, E_{22}, $ and $E_{12}$ (=$E_{21}$).
  The low-symmetry rectangular lattices are derived from the square lattice by homogeneous strain.
  For describing the structural changes, it is more convenient to introduce three reparameterized strains, which are based on the components of $\BE$ and defined as $e_1 = (E_{11} + E_{22})/\sqrt{2}$, $e_2 = (E_{11} - E_{22})/\sqrt{2}$, and $e_3 = \sqrt{2}E_{12}$. Here, $e_1$ and $e_3$ represent the dilatation and shear strain, respectively, in the infinitesimal strain limit. 
  The reparameterized strain $e_2$ uniquely distinguishes the square lattice (when $e_2 = 0$) and its two rectangular variants: the ``positive'' rectangle ($e_2>0$) with elongated lattice in the global $X_1$ direction and the negative rectangle ($e_2<0$) with elongated lattice in the global $X_2$ direction. It thus serves as a structural order parameter. The reader is directed to Rudraraju et al. \cite{Garikipati2016Rudraraju-NPJ} for details of the systematic reduction from three dimensions yielding the above simple formulas for $e_1, e_2$ and $e_3$.\footnote{Note that $e_3$ in this communication is the reparameterized strain $e_6$ in Ref. \cite{Garikipati2016Rudraraju-NPJ}.}
  The composition $c$, which varies between 0 and 1, is the  order parameter controlling the chemistry, with $c\sim0$ denoting the composition state with the stable square phase and $c\sim1$ denoting the composition state with two unstable rectangular phases, as illustrated in Fig.~\ref{fig:free-energy}.

At low temperature, the coupled diffusional and structural phase transformation is triggered by instabilities with respect to both the compositional parameter $c$ and the structural order parameter $e_2$.
This coupled phase transformation can be described by a non-convex free energy density function $\psi$, 
  \begin{equation}
    \psi (c, \Be, \nabla c, \nabla\Be) = \scrF(c, \Be) + \scrG(\nabla c, \nabla \Be),
  \label{eq:general-free-energy}
  \end{equation}
  with $\scrF$ representing a homogeneous contribution from both composition and strain as illustrated in a reduced $e_2-c$ space in Fig.~\ref{fig:free-energy}, and $\scrG$ being a gradient-dependent, inhomogeneous contribution to regularize the free energy density.
  In \eref{eq:general-free-energy}, $\Be$ is a vector with $e_1$, $e_2$, and $e_3$ as its components.
  In the DNS, the following specific forms of the homogeneous and inhomogeneous components of $\psi$ are used to generate two-dimensional microstructures 
   \begin{subequations}\label{eq:2d-psi}
     \begin{alignat}{2}
      \scrF(c,\Be)
     & =  \underbrace{16 d_c c^4 - 32 d_c c^3 + 16 d_c c^2}_\text{\tiny{chemical}}+ \underbrace{\frac{2d_e}{s_e^2}(e_1^2 + e_3^2) + \frac{d_e}{s_e^4}e_2^4}_\text{\tiny{mechanical}} + \underbrace{(1-2c) \frac{2d_e}{s_e^2}e_2^2}_\text{\tiny{mechanochemical}} \label{eq:2d-F}\\
    \scrG(\nabla c, \nabla \Be)
    & = \underbrace{\frac{1}{2} \nabla c \cdot \kappa \nabla c}_\text{\tiny{chemical}} + \underbrace{\frac{1}{2} \nabla e_2 \cdot \lambda_e \nabla e_2}_\text{\tiny{mechanical}} \label{eq:2d-G}
     \end{alignat}
   \end{subequations}
   where $d_c$, $d_e$, $s_e$, $\kappa$, and $\lambda_e$ are material parameters with $d_c=2.0$, $d_e=0.1$, $s_e=0.1$, $\kappa=10^{-6}$, and $\lambda_e=10^{-6}$ \cite{Garikipati2016Rudraraju-NPJ}.
   As shown, the homogeneous free energy density function $\scrF$ in \eref{eq:2d-F} consists of contributions from chemistry, mechanics and coupled mechano-chemistry, while the gradient-dependent $\scrG$ has distinct chemical and mechanical contributions.

\subsection{Governing equations}

Based on a generalized, Landau-type free energy density function in \eref{eq:general-free-energy} that couples strain and composition instability, mechanochemical spinodal decomposition can be described by a set of equations that couple the classical Cahn-Hilliard formulation and nonlinear gradient elasticity.
The non-equilibrium chemistry in this coupled system is governed by
  \begin{equation}
    \begin{aligned}
      \frac{\partial c}{\partial t} + \nabla \cdot \BJ & = 0 
      \quad \text{with} \quad
      \BJ= -\BL(c, \Be) \nabla \mu \\ 
    \end{aligned}
    \label{eq:govern-chemistry}
  \end{equation}
  where $\BL$ is a transport tensor related to mobility. 
  In \eref{eq:govern-chemistry}, $\mu$ is the chemical potential, which is obtained as a variational derivative of \eref{eq:general-free-energy}
  \begin{equation}
    \begin{aligned}
      \mu & = \frac{\partial \scrF}{\partial c} + \frac{\partial \scrG}{\partial c} - \nabla \cdot \left[ \frac{\partial \scrG}{\partial(\nabla c)} \right]. \\ 
    \end{aligned}
    \label{eq:mu}
  \end{equation}
  On substituting \eref{eq:2d-F} and \eref{eq:2d-G} in \eref{eq:mu}, and then back in \eref{eq:govern-chemistry}, the fourth-order mechanochemically coupled extension of Cahn-Hilliard dynamics becomes clear. Mechanical equilibrium in the setting of strain gradient elasticity (most transparently written in coordinate notation) is governed by:
  \begin{equation}
    \begin{aligned}
      P_{iJ,J} - B_{iJK,JK} & = 0 
    \label{eq:govern-mechanical}
    \end{aligned}
  \end{equation}
  where $\BP$ and $\BB$ are the stress tensors, conjugate to the deformation gradient $\BF$ and the gradient of the deformation gradient $\nabla \BF$, respectively,  whose forms are given as  
  \begin{equation}
    \begin{aligned}
      P_{iJ} & = \sum_\alpha \frac{\partial \scrF}{\partial e_\alpha} \frac{\partial e_\alpha}{\partial F_{iJ}}
      + \sum_a \frac{\partial \scrG}{\partial e_{\alpha,I}} \frac{\partial e_{\alpha,I}}{\partial F_{iJ}} \\ 
      B_{iJK} & = \sum_a \frac{\partial \scrG}{\partial e_{\alpha,I}} \frac{\partial e_{\alpha,I}}{\partial F_{iJ,K}}. \\ 
    \end{aligned}
  \end{equation} 
  See Refs \cite{Toupin1964,Rudraraju2014-IGA-grad-elasticity,Garikipati2016Rudraraju-NPJ,Garikipati2016Sagiyama-Unconditionally,Garikipati2016Wang-Toupin,Garikipati2018Sagiyama-Martensitic} for details. With appropriate initial conditions and boundary conditions, the composition and deformation fields are obtained by solving equations \eref{eq:govern-chemistry} and \eref{eq:govern-mechanical}. Our implementation is with the \texttt{mechanoChemIGA} code\footnote{Code available at \href{https://github.com/mechanoChem/mechanoChemIGA}{github.com/mechanoChem/mechanoChemIGA}} , which is a publicly available and highly parallelized multiphysics code developed based on \texttt{PETSc} \cite{petsc-efficient, petsc-user-ref}, \texttt{Trilinos} \cite{Trilinos2005, Trilinos-Overview}, and \texttt{PetIGA} \cite{PetIGA} libraries within the Isogeometric Analysis (IGA) framework.

\subsection{Homogenized mechanical properties for heterogeneous microstructures}
The microstructures obtained by solving \eref{eq:govern-chemistry} and \eref{eq:govern-mechanical} are highly heterogeneous, as illustrated in Fig.~\ref{fig:dns-results}.
To describe their macroscopic mechanical responses, the averaged deformation gradient $\BF^\text{avg}$ and the total mechanical free energy $\Psi_\text{mech}$ are used, which are computed as 
\begin{equation}
  \BF^\text{avg} = \frac{1}{\bar{V}} \int_{\Omega} \BF~dV \quad \text{and} \quad \Psi_\text{mech} = \int_{\Omega} \psi_\text{mech}(c, \Be, \nabla\Be)~dV,
  \label{eq:avg-F-psi}
\end{equation}
with $\Omega$ and $\bar{V}$ representing the domain of interest and the total volume of the domain, respectively.
In \eref{eq:avg-F-psi}, $\psi_\text{mech}(c, \Be, \nabla\Be)$ is the total elastic free energy density that consists of the purely mechanical and mechanochemical terms in \eref{eq:2d-F} and \eref{eq:2d-G} as
\begin{equation}
  \psi_\text{mech}(c, \Be, \nabla\Be) =
  \frac{2d_e}{s_e^2}(e_1^2 + e_3^2) + \frac{d_e}{s_e^4}e_2^4 + \frac{1}{2} \nabla e_2 \cdot \lambda_e l_e^2 \nabla e_2 
 + (1-2c) \frac{2d_e}{s_e^2}e_2^2.
  \label{eq:2d-psi-mech}
\end{equation}
The macroscopic first Piola-Kirchhoff stress tensor $\BP^\text{avg}$ is computed as   
\begin{equation}
  P^\text{avg}_{iJ} = \frac{1}{\bar{A}_J} \int_{\Gamma} P_{iK}N_{K}~dA_{J} 
  \label{eq:avg-P}
\end{equation}
by averaging the surface traction components ($T_i=P_{iK}N_K$)  on a given surface $\Gamma \subset\partial\Omega$ with unit outward normal $\BN$  \cite{Garikipati2019Sagiyama-ML-Martensitic}, where $\bar{A}_J$ represents the area of the corresponding surface.

\section{Neural networks} \label{sec:NN}
In this section, the architectures of DNNs, CNNs, and KBNNs to be used in Section~\ref{sec:num-example} are laid out. 

\subsection{DNN} \label{sec:NN-dnn}
A DNN consists of multiple layers with one input layer, one output layer, and several hidden layers in between.   
The inputs and outputs are called features and labels, respectively.
The optimal architecture of a DNN for a specific problem is unknown \emph{a priori}. 
Users need to select the type and structure of each layer and the number of hidden layers.
In this work, DNNs specifically refer to neural networks made up of fully connected (FC) layers, to distinguish from CNNs discussed in Section~\ref{sec:NN-cnn}.
A FC layer consists of multiple neurons, which take a group of weighted values and a bias as inputs and return the output by applying an activation function to their summation.
In DNNs, the weights and biases are variables subject to global optimization.
The architecture of DNNs is determined by the total number of hidden layers and the number of neurons per layer, which are collectively referred to as ``hyperparameters''.

\subsection{CNN}\label{sec:NN-cnn}

A CNN is a versatile type of neural network developed originally to analyze image data for tasks such as pattern detection or feature selection \cite{Krizhevsky2012imagenet}. 
As discussed in the introduction, it has recently become a very useful tool for the study of material microstructure-property relationships in situations where data from both experiments and computational materials physics simulations are available as easily visualizable images.  
A CNN often is a mixture of convolutional layers, pooling layers, and FC layers.
It can significantly reduce the dimensionality of the representation. A CNN typically requires  far fewer variables than a DNN with only FC layers does for the same task.
The structure of a convolutional layer is defined by hyperparameters, such as the size and number of filters, choices of paddings, and the stride numbers.
In a convolutional layer, the biases and the kernels of filters are variables subject to global optimization.
A pooling layer has the filter size, paddings, and stride number as hyperparameters but with no global variables.

\subsection{KBNN} \label{sec:NN-kbnn}

\begin{figure}[t]
  \centering
  {\small
  \includegraphics[width=0.45\textwidth]{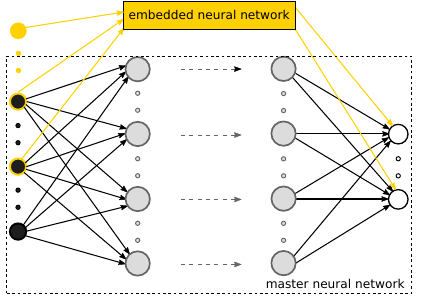}}
  \caption{Illustration of the architecture of a KBNN that consists of a master neural network (MNN) and an embedded neural network (ENN). 
    The MNN and ENN could share some of the features and also have their own features.
    The ENN is pre-trained; i.e., its variables, e.g.\ weights, biases, or kernels are not re-trained with the MNN.}
  \label{fig:kbnn}
\end{figure}

A KBNN utilizes information from pre-trained models, as illustrated in Fig.~\ref{fig:kbnn}.
Whether or not to use a KBNN depends on the nature of the available data.
For example, when the available data include abundant, less accurate data as well as expensive, scarce, highly accurate data, one can use the so-called multi-fidelity model.
  One such example is given in \cite{Garikipati2019Teichert-ML-SurrogateOpt}, where a surrogate model was constructed to represent the total energy of precipitates. In that work, a DNN, with some predefined features as the input and the energy of the precipitates as the label, was first trained with abundant low-fidelity data that were generated from less time-consuming DNSs with coarse meshes. A KBNN was then built upon the pre-trained, low-fidelity DNN. The KBNN contained additional hidden layers and neurons, which were further trained with fewer high-fidelity data generated from computationally expensive DNSs with finer meshes.
The pre-trained DNN itself might have inaccurate prediction of the energy of precipitates. 
But the accuracy of the KBNN prediction was improved as greater expressiveness was gained for its representation from the high-fidelity data.
Such an approach can significantly reduce the required amount of expensive and high-fidelity data/simulations, but still achieve the desired model accuracy.
The data itself may also have a multi-resolution structure, for which one neural network may be incapable of capturing all the information.
In such a scenario, one NN can  be trained first to describe the dominant characteristics of the data. Next, a KBNN can be built upon this pre-trained model with other free variables to be trained on the same dataset. The additional variables are used to resolve other details in the data, not well-delineated by the pre-trained model. 
In this work, the main neural network of the KBNN is named the master neural network (MNN), and the pre-trained neural network is called the embedded neural network (ENN). 
The variables in the MNN need to be optimized, whereas those in the ENN are untrainable;
i.e., variables in the ENN are fixed while training the MNN.

\textbf{Remark 1:}
For NNs, their global parameters are optimized via a back-propagation algorithm during the training process to drive down a loss function.
The hyperparameters, which define the optimal architecture of NNs, need to be chosen by a separate process that usually involves cross-validation.
For a given NN architecture, one further needs to adjust the learning rate to obtain the optimal weights and biases.
A full-fledged discussion on avoidance of model underfitting or overfitting is beyond the scope of this work. 

\textbf{Remark 2:}
The open source library \texttt{TensorFlow} \cite{tensorflow2015-whitepaper} has been used to create different neural network structures in this work.
When NNs are used to learn a mathematical relationship with a unique physical meaning, the NNs are considered accurate only when both the label(s) and other physically meaningful quantities, usually involving the derivatives of the label(s), are accurate. 
For example, a DNN with fully connected layers is trained to learn the free energy density function of a Neo-Hookean hyperelastic material in \cite{Garikipati2019Sagiyama-ML-Martensitic}. 
For such problem, a NN is required not only to accurately represent the free energy function, but also its derivatives with respect to its features. 
In that specific problem, the features are the strain components and the derivatives of the NN are the stress fields.
In this work, we evaluate the performance of NNs primarily based on the loss function, but also consider their derivatives whenever necessary.
The standard automatic differentiation API from \texttt{TensorFlow} is utilized to compute the derivatives of NNs.

\section{Data generation, feature selection, and hyperparameter search} \label{sec:data}
In this section, we first present detailed simulation procedures to generate synthetic microstructures based on the computational framework outlined in Section~\ref{sec:spinodal-framework}.
Next, several pre-defined features for DNNs to be used in Section~\ref{sec:num-example} are discussed.
The hyperparameter search procedure for DNNs, CNNs, and KBNNs is covered in Section~\ref{sec:hyperparameter}.

\subsection{Microstructure generation}
\begin{figure}[t]
  \centering
  \subfloat[Setup]{\includegraphics[trim={0cm 0cm 0cm 0cm}, clip, width=0.25\linewidth]{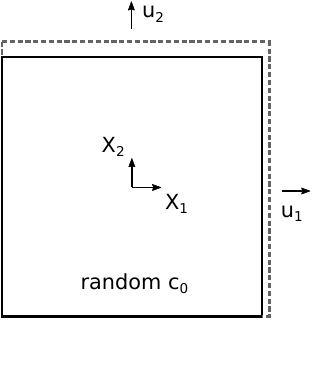}}
  \hspace{1.5cm}
  \subfloat[Total free energy $\Psi$]{\includegraphics[trim={0cm 0cm 0cm 0cm}, clip, height=48mm]{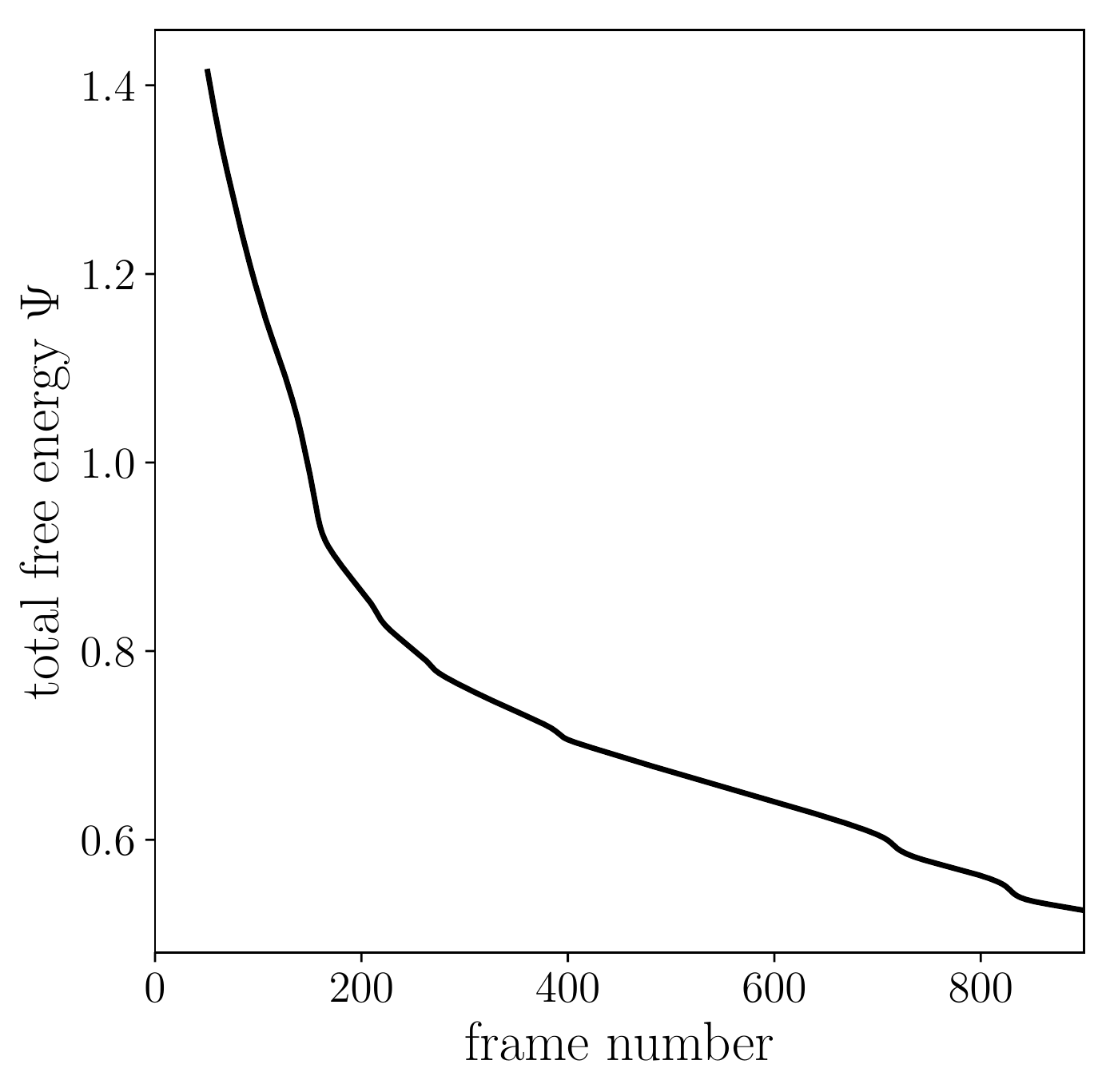}}
  \caption{(a) Illustration of the setup of DNSs with imposed Dirichlet boundary conditions, $u_1, u_2$ on the displacement, zero flux boundary conditions and a random initial concentration, $c_0$. (b) Evolution of the total free energy $\Psi$.}
  \label{fig:dns-setup-psi}
\end{figure}

\begin{figure}[ht]
  \centering
  \subfloat[$c$ at step 51]{\includegraphics[trim={15cm 3.5cm 15cm 3.5cm}, clip, width=0.24\linewidth]{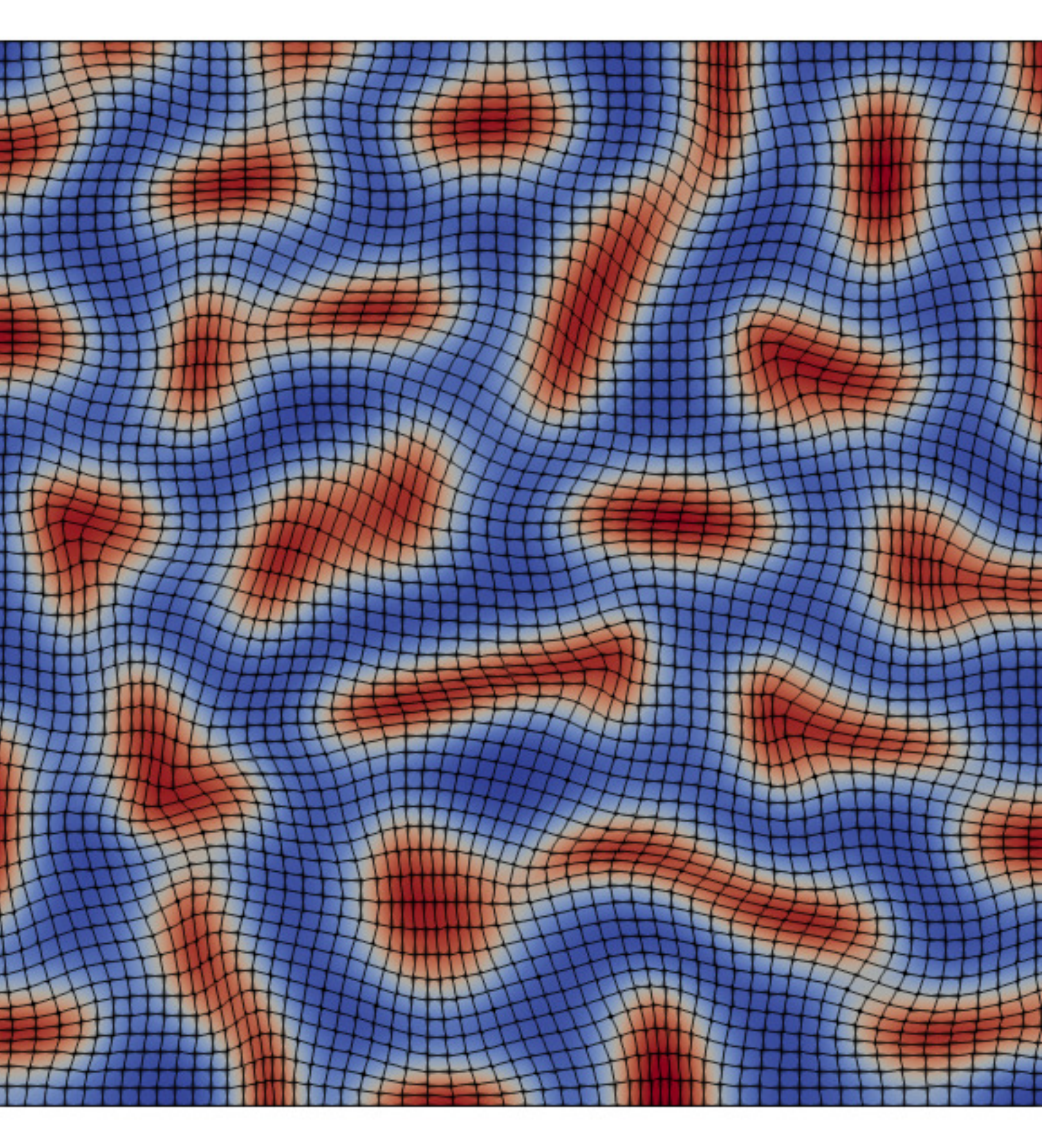}} \hfill
  \subfloat[$c$ at step 150]{\includegraphics[trim={15cm 3.5cm 15cm 3.5cm}, clip, width=0.24\linewidth]{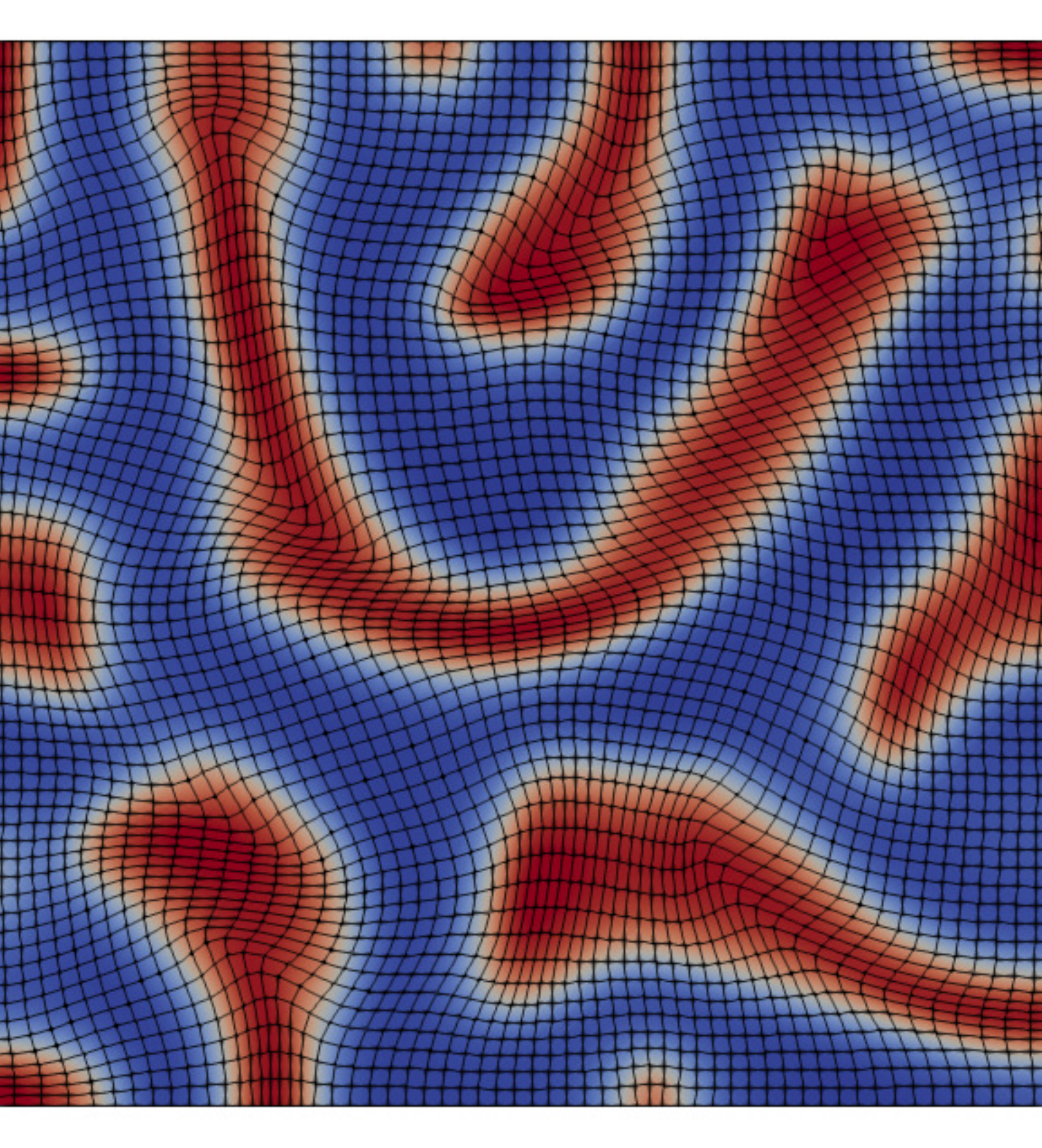}} \hfill
  \subfloat[$c$ at step 400]{\includegraphics[trim={15cm 3.5cm 15cm 3.5cm}, clip, width=0.24\linewidth]{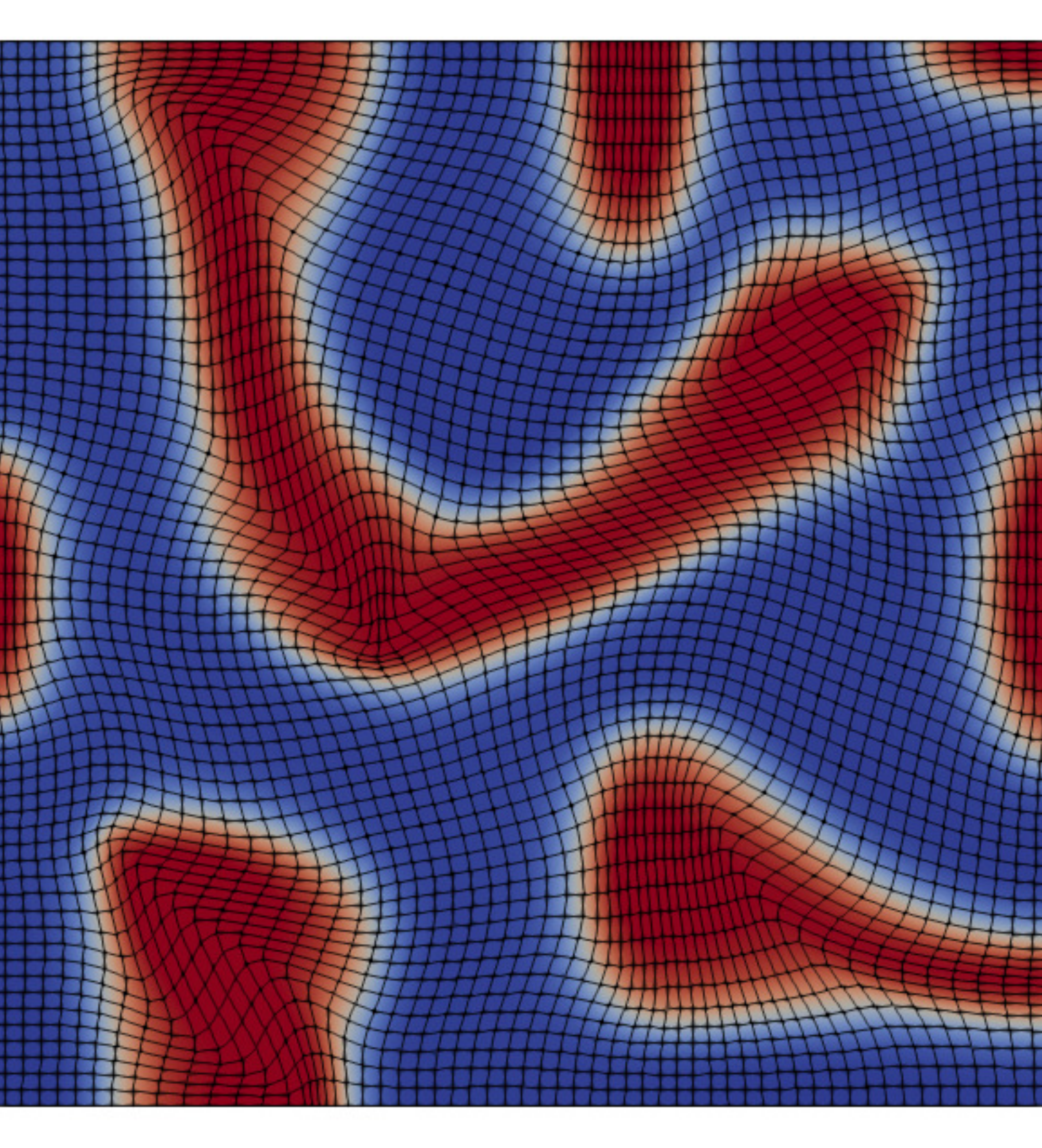}} \hfill
  \subfloat[$c$ at step 900]{\includegraphics[trim={15cm 3.5cm 15cm 3.5cm}, clip, width=0.24\linewidth]{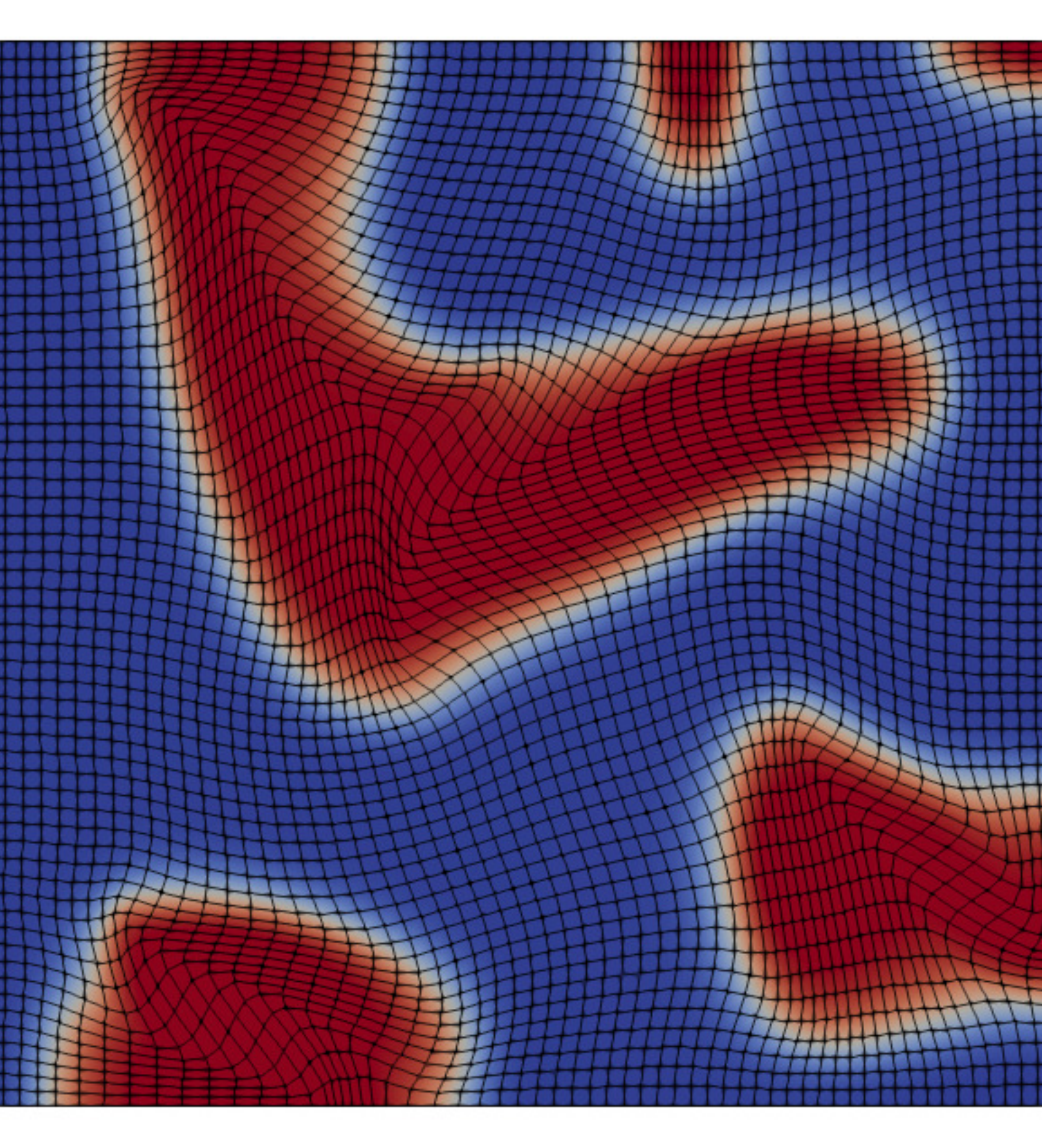}} \\
  \subfloat[$e_2$ at step 51]{\includegraphics[trim={15cm 3.5cm 15cm 3.5cm}, clip, width=0.24\linewidth]{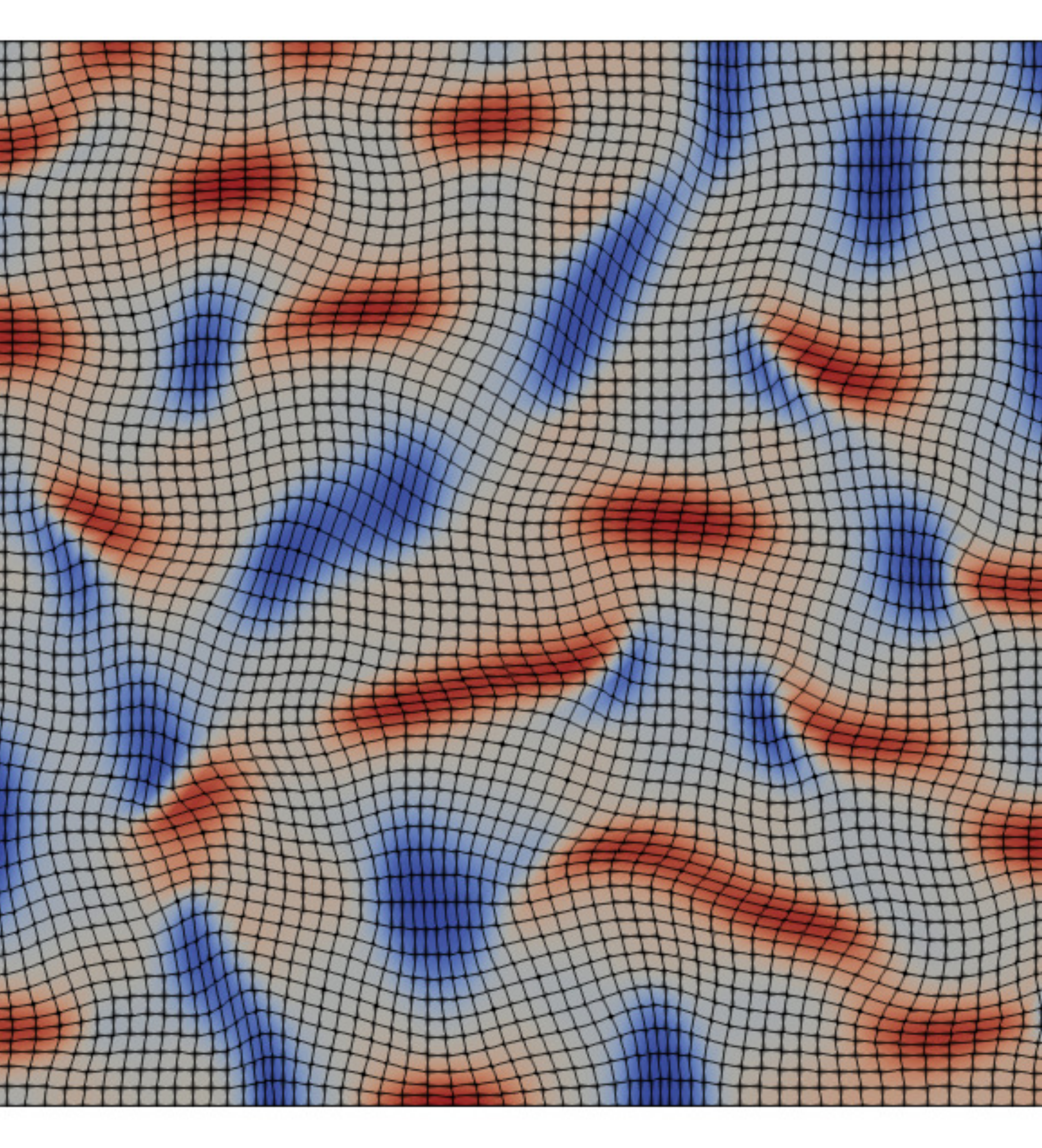}} \hfill
  \subfloat[$e_2$ at step 150]{\includegraphics[trim={15cm 3.5cm 15cm 3.5cm}, clip, width=0.24\linewidth]{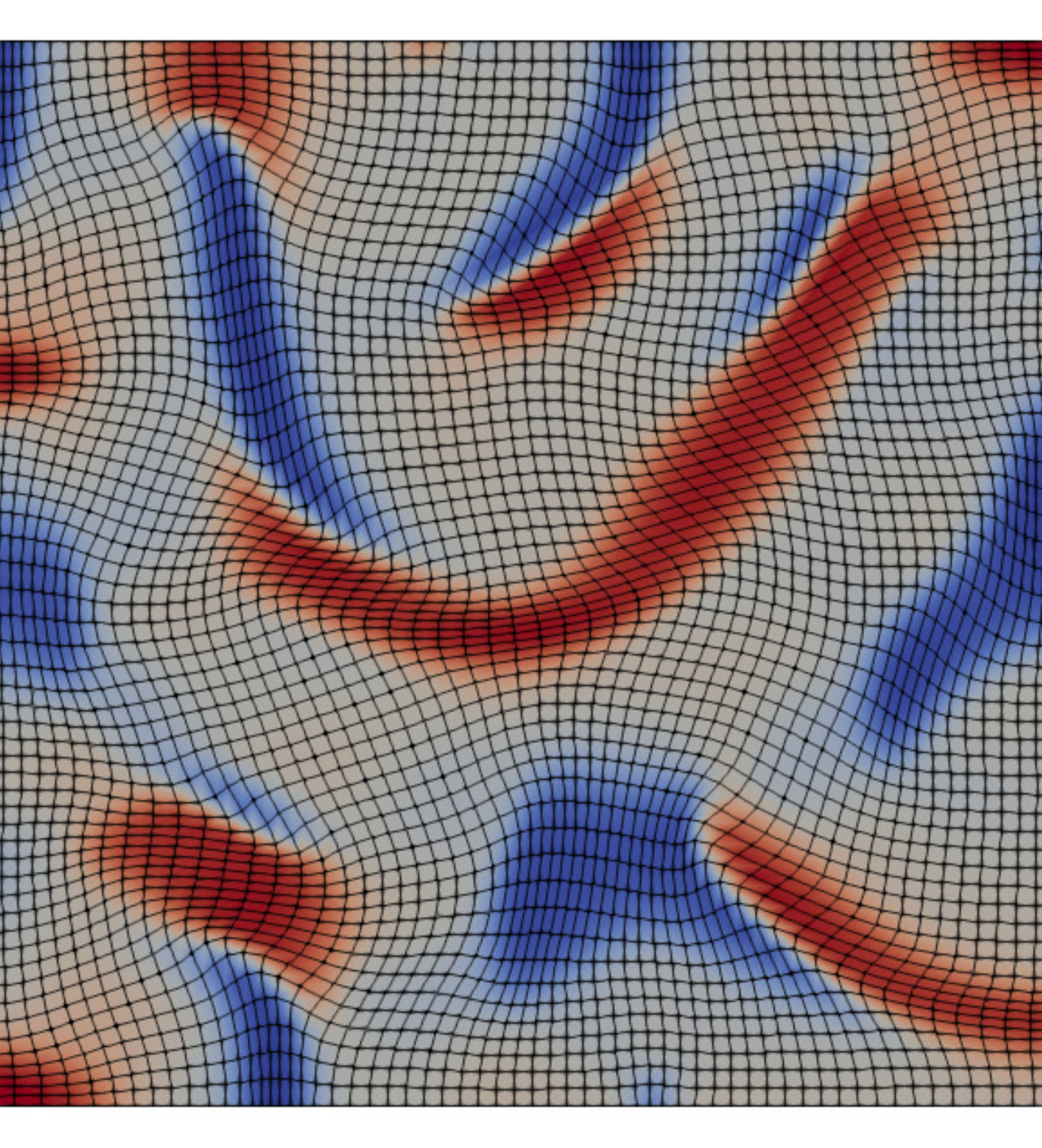}} \hfill
  \subfloat[$e_2$ at step 400]{\includegraphics[trim={15cm 3.5cm 15cm 3.5cm}, clip, width=0.24\linewidth]{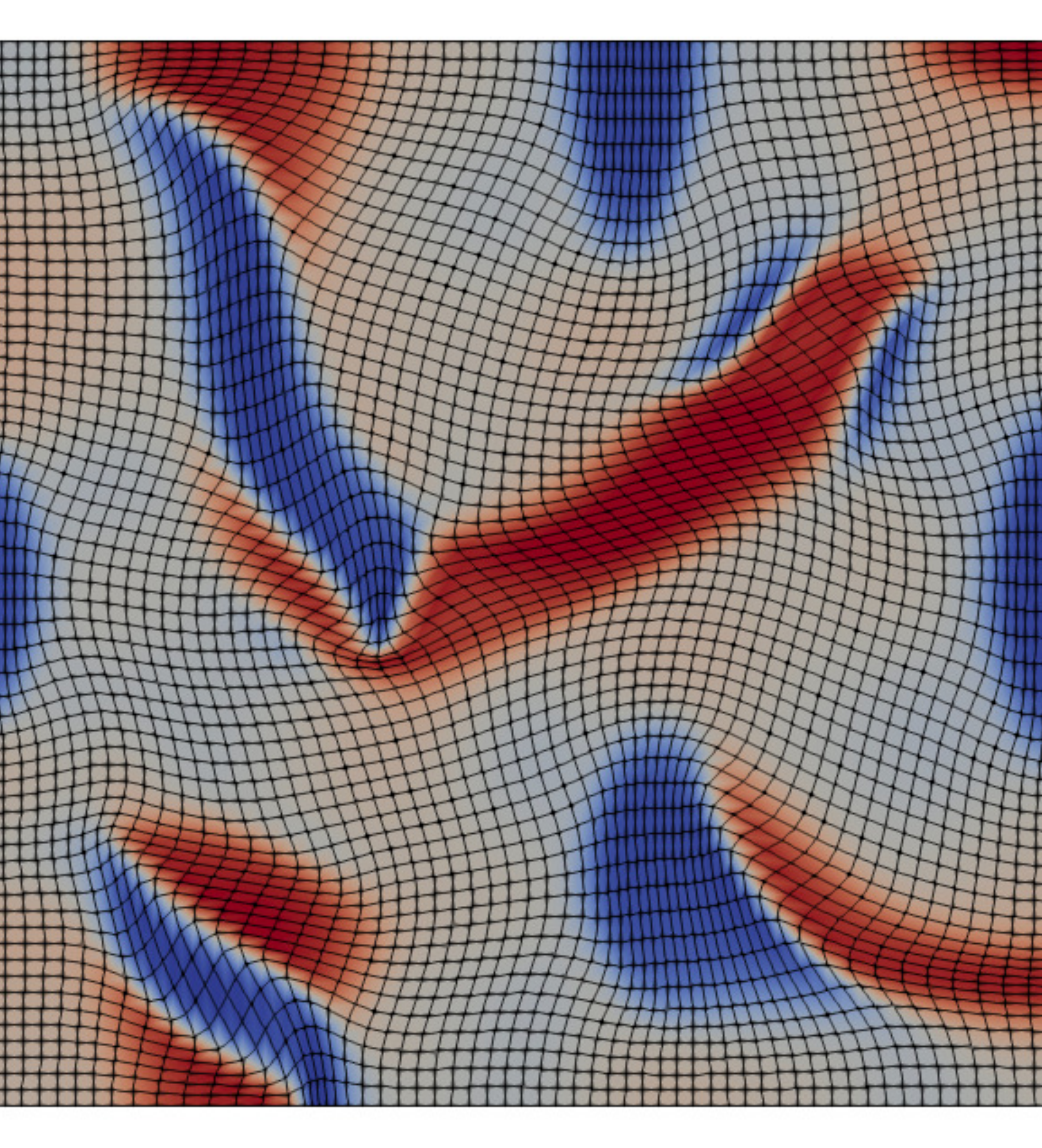}} \hfill
  \subfloat[$e_2$ at step 900]{\includegraphics[trim={15cm 3.5cm 15cm 3.5cm}, clip, width=0.24\linewidth]{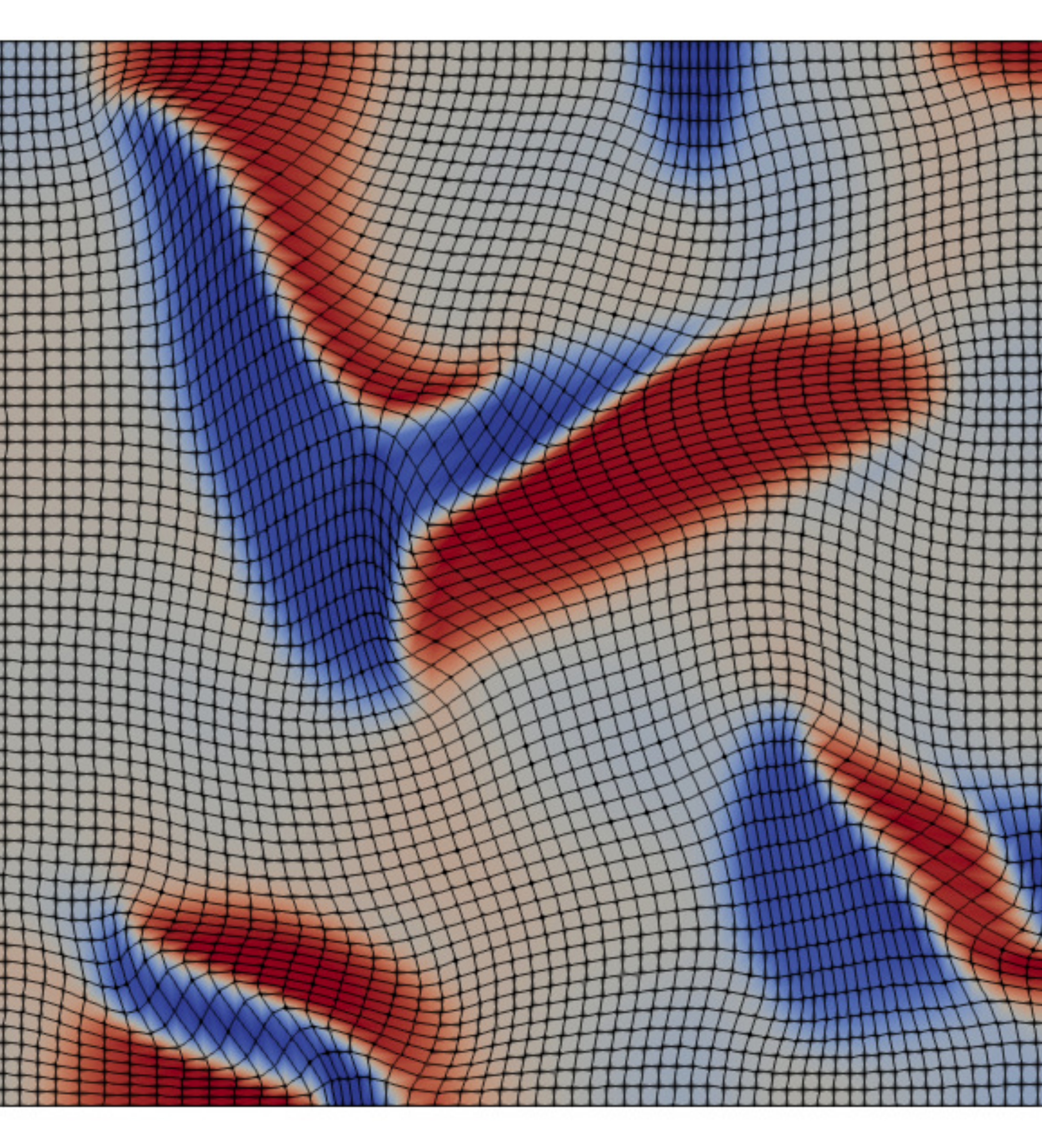}}
  \caption{Snapshots from one of the DNSs in the deformed configuration (scaled $10\times$ to make distortion discernible) at different time steps.
  (a-d) Composition field with red color for $c=1$ and blue color for $c=0$.
  (e-h) $e_2$ field with red color for $e_2=0.1$ (``positive'' rectangle phase) and blue color for $e_2=-0.1$ (``negative'' rectangle phase). 
In the region where $c=0$, $e_2$ has a value of $0.0$, corresponding to the square phase.}
  \label{fig:dns-results}
\end{figure}

A solid subjected to plane strain conditions in a two-dimensional domain $\Omega = (0,0.01)\times (0,0.01)$  is studied with a mesh size of $60\times60$.
  The solid is loaded by a steady biaxial Dirichlet boundary conditions, as shown in Fig.~\ref{fig:dns-setup-psi}(a). 
  The solid has a randomly fluctuating initial composition in the range of $c=0.46 \pm 0.05$ with a uniform initial $e_2$ field ($e_2=0$), which corresponds to a single square phase that exists at high temperature.
  Zero chemical flux boundary conditions are applied to the solid.
  The solid is governed by a non-convex free energy density, as illustrated in Fig.~\ref{fig:free-energy} and expressed in \eref{eq:2d-psi}, which describes the behavior of microstructures at low temperature.
  This initial and boundary value problem resembles a quenching process, during which mechanochemical spinodal decomposition occurs, and the microstructure evolves from a single square phase to one with coexistence of square/rectangle phases.
  The biaxial Dirichlet boundary conditions remain unchanged during the mechanical spinodal decomposition, which is run for 900 simulation time steps.

We conducted 20 phase evolution DNSs, with each of them starting from different random initial compositions and mechanical boundary conditions.
For each DNS, the values of the imposed $u_1$ and $u_2$ are in the range of $[-1.0 \times 10^{-5}, 3.0 \times 10^{-5}]$, which is equivalent to a value of the Green-Lagrange strain tensor components $E_{11} $ and $E_{22}$ approximately in the range of $[-1.0 \times 10^{-3}, 3 \times 10^{-3}]$. Such biaxial loading results in a value of $E_{12}~(\text{or}~E_{21})$ approximately in the range of $[-2 \times 10^{-3}, 2 \times 10^{-3}]$.
Throughout each DNS of phase evolution, the total free energy of the solid and its mechanical part are driven by the second law of thermodynamics.\footnote{If the mechanical boundary conditions do no incremental work during the phase evolution, and if boundary fluxes vanish, the coupling of the first-order Cahn-Hilliard dynamics and gradient elasticity obeys the second law of thermodynamics, and the total free energy decreases. However, the use of time-varying Dirichlet boundary conditions on the mechanics translates to work done on the system, and the free energy may increase.}
Results from one of the 20 DNSs are shown in Figs.~\ref{fig:dns-setup-psi}(b) and~\ref{fig:elastic-free-energy-ossilcation}(a).
Selected snapshots of the composition $c$ and the strain order parameter $e_2$ at different states from this particular simulation are shown in Fig.~\ref{fig:dns-results}, in which the coexistence of the square phase, the positive rectangle phase, and the negative rectangle phase is observed.

Each DNS takes 900 time steps.  We refer to the solution at each time step as a frame.
The homogenized deformation gradient $\BF^\text{avg}$ in \eref{eq:avg-F-psi}, homogenized first Piola-Kirchhoff stress $\BP^\text{avg}$ in \eref{eq:avg-P}, and total mechanical free energy {$\Psi_\text{mech}$} in \eref{eq:avg-F-psi} are computed for each frame of every DNS. 
Since each frame has a different volume ratio and different spatial distribution of the three phases, it is considered as a unique microstructure, whose effective mechanical behavior differs from those of the other microstructures.
Thus, each DNS generates 900 microstructures.
We discard the first 50 frames of each simulation, as phase separation, with well-defined interfaces, is not yet fully developed at this stage.
In this work, $17000$ microstructures have been generated from the 20 DNSs.

\subsection{Microstructure feature selection} \label{sec:feature-selection}

\begin{figure}[t]
  \centering
  \subfloat[$l_s^r$ ]{\includegraphics[trim={19cm 5.0cm 19cm 5.0cm}, clip, width=0.32\linewidth]{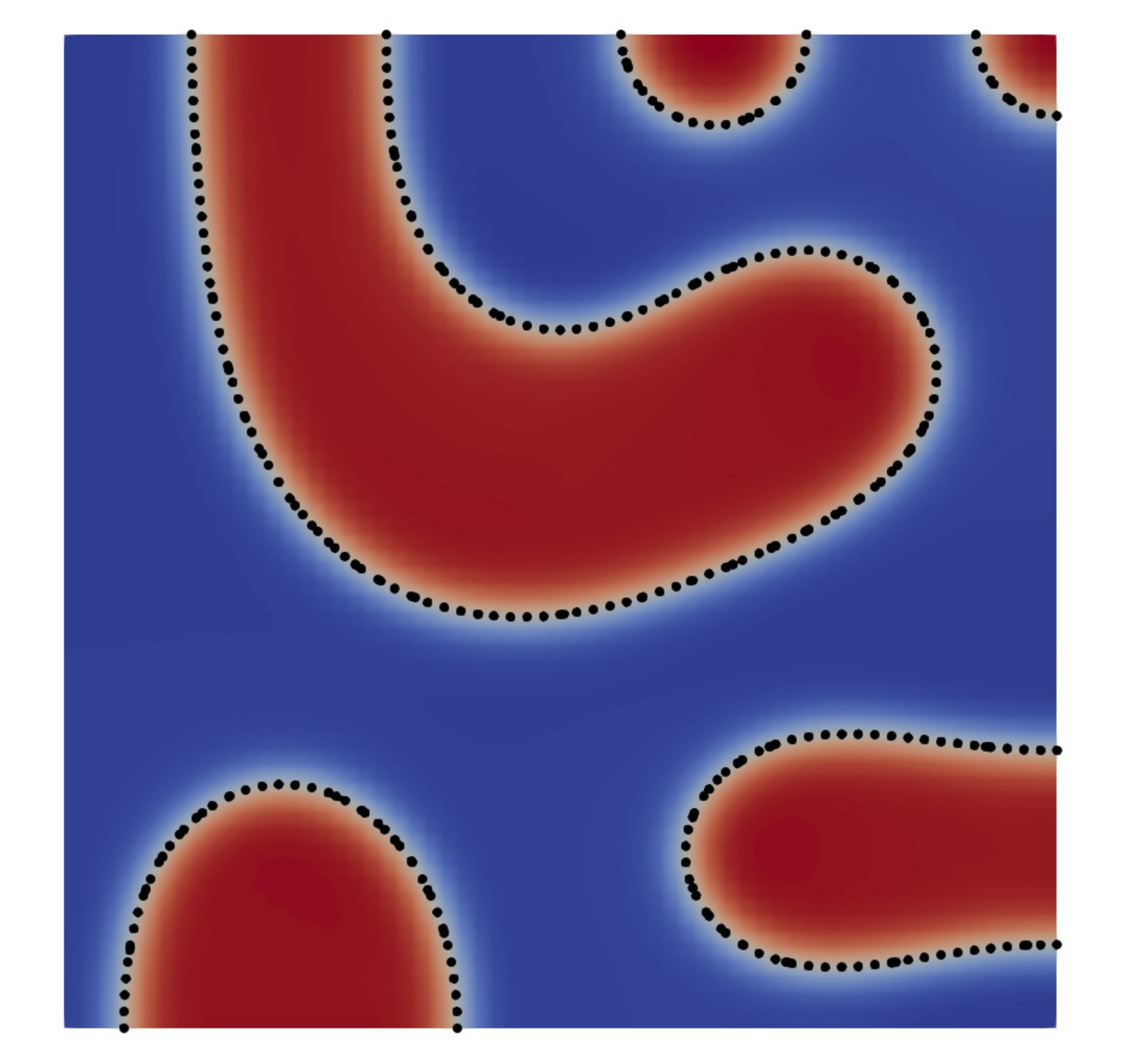}}
  \subfloat[$l^{r+}$ ]{\includegraphics[trim={19cm 5.0cm 19cm 5.0cm}, clip, width=0.32\linewidth]{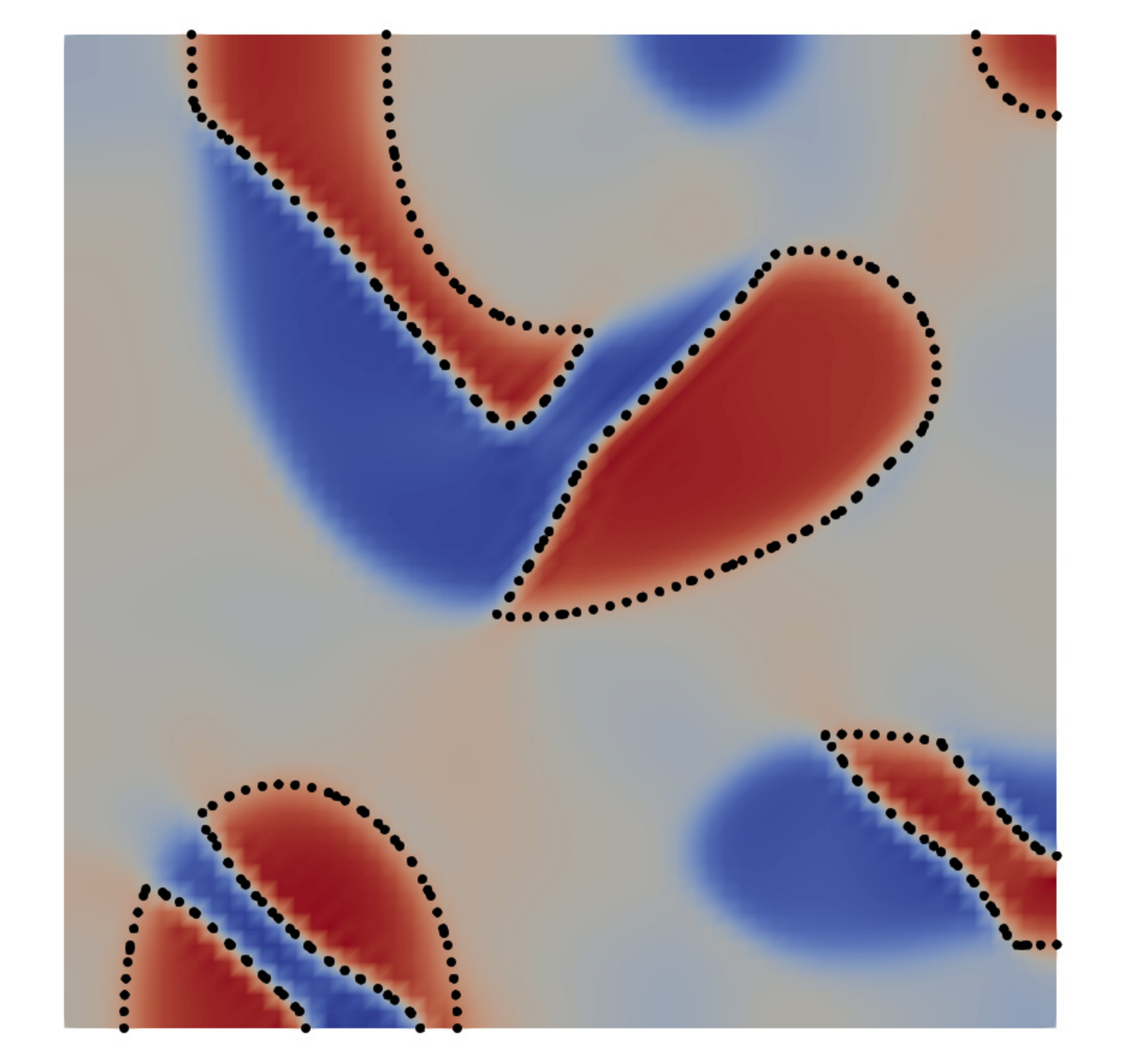}}
  \subfloat[$l^{r-}$ ]{\includegraphics[trim={19cm 5.0cm 19cm 5.0cm}, clip, width=0.32\linewidth]{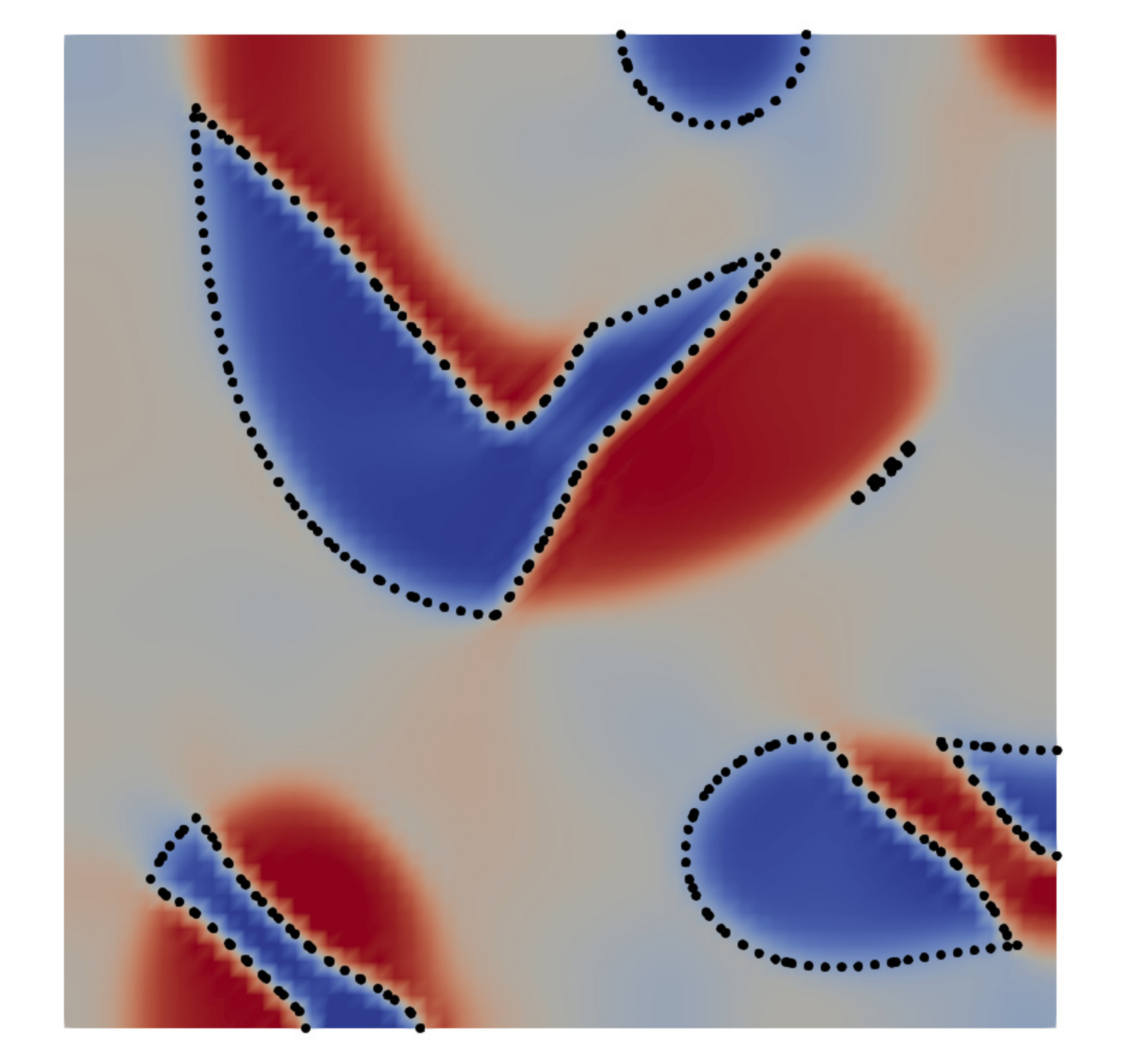}}
  \caption{Illustration of the interfacial length: (a) between the square and the rectangle phases for the concentration; (b) between the positive rectangle phase (red) and the other structures for the $e_2$ field; (c) between the negative rectangle phase (blue) and the other structures for the $e_2$ field. }
  \label{fig:interfacial-length}
\end{figure}

To differentiate microstructures from each other, several features were selected. The phase volume fractions $\phi_r^+$ and $\phi_r^-$ were chosen, representing the positive and negative rectangular phases. 
The volume fraction of the square phase was not selected as an independent feature because it can be calculated as $\phi_s = 1-\phi_r^+-\phi_r^-$.
Other selected features include the interfacial length between the square phase and the rectangle phases $l_s^r$, as shown in Fig.~\ref{fig:interfacial-length}(a),  
the interfacial length of the positive rectangle phase $l^{r+}$, as shown in Fig.~\ref{fig:interfacial-length}(b),
and the interfacial length of the negative rectangle phase $l^{r-}$, as shown in Fig.~\ref{fig:interfacial-length}(c). 
These input features are selected based on the established understanding in materials physics that phase volume fractions, interface areas and effective strains must determine the homogenized elastic response.
To compute the interfacial length,  we use the \emph{Contour} filter of ParaView \cite{paraview2005ahrens} to extract the contours for $c=0.5$ and $e_2=0.0$, which define these different interfaces. The contour data is then exported in to a separate VTK file. A customized Python script is created to select data points that define $l_s^r$, $l^{r+}$, and $l^{r-}$ and calculate their lengths.
Together, these lengths represent the three types of interfaces possible between the square and two rectangular phases.

\subsection{Data preparation} \label{sec:data-preparation}

Nine microstructures were uniformly sampled from each DNS to evaluate their relation to the macroscopic mechanical behavior of solids. Thus, 180 microstructures were sampled in total.
Combinations of different random shear and biaxial mechanical loadings were applied to each sampled microstructure, with the resulting $\Delta E_{11}$ and $\Delta E_{22}$ that are in the range of $[-5\times 10^{-5}, 5\times 10^{-5} ]$ and the resulting $\Delta E_{12}$ (or $\Delta E_{21}$) that is in the range of $[-3\times 10^{-4}, 3\times 10^{-4} ]$.
The newly applied mechanical loadings were much smaller than the initially applied ones for microstructure generation, hence the microstructures themselves were not altered during this mechanical testing protocol.
The quantities $\BF^\text{avg}$, $\BP^\text{avg}$, and {$\Psi_\text{mech}$}, were collected for each test.

\begin{figure}[t]
  \centering
  \subfloat[Base mechanical free energy $\Psi_\text{mech}^0$ from DNS\protect\footnotemark]{\includegraphics[trim={0cm 0cm 0cm 0cm}, clip, height=48mm]{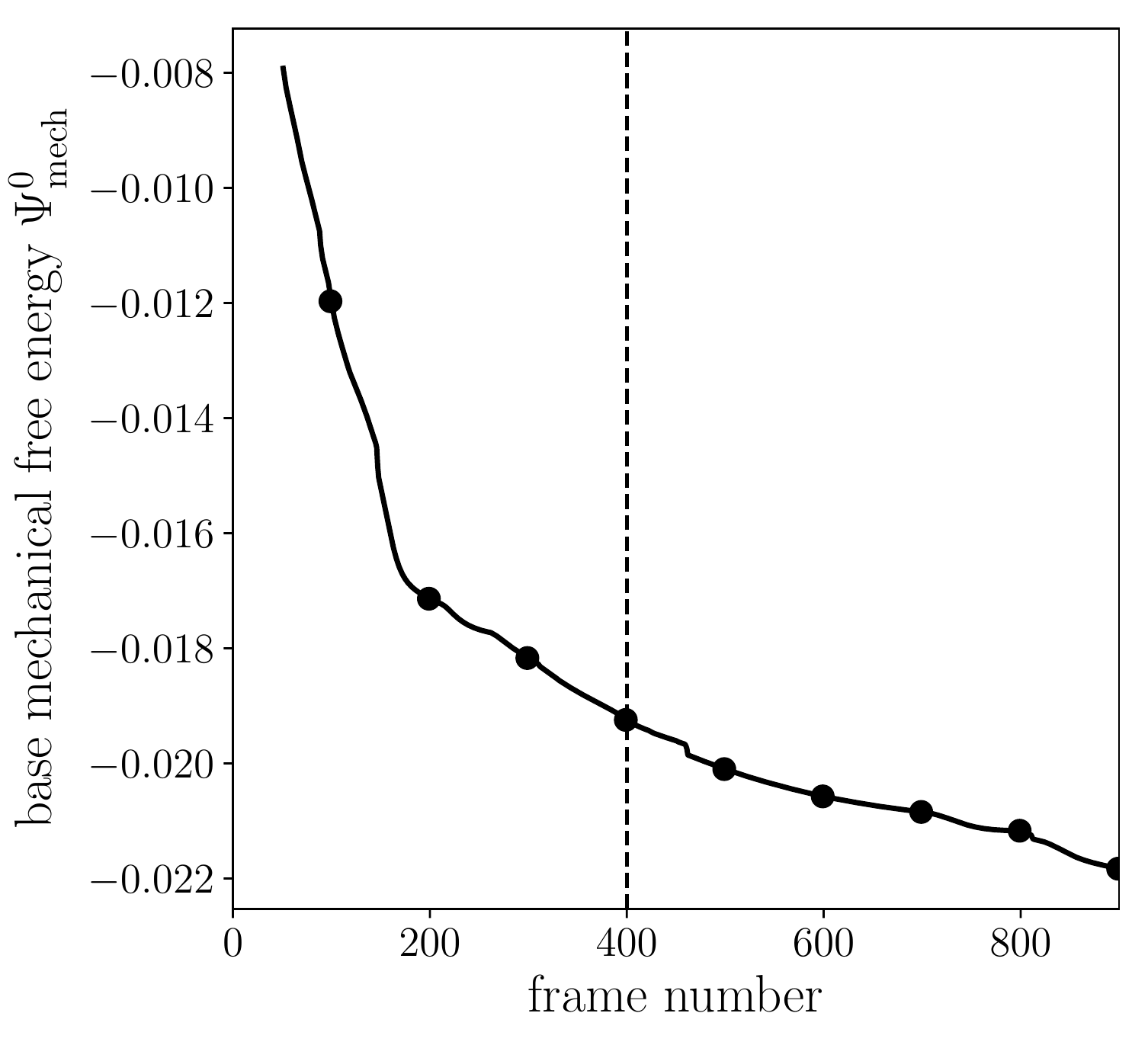}}
  \hspace{1.5cm}
  \subfloat[$\Psi_\text{mech}$ of a single microstructure under different tests]{\includegraphics[trim={0cm 0cm 0cm 0cm}, clip, height=48mm]{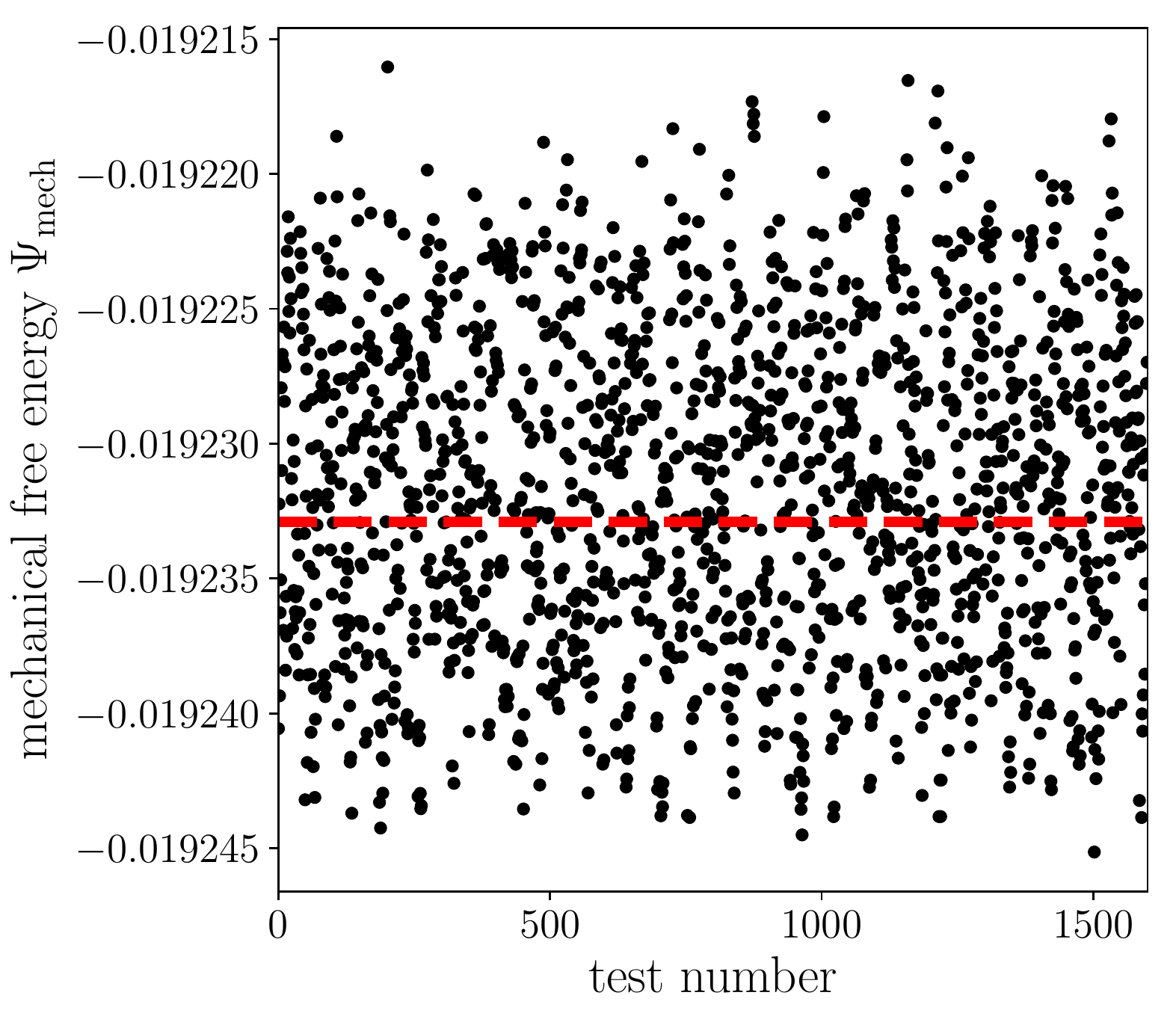}}
  \caption{Illustration of (a) the evolution of the elastic free energy from one DNS with the dashed line indicating frame 400 for dataset $\text{D}_\text{III}$ and dots indicating the uniformly sampled nine microstructures for dataset $\text{D}_\text{IV}$,  and (b) the elastic free energy in  dataset $\text{D}_\text{III}$, where a microstructure (frame 400) was tested under 1600 combinations of mechanical loading.
  One can observe that $\Psi_\text{mech}$ in (b) is oscillating around the red dashed line, which represents the base elastic free energy (a) resulting from the DNS.}
  \label{fig:elastic-free-energy-ossilcation}
\end{figure}

Four datasets were created in this work, which are summarized in Table \ref{tab:dataset}.
Datasets $\text{D}_\text{I}$ and $\text{D}_\text{II}$, which contained microstructure features defined in Section~\ref{sec:feature-selection}, the $e_2$ solution and the base mechanical free energy $\Psi_\text{mech}^0$ from DNS, were created for microstructures from a single DNS and from 20 different DNSs, respectively.
Datasets $\text{D}_\text{III}$ and $\text{D}_\text{IV}$ contained mechanical testing information for a single microstructure and all the sampled 180 microstructures, respectively.
Specifically, in dataset $\text{D}_\text{III}$, the microstructure at frame 400 from one particular DNS, as shown in Fig.~\ref{fig:elastic-free-energy-ossilcation}(a), was tested with 1600 different combinations of mechanical loading. $\text{D}_\text{III}$ has the components of $\BE^\text{avg}$, pre-defined microstructure features, and the $e_2$ solution as features, and the elastic free energy $\Psi_\text{mech}$ as the label. $\text{D}_\text{III}$ further contains the components of  $\BF^\text{avg}$ and $\BP^\text{avg}$ as  auxiliary information, and   $\BE^\text{avg}$ is computed as $\frac{1}{2} ( {\BF^\text{avg}}^T \BF^\text{avg} - \mathbf{1})$. Recall that use of $\BE^\text{avg}$ as the input deformation measure ensures frame invariance of the NN representations.
The $\Psi_\text{mech}$ from all the 1600 tests was plotted in Fig.~\ref{fig:elastic-free-energy-ossilcation}(b), where $\Psi_\text{mech}$ oscillates around the base elastic free energy $\Psi_\text{mech}^0 = -0.01923$.
Recall that here, $\Psi_\text{mech}^0$ refers to the elastic free energy stored in the microstructure during phase evolution shown in Figs \ref{fig:dns-setup-psi}b and \ref{fig:dns-results}, which is before the mechanical tests. The fine scale oscillations in  $\Psi_\text{mech}$ in
Fig.~\ref{fig:elastic-free-energy-ossilcation}(b) further confirm that the magnitude of the applied mechanical loadings is very small.
In dataset $\text{D}_\text{IV}$, all the 180 sampled microstructures, nine of which come from one specific DNS and whose mechanical free energies are shown in Fig.~\ref{fig:elastic-free-energy-ossilcation}(a), are tested under different mechanical loadings with 60366 data points collected.
$\text{D}_\text{IV}$ has the components of $\BE^\text{avg}$, pre-defined microstructure features, and the $e_2$ solution, and the perturbed $e_2$ solution as features, and the elastic free energy $\Psi_\text{mech}$ as the label. Similar to $\text{D}_\text{III}$, $\text{D}_\text{IV}$ has the components of  $\BF^\text{avg}$ and $\BP^\text{avg}$ as auxiliary information. The superscript of the averaged quantities ($\BE^\text{avg}, ~\BF^\text{avg}, ~\text{and}~\BP^\text{avg}$) is dropped to simplify notation.

\begin{table}
  \centering
  \begin{tabular}{l | l | l | l }
    \hline
      & Features & Labels & Description \\ \hline
    $\text{D}_\text{I}$   & DNN: $\{\phi_r^+,~\phi_r^-,~l_s^r,~l^{r+},~l^{r-}\}$   & $\Psi_\text{mech}^0$ & 850 data points from a single DNS\\
    &  CNN: $\{e_2~\text{solution}\}$ &  $\Psi_\text{mech}^0$ & \\ \hline

    $\text{D}_\text{II}$  & DNN: $\{\phi_r^+,~\phi_r^-,~l_s^r,~l^{r+},~l^{r-}\}$  &$\Psi_\text{mech}^0$ & 17000 data points from 20 DNSs\\
    &  CNN: $\{e_2~\text{solution}\}$ & $\Psi_\text{mech}^0$&  \\ \hline

    $\text{D}_\text{III}$ & DNN-based KBNN: & $\Psi_\text{mech}$ & 1600 mechanical testing data points \\
     & $\{E_{11},~E_{12},~E_{22},~\phi_r^+,~\phi_r^-,~l_s^r,~l^{r+},~l^{r-}\}$ &   &  from one microstructure with\\

   &  CNN-based KBNN:  & $\Psi_\text{mech}$  &  auxiliary data: $\{P_{11},~P_{12},~P_{21},~P_{22},$  \\    
       &  $\{E_{11},~E_{12},~E_{22},~e_2~\text{solution}\}$ &   & $F_{11},~F_{12},~F_{21},~F_{22}\}$ \\  \hline 

    $\text{D}_\text{IV}$  & CNN enhanced KBNN: & $\Psi_\text{mech}$ & 60366  mechanical testing data  points \\
      & $\{E_{11},~E_{12},~E_{22},~\phi_r^+,~\phi_r^-,~l_s^r,~l^{r+},~l^{r-}$  & &  from 180 microstructures with \\
    & $\text{perturbed}~e_2~\text{solution}$\} & &  auxiliary data: $\{P_{11},~P_{12},~P_{21},~P_{22},$  \\
     &CNN enhanced KBNN with penalization: & $\{\Psi_\text{mech}, $& $F_{11},~F_{12},~F_{21},~F_{22}\}$   \\
      & $\{E_{11},~E_{12},~E_{22},~\phi_r^+,~\phi_r^-,~l_s^r,~l^{r+},~l^{r-}$  & $ P_{11}, P_{12}$&  \\
    & $\text{original}~e_2~\text{solution}$\} &  $P_{21}, P_{22} \}$&  \\

	\hline
  \end{tabular}
  \caption{Summary of the four datasets used in Section \ref{sec:num-example}.}
  \label{tab:dataset}
\end{table}

\footnotetext{As defined in \eref{eq:2d-psi-mech}, $\psi_\text{mech}$ consists of a purely mechanical term, which is always positive, and a mechanochemical term, which can be either positive or negative depending on the value of the composition variable $c$. Because of this mechanochemical term, the base mechanical free energy $\Psi_\text{mech}^0$ could have a negative value.}

\subsection{Hyperparameter search} \label{sec:hyperparameter}
As discussed in Section~\ref{sec:NN}, the optimal architecture of NNs is unknown \emph{a priori}.
Hyperparameters can be selected via either manual tuning or automatic optimization algorithms, such as grid search or random search \cite{goodfellow2016deep}. 
In this work, grid search is performed for all the NNs.
For DNNs and the MNN of KBNNs, we search for the number of hidden layers ($N_\text{HL}$) and the number of neurons per layer ($N_\text{NPL}$). 
In our search space, $N_\text{HL}$ varies between 1 and 10 with a step of 1.
An identical $N_\text{NPL}$ is assumed for each hidden layer with its value varying between 2 and 256 with a step of 2.
For CNNs, a kernel size of $(3,3)$ and a stride size of $(1,1)$ are pre-chosen.
We only search for $N_\text{HL}$ and the number of filters per layer ($N_\text{FPL}$), with $N_\text{HL}$ varying from 1 to 10 with a step of 1 and $N_\text{FPL}$ varying from 2 to 32 with a step of 1.
Unlike the case of $N_\text{NPL}$ for DNNs/MNNs, $N_\text{FPL}$ is not identical for each layer. Its value increases with the depth of the hidden layer. 
The exponentially decaying learning rate implemented in \texttt{Tensorflow}, which follows a staircase function, is used 
\begin{equation}
  \text{lr} = \text{lr}_0 \cdot \text{pow}\left(  v_\text{decay}, \frac{N_\text{total}}{N_\text{decay}}\right)
  \label{eq:lr-step}
\end{equation}
with an initial learning rate $\text{lr}_0 = 0.001$, a decay rate $v_\text{decay} = 0.7$, a decay step $N_\text{decay} = 100$, and a final $N_\text{total} = 2000$ epochs.
The dataset is randomly split into a set consisting of $90\%$ for training and validation and a set of $10\%$ for testing.
A $K$-fold cross-validation procedure (with $k=5$) \cite{goodfellow2016deep} is performed on the set consisting of $90\%$ of the data to train and evaluate different NN models.
Feature normalization and label scaling are used to improve the accuracy of NNs during training.

When performing the hyperparameter search, first, the total number of variables of each possible NN architecture in our search space is computed and sorted in an ascending order. 
Those NNs with a total variable number larger than the size of the dataset are excluded from the search space. 
Then, a grid search based on the total number of variables of the NNs is performed.
The performance of each NN is evaluated based on the averaged validation loss and is sorted in ascending order.
The total number of variables of the top performing 30\%  NNs defines a refined search space, in which a new grid search is performed.  
The grid search is repeated three times in total.
The model with the smallest averaged validation loss is selected as the best one. 
The hyperparameter search procedure is summarized in the Algorithm~\ref{algo:hyper-search}.

\begin{algorithm}[ht]
  \caption{Hyperparameter search procedure. \label{algo:hyper-search}}
  \begin{algorithmic}[1]
    \STATE Create a set $S$ containing all possible NN structures that lie in the search space defined by hyperparameters ($N_\text{HL}$, $N_\text{NPL}$, or $N_\text{FPL}$), with NNs in $S$ being sorted in an ascending order based on the total number of variables  ($V_\text{total}$) of each NN.
    \STATE Grid search of hyperparameters in $S$ based on $V_\text{total}$.

    \STATE Define an initial lower limit and an initial upper limit of $V_\text{total}$ with $V_\text{total}^\text{min} = 0$ and $V_\text{total}^\text{max} = \text{size of (dataset $D$)}$.
    \FOR{$s$ in multiple sampling steps $(=3,~\text{in this work})$}
    \STATE Uniformly sample multiple $(=25,~\text{in this work})$ NNs out of all NNs, where each NN has $ V_\text{total}^\text{min} \le V_\text{total} \le V_\text{total}^\text{max}$, to form a subset $\bar{S}$.
      \STATE Perform $K$-fold cross-validation for each NN in $\bar{S}$.
      \FOR{each model $M_i$ in $\bar{S}$} 
      \STATE Split $D$ into $K$ mutually exclusive subsets $D_k$
      \FOR{$k$ from $1$ to $K(=5,~\text{in this work})$}
        \STATE Train $M_i$ with $D\backslash D_k$ 
        \STATE Evaluate (validate) $M_i$ with $D_k$ to get the loss $\mathcal{L}_i^k$.
        \ENDFOR
        \STATE Compute the averaged validation loss $\bar{\mathcal{L}}_i$ for $M_i$.
      \ENDFOR
      \STATE Sort models in $\bar{S}$ based on $\bar{\mathcal{L}}_i$ in ascending order.
      \STATE Refine the search space by updating $V_\text{total}^\text{min}$ and $V_\text{total}^\text{max}$, where $V_\text{total}^\text{min} = \text{min} ( V_\text{total})$ and $V_\text{total}^\text{max} = \text{max} ( V_\text{total})$ in $\bar{S}_{30}$, with $\bar{S}_{30}$ representing a subset of $\bar{S}$ that contains the top $30\%$ (an user-defined threshold value) performed models.
    \ENDFOR
    \STATE Select the best model $M$ with the smallest $\bar{\mathcal{L}}$. 
  \end{algorithmic}
\end{algorithm}

\noindent\textbf{Remark 3:} In this work, we chose to tune hyperparameters for NNs with a total number of variables less than the size of the available datasets. In our study, we also explored NNs with high capacity, which could achieve comparable performance as the well-tuned NNs. However, as is well-known, high capacity models with limited data points can yield overfitting. Extra effort, such as using different regularization techniques, will be needed to prevent overfitting when using such models \cite{goodfellow2016deep}.

\section{Numerical examples} \label{sec:num-example}

In this section, we explore different NNs to predict the homogenized mechanical behavior of synthetically generated heterogeneous microstructures\footnote{Code available at \href{https://github.com/mechanoChem/dataDrivenHomogenization}{github.com/mechanoChem/dataDrivenHomogenization}}.
Specifically, the base elastic free energy of microstructures from a single DNS is studied in Section~\ref{sec:sim-base-free-energy}, and from multiple DNSs in Section~\ref{sec:sim-base-free-energy-m-dns} with both CNNs and DNNs.
The homogenized mechanical behavior of a single microstructure is studied with KBNNs in Section~\ref{sec:sim-kbnn-one-microstructure}. 
Finally, CNN-enhanced KBNNs are trained to predict the homogenized mechanical behavior of different microstructures from multiple DNS in Section~\ref{sec:sim-kbnn-multi-DNS}.
The features and labels for each NN used in this Section are summarized in Table \ref{tab:dataset}. The wall-time required for training each NN is given in Table \ref{tab:wall-time}.

\subsection{Base mechanical free energy for one DNS} \label{sec:sim-base-free-energy}
As revealed in Figs.~\ref{fig:dns-setup-psi} and~\ref{fig:elastic-free-energy-ossilcation}, the elastic free energy $\Psi_\text{mech}$ stored in microstructures due to phase evolution is of a sharply multi-resolution nature. It has $\Psi_\text{mech}^0$ from microstructure phase evolution as the dominant {characteristic} and $\Delta \Psi_\text{mech} := \Psi_\text{mech} - \Psi_\text{mech}^0$ from mechanical testing as the detailed {characteristic}.
It is challenging to capture both {characteristics} in a single NN, because the weights emphasize the dominant {characteristic} over the detailed {characteristic} during the training process.
As in Ref. \cite{Perrone93whennetworks}, we leverage the discrepancy between networks in functional rather than in parameter space. However, our treatment differs in that we directly use as labels the discrepancy rather than average.
To overcome this challenge, we use KBNNs, as discussed in Section~\ref{sec:NN-kbnn}, to represent this multi-resolution data.
The ENN is trained to learn the base free energy $\Psi_\text{mech}^0$ (dataset $\text{D}_\text{I}$) in this section with both DNNs and CNNs being explored. 

\subsubsection{Base mechanical free energy represented by DNNs}\label{sec:label-shift-dnn}

\begin{figure}[t]
  \centering
  \subfloat[DNN: learning curve ]{\includegraphics[height=50mm]{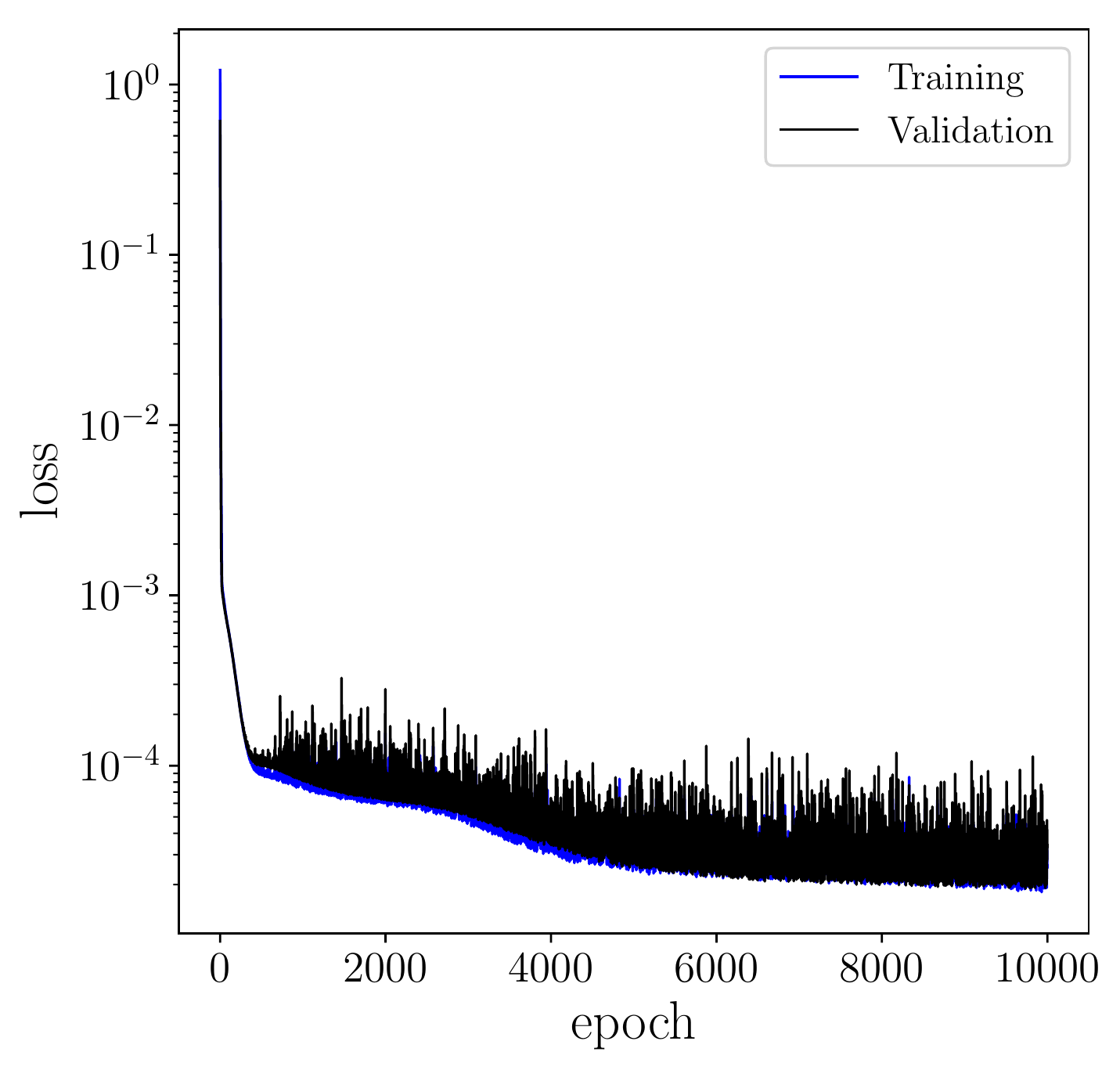}} \hfill
  \subfloat[DNN: test dataset prediction]{\includegraphics[height=50mm]{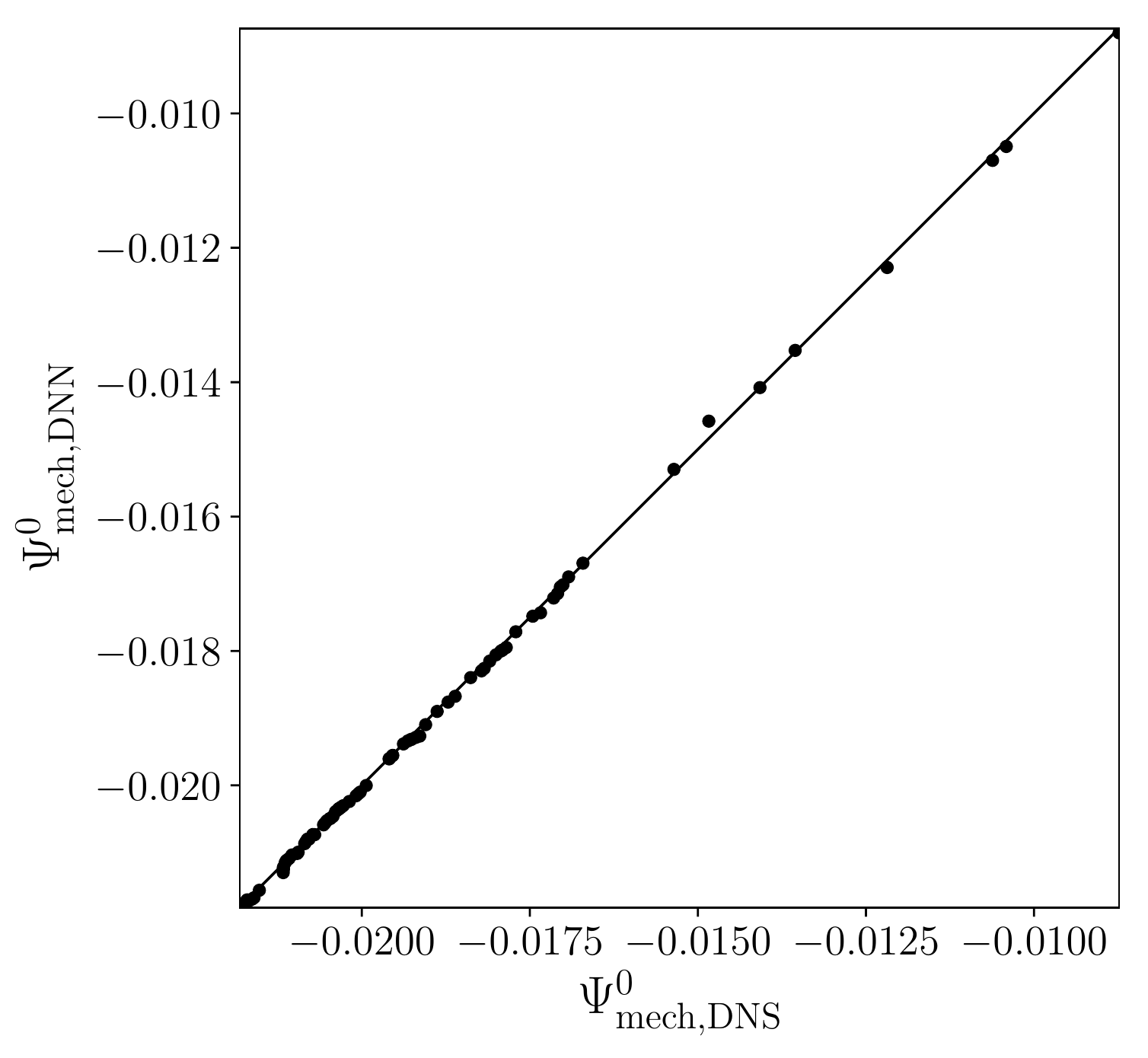}}\hfill
  \subfloat[DNN: NN predicted $\Psi_\text{mech}^0$ ]{\includegraphics[height=50mm]{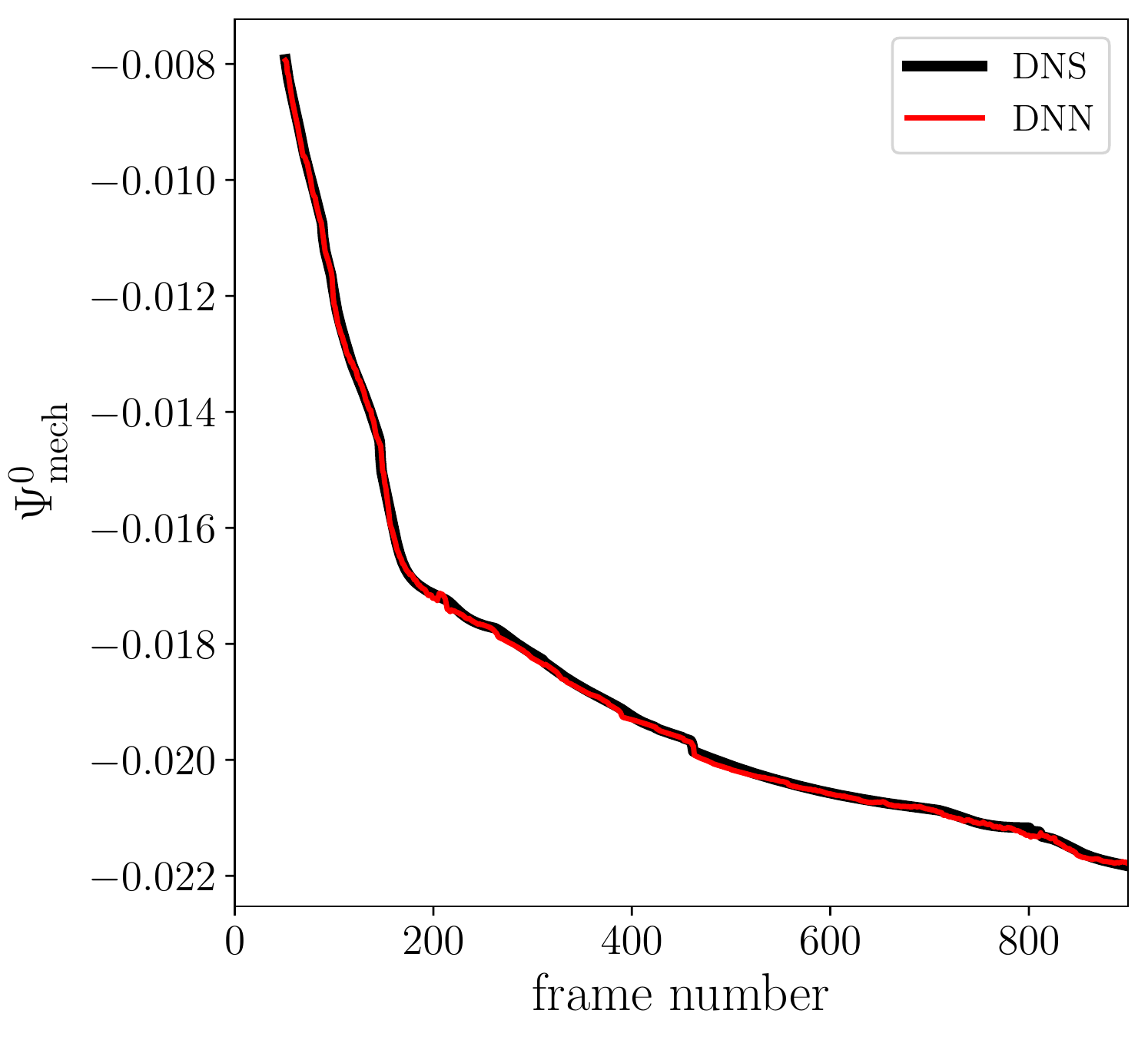}} \\
  \subfloat[CNN: learning curve ]{\includegraphics[height=50mm]{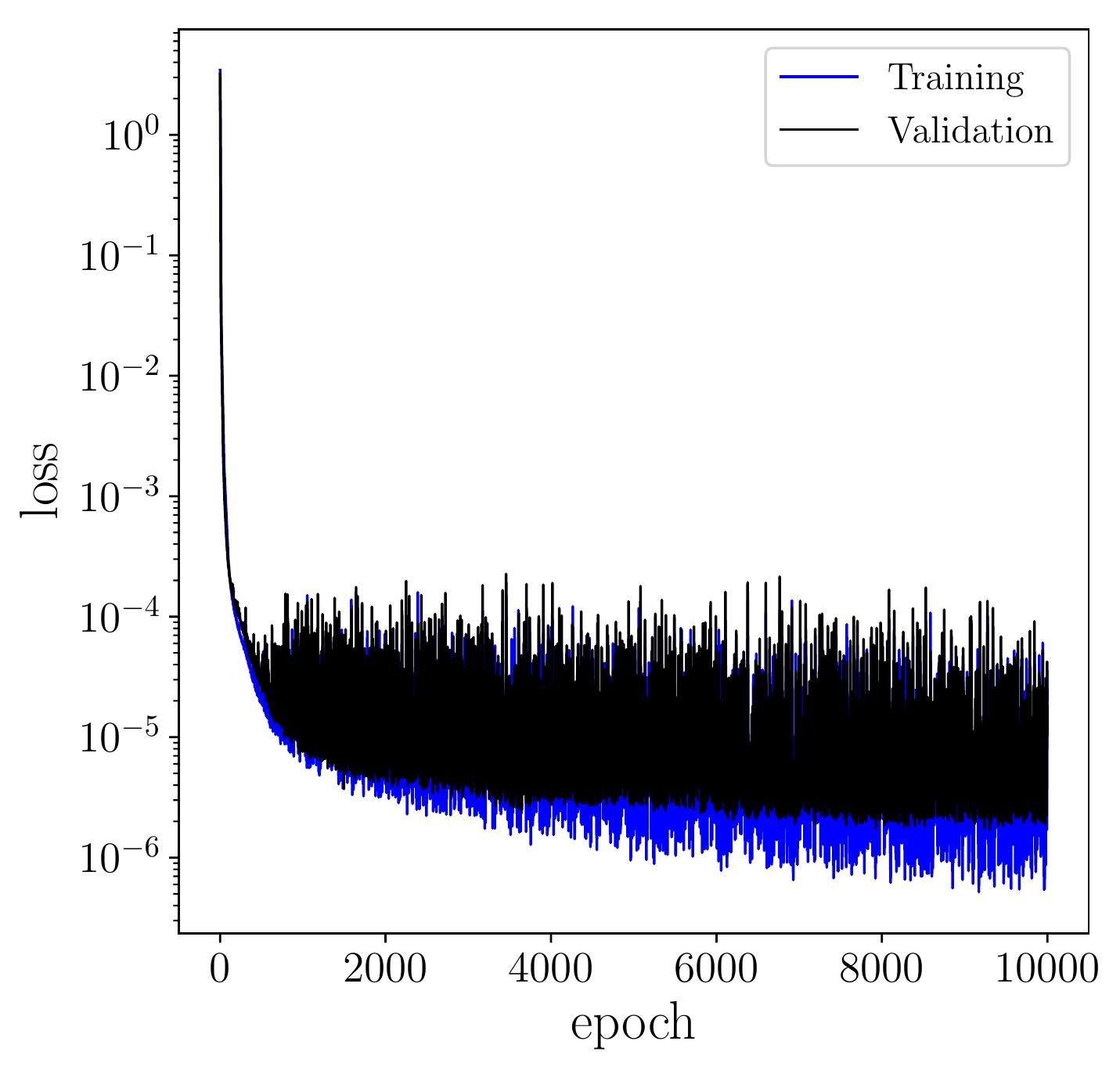}} \hfill
  \subfloat[CNN: test dataset prediction]{\includegraphics[height=50mm]{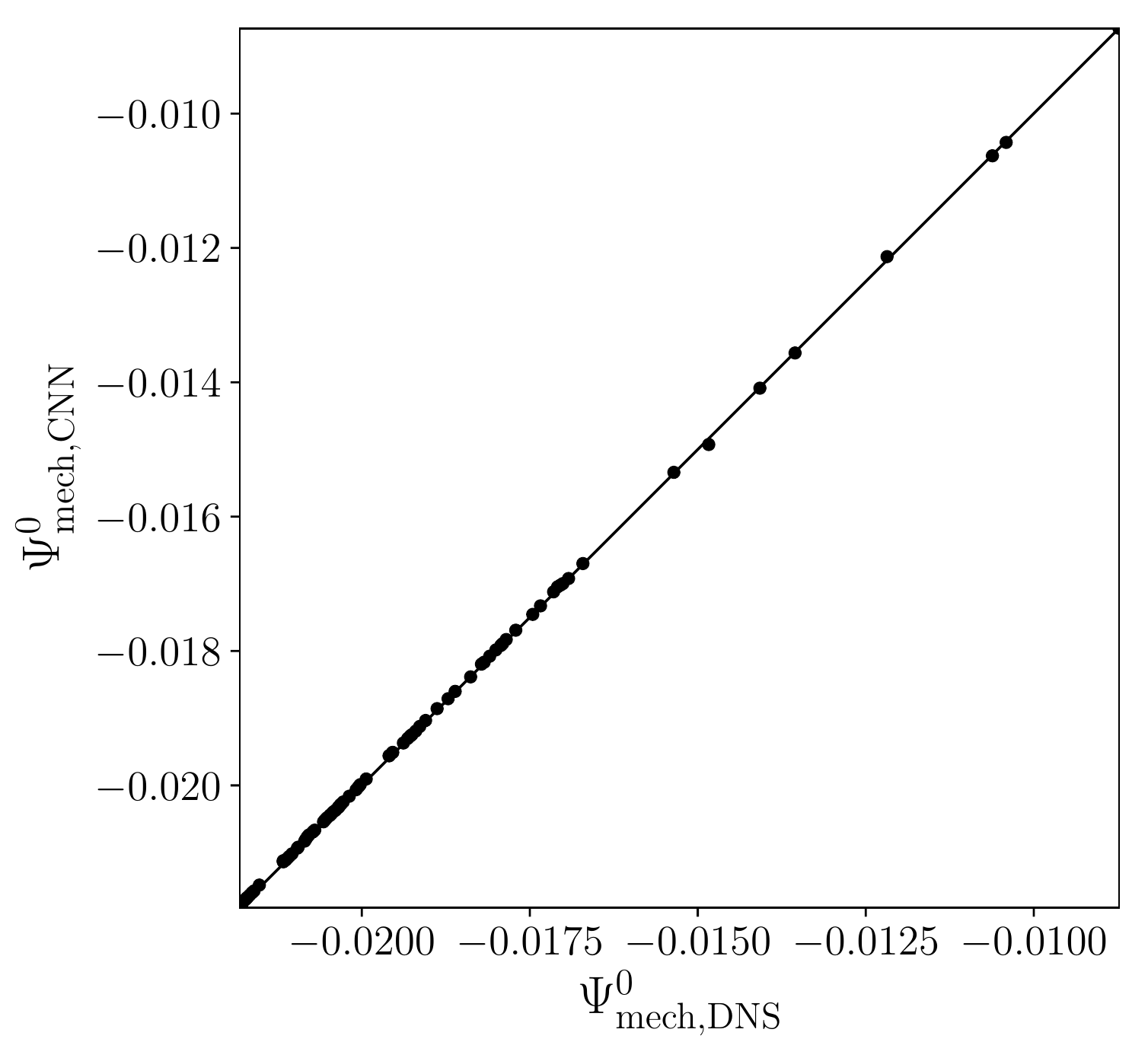}}\hfill
  \subfloat[CNN: NN predicted $\Psi_\text{mech}^0$ ]{\includegraphics[height=50mm]{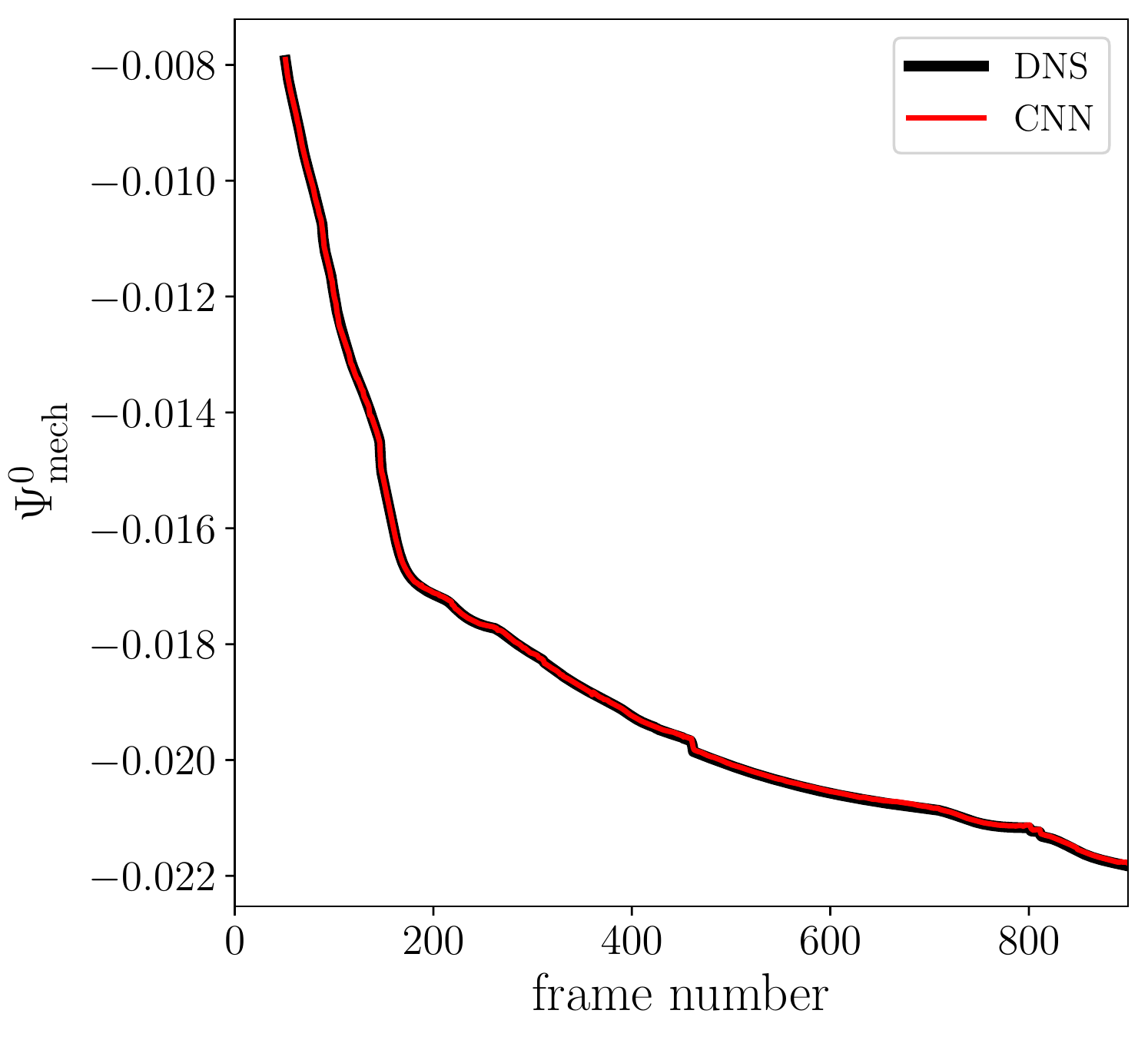}}
  \caption{Representation of base free energy for single DNS.
    (a, d) Learning curve for $\Psi_\text{mech}^0$; 
    (b, e) NN model predicted $\Psi_\text{mech}^0$ vs $\Psi_\text{mech}^0$ from DNS on the test dataset of $\text{D}_\text{I}$; 
    (c, f) Comparison between $\Psi_\text{mech,NN}^0$ and $\Psi_\text{mech}^0$ vs frame numbers for the whole dataset of $\text{D}_\text{I}$.}
  \label{fig:psi-label-shift-dnn-cnn}
\end{figure}

A DNN using the mean squared error (MSE) loss function is trained to predict the base elastic free energy $\Psi_\text{mech}^0$.
The Softplus activation function is used for all the layers. 
The DNN has $\phi_r^+$, $\phi_r^-$, $l_s^r$, $l^{r+}$, and $l^{r-}$ as its features and $\Psi_\text{mech}^0$ as its label. 
A grid search of the hyperparameters $\{N_\text{HL},~N_\text{NPL}\}$ for the DNN is conducted by following the procedure discussed in Section~\ref{sec:hyperparameter}, with an obtained optimal structure of $N_\text{HL} = 1$, $N_\text{NPL}= 76$, and a total  variable number of $533$.
The model is trained with the Adam optimizer for 10000 epochs with the exponentially decaying learning rate given in \eref{eq:lr-step} where $v_\text{decay} = 0.92$.
The learning curve for the DNN is plotted in Fig.~\ref{fig:psi-label-shift-dnn-cnn}(a), where neither overfitting nor underfitting is observed.
Figs.~\ref{fig:psi-label-shift-dnn-cnn}(b,c) show that the model can predict $\Psi_\text{mech}^0$ to high accuracy. 
The value of $\Psi_\text{mech}^0$ computed from the DNN is denoted as $\Psi_\text{mech,DNN}^0$.  

\subsubsection{Base mechanical free energy represented by CNNs}\label{sec:label-shift-cnn}

\begin{table}
  \centering
  \begin{tabular}{l | l | l}
    \hline
    Layer type &  & Notes \\ \hline
    Input & $e_2$ field & \\
    Conv2D & filters = 2 & kernel (3,3), stride (1,1), padding (2,2), ReLU \\
    MaxPooling2D & - & kernel (2,2), stride (1,1), padding (1,1)\\
    Conv2D & filters = 3 & kernel (3,3), stride (1,1), padding (2,2), ReLU \\
    MaxPooling2D & - & kernel (2,2), stride (1,1), padding (1,1)\\
    Conv2D & filters = 5 & kernel (3,3), stride (1,1), padding (2,2), ReLU \\
    MaxPooling2D & - & kernel (2,2), stride (1,1), padding (1,1)\\
    Conv2D & filters = 6 & kernel (3,3), stride (1,1), padding (2,2), ReLU \\
    MaxPooling2D & - & kernel (2,2), stride (1,1), padding (1,1)\\
    Flatten & - & - \\
    Output Dense Layer & label =1 & Linear \\
  \end{tabular}
  \caption{Detail of the CNN architecture for representing $\Psi_\text{mech}^0$ of single DNS.}
  \label{tab:cnn-base-psi-1-dns}
\end{table}

The features selected in Section~\ref{sec:feature-selection} ($\phi_r^+$, $\phi_r^-$, $l_s^r$, $l^{r+}$, and $l^{r-}$) are the interpretation of image data based {on the accepted understanding of the global quantities that distinguish microstructures (these represent ``domain knowledge'').}
Alternately, we can train CNNs to automatically identify features to represent microstructures. 
Such an approach underlies the treatment of this section with the goal of investigating the existence of any advantage for CNNs over DNNs for computational materials physics simulations. 

A CNN consisting of multiple convolutional layers, multiple pooling layers, and one dense layer was trained to predict the base elastic free energy in Fig.~\ref{fig:elastic-free-energy-ossilcation}(a).
{The CNN takes the whole $e_2$ field solution from DNSs as input, in order to discern the square and two rectangular phases.\footnote{We remark that Digital Image Correlation (DIC) techniques are capable of reporting full-field strains from real (non-synthetic) microstructures, from which $e_2$ could be computed.} Note that this information is provided to the DNNs as the feature set $\phi_r^+$, $\phi_r^-$, $l_s^r$, $l^{r+}$, and $l^{r-}$. The $c$ field alone does not provide information on the rectangular phases. Image data is fed to the CNNs with a pixel resolution of $61\times 61$, and $\Psi_\text{mech}^0$ as its label. }
A hyperparameter search was conducted by following the procedure discussed in Section~\ref{sec:hyperparameter}, with the best architecture of the CNN given in Table~\ref{tab:cnn-base-psi-1-dns} with a total variable number of 590.  
The model was trained with the Adam optimizer for 10000 epochs with the exponentially decaying learning rate given in \eref{eq:lr-step} where $v_\text{decay} = 0.92$.
The learning curve for the CNN is plotted in Fig.~\ref{fig:psi-label-shift-dnn-cnn}(d). 
The model can accurately predict $\Psi_\text{mech}^0$, as plotted in Figs. \ref{fig:psi-label-shift-dnn-cnn}(e,f), which show essentially the same high accuracy as the DNN results in Figs. \ref{fig:psi-label-shift-dnn-cnn}(b,c).

\subsection{Base elastic free energy for multiple DNS} \label{sec:sim-base-free-energy-m-dns}

\begin{figure}[t]
  \centering
  \subfloat[DNN: learning curve ]{\includegraphics[height=50mm]{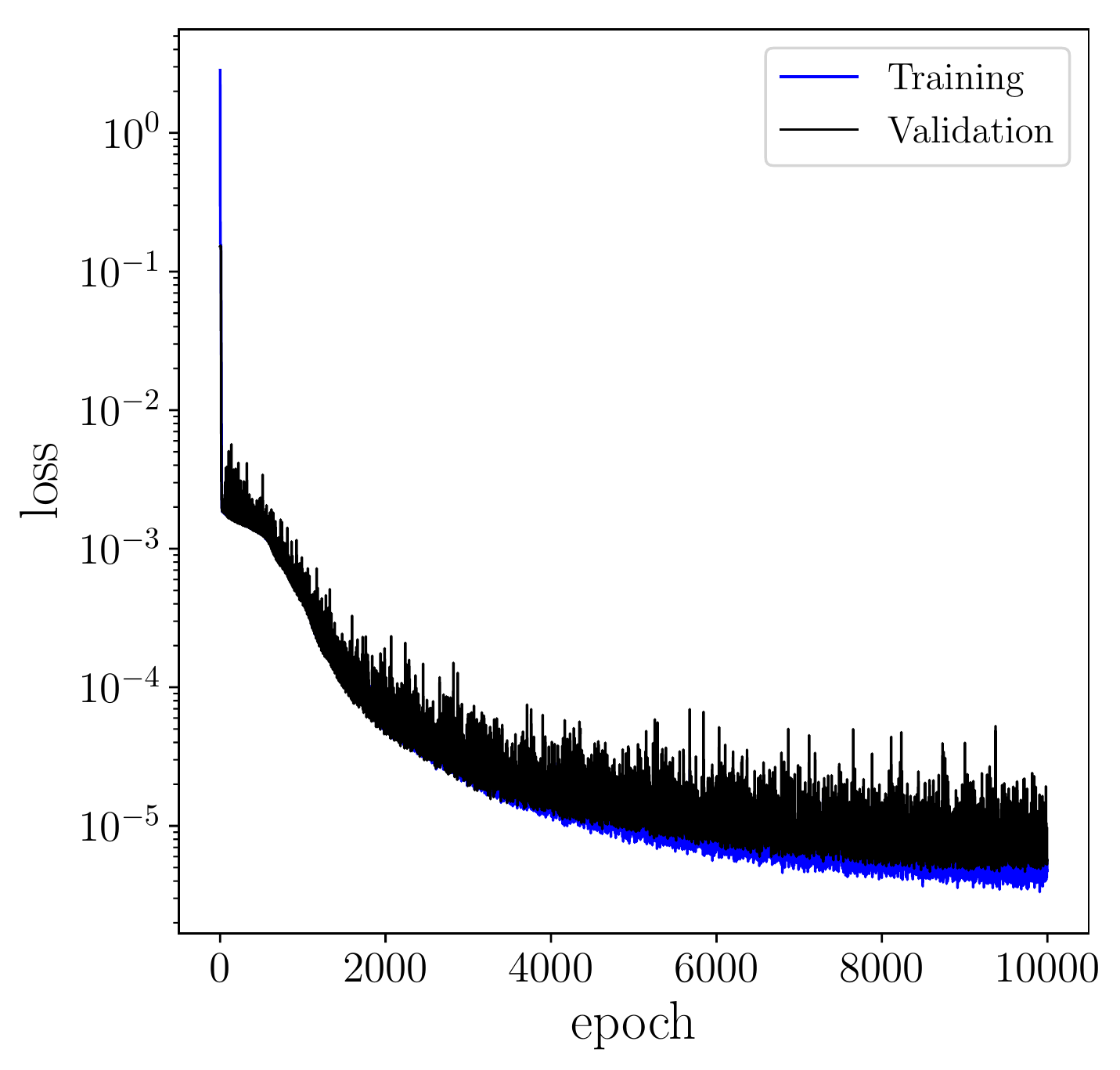}} \hfill
  \subfloat[DNN: test dataset prediction]{\includegraphics[height=50mm]{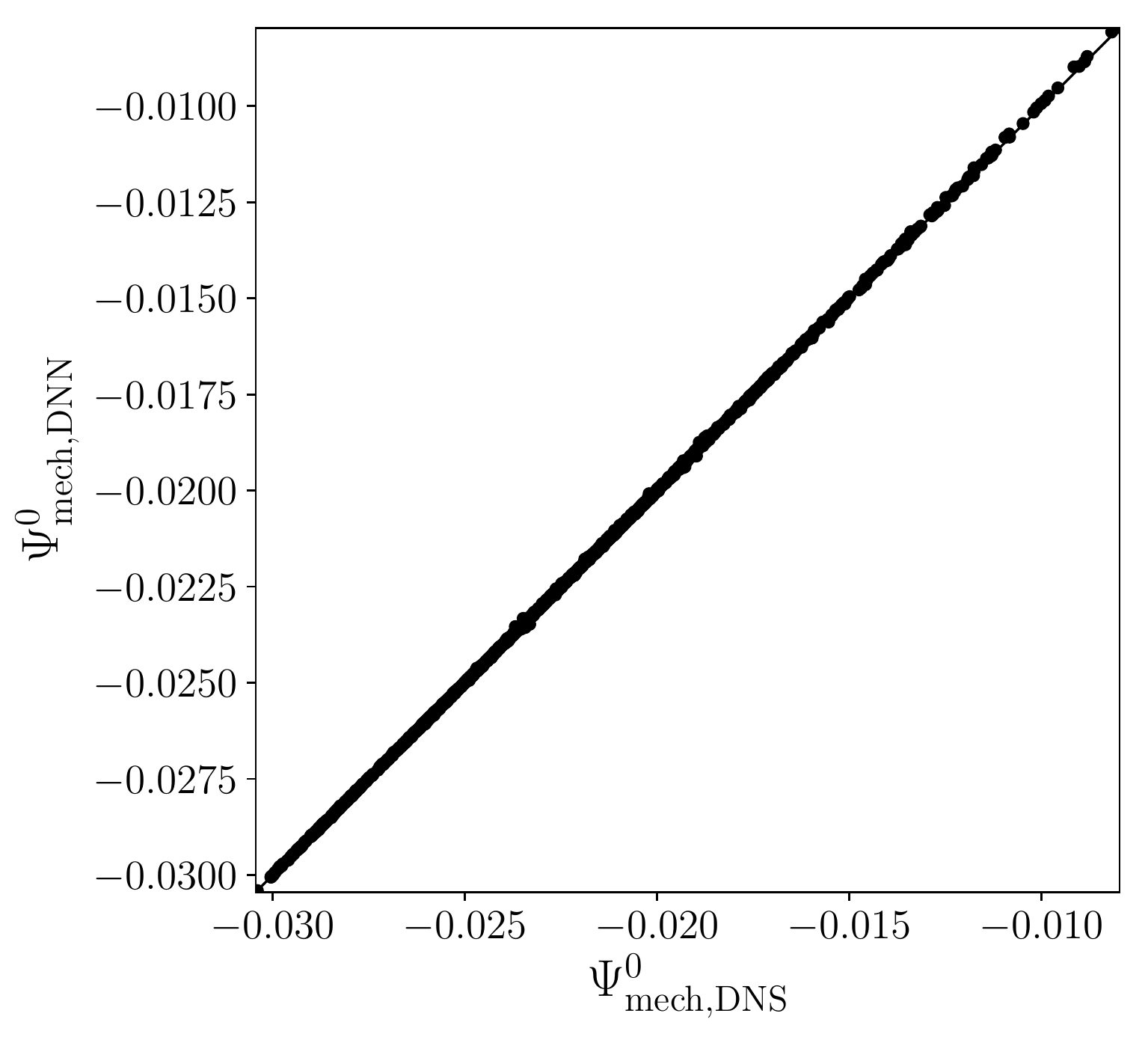}}\hfill
  \subfloat[DNN: NN predicted $\Psi_\text{mech}^0$ ]{\includegraphics[height=50mm]{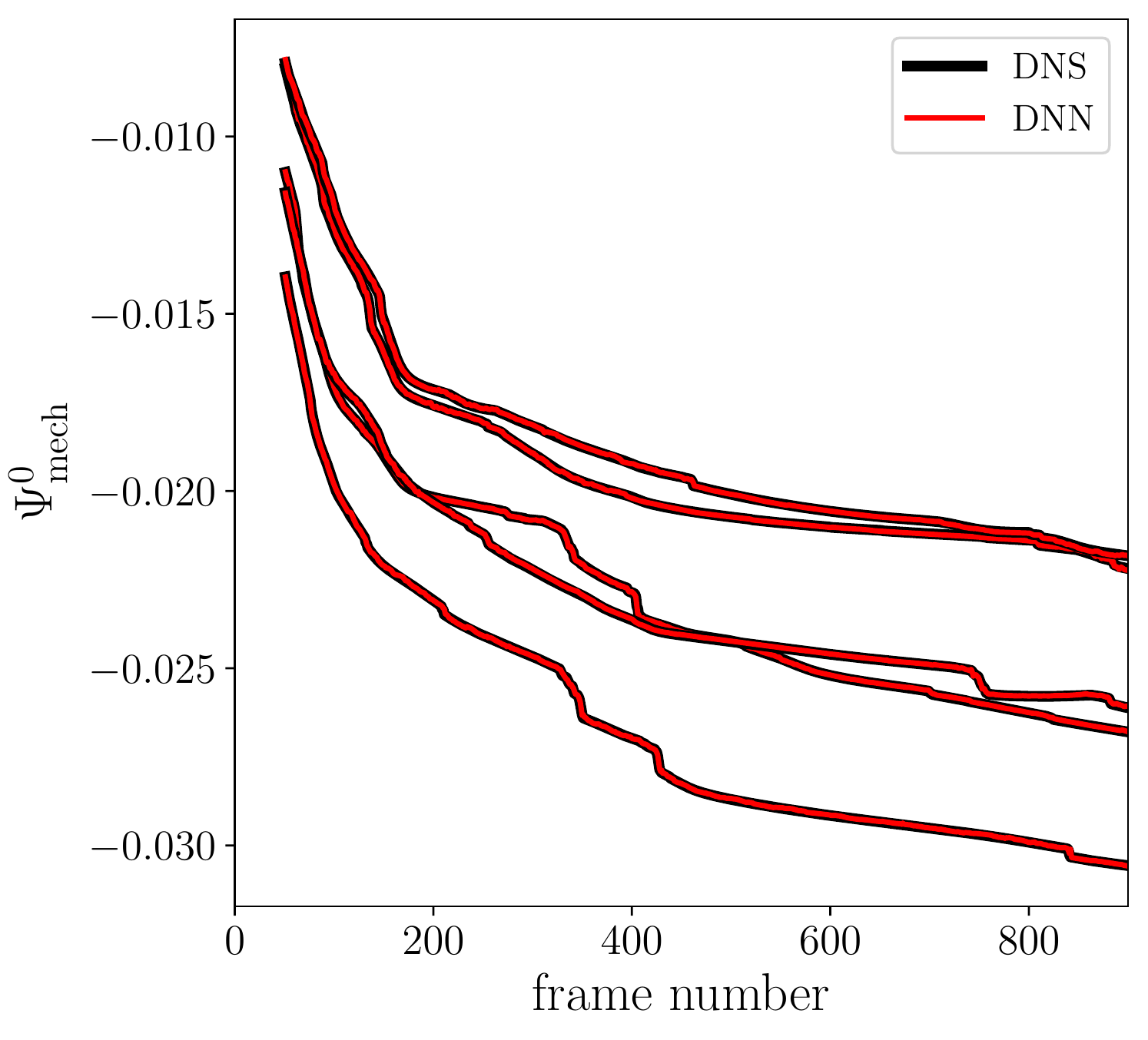}} \\
  \subfloat[CNN: learning curve ]{\includegraphics[height=50mm]{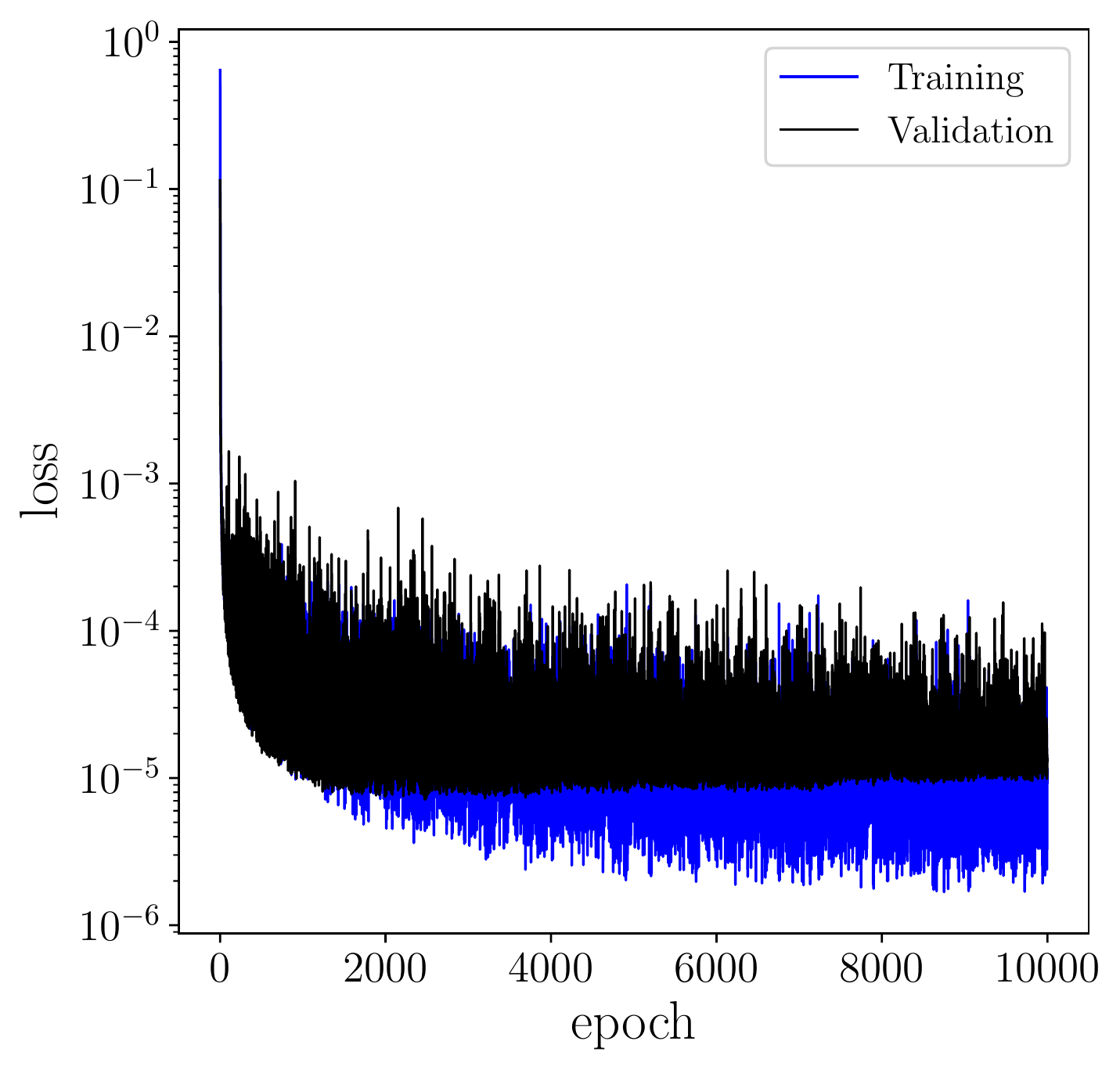}} \hfill
  \subfloat[CNN: test dataset prediction]{\includegraphics[height=50mm]{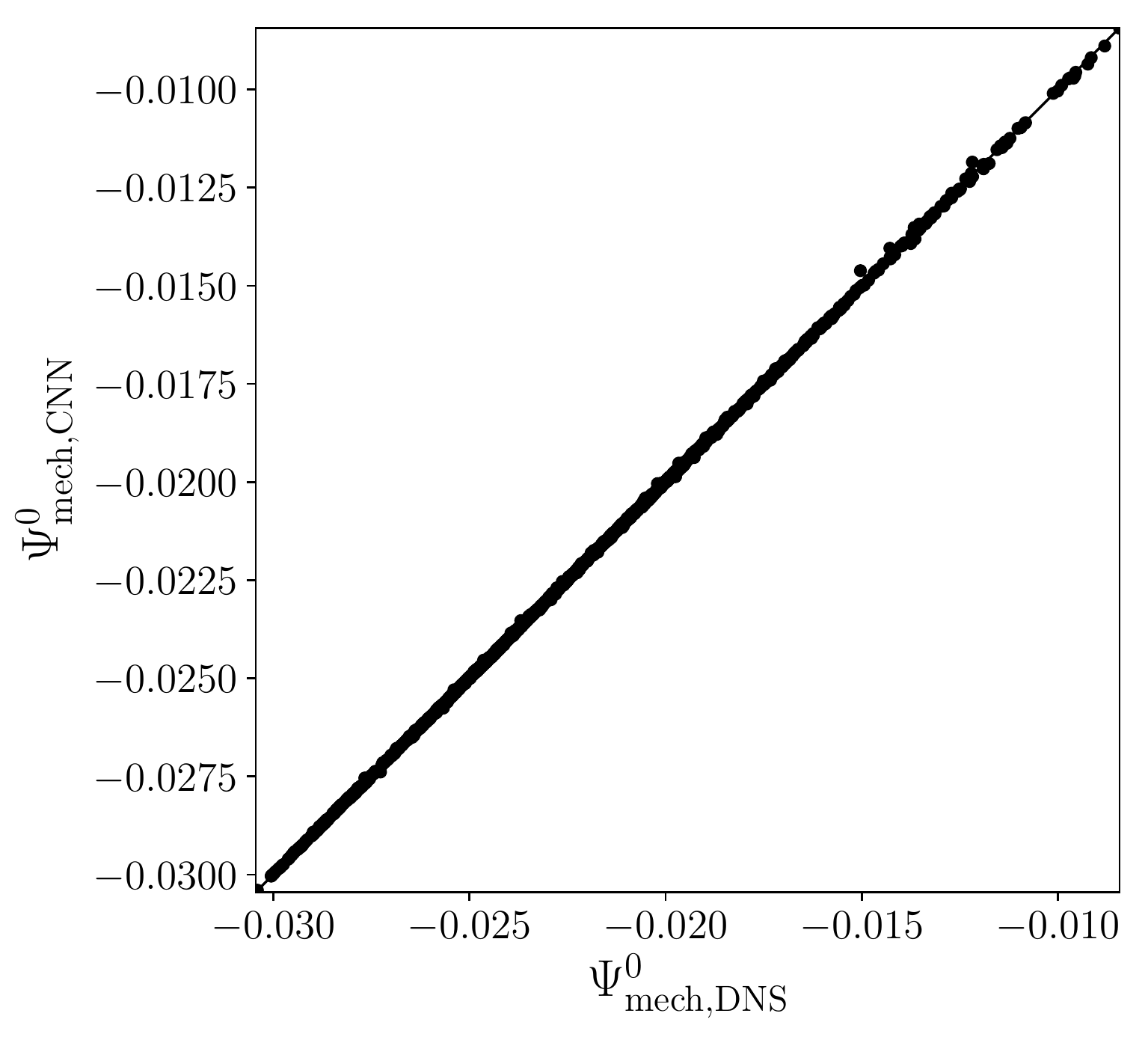}}\hfill
  \subfloat[CNN: NN predicted $\Psi_\text{mech}^0$ ]{\includegraphics[height=50mm]{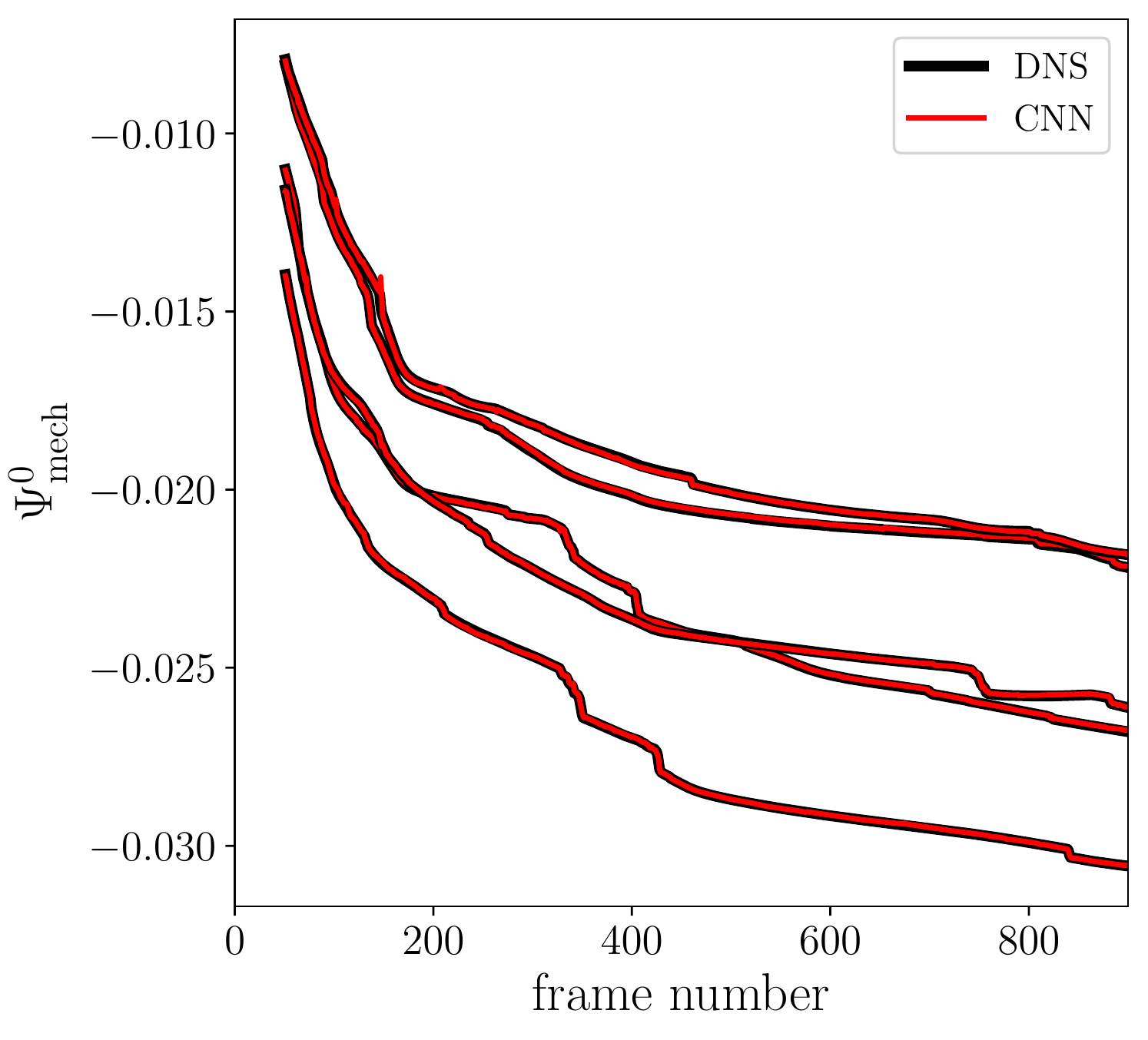}}
  \caption{Representation of base free energy for multiple DNSs.
    (a, d) Learning curve for $\Psi_\text{mech}^0$; 
    (b, e) NN model predicted $\Psi_\text{mech}^0$ vs $\Psi_\text{mech}^0$ from DNS on the test dataset of $\text{D}_\text{II}$; 
  (c, f) Comparison between $\Psi_\text{mech,NN}^0$ and $\Psi_\text{mech}^0$ vs frame numbers for five DNS from $\text{D}_\text{II}$. }
  \label{fig:psi-label-shift-dnn-cnn-m-dns}
\end{figure}

\begin{table}
  \centering
  \begin{tabular}{l | l | l}
    \hline
    Layer type &  & Notes \\ \hline
    Input & $e_2$ field & \\
    Conv2D & filters = 9 & kernel (3,3), stride (1,1), padding (2,2), ReLU \\
    MaxPooling2D & - & kernel (2,2), stride (1,1), padding (1,1)\\
    Conv2D & filters = 15 & kernel (3,3), stride (1,1), padding (2,2), ReLU \\
    MaxPooling2D & - & kernel (2,2), stride (1,1), padding (1,1)\\
    Conv2D & filters = 16 & kernel (3,3), stride (1,1), padding (2,2), ReLU \\
    MaxPooling2D & - & kernel (2,2), stride (1,1), padding (1,1)\\
    Flatten & - & - \\
    Output Dense Layer & label = 1 & Linear \\
  \end{tabular}
  \caption{Detail of the CNN architecture for representing $\Psi_\text{mech}^0$ of multiple DNSs.}
  \label{tab:cnn-base-psi-m-dns}
\end{table}

In this section, both DNNs and CNNs are explored to represent the base free energy $\Psi_\text{mech}^0$ from multiple DNS for the ENN (dataset $\text{D}_\text{II}$).
As in Section~\ref{sec:label-shift-dnn}, {the DNN and the CNN take \{$\phi_r^+$, $\phi_r^-$, $l_s^r$, $l^{r+}$, $l^{r-}$\} and the whole $e_2$ field solution as their features, respectively, with $\Psi_\text{mech}^0$ as the label for both cases.}
The results with an optimal DNN structure obtained from the hyperparameter search, which has $N_\text{HL} = 6$, $N_\text{NPL}= 46$, and $V_\text{total} = 11133$, are shown in Fig.~\ref{fig:psi-label-shift-dnn-cnn-m-dns}(a-c).
The results with an optimal CNN structure, whose architecture is given in Table~\ref{tab:cnn-base-psi-m-dns} with $V_\text{total} = 4521$, are shown in Fig.~\ref{fig:psi-label-shift-dnn-cnn-m-dns}(d-f).
From Fig.~\ref{fig:psi-label-shift-dnn-cnn-m-dns}, one can observe that both the DNN and the CNN show a good representation of the base free energy from multiple DNSs with different initial and boundary conditions. 

{
\noindent\textbf{Remark 4:} We remark that in Figs. \ref{fig:psi-label-shift-dnn-cnn} and \ref{fig:psi-label-shift-dnn-cnn-m-dns} only the mechanical and mechanochemical contributions to the base free energy, defined in Eq. \eref{eq:2d-psi-mech} have been used as labels to train DNN and CNN representations on single and multiples DNSs. This is because in the sections to come, our interest is focused on representing the homogenized mechanical response rather than the full mechanochemical response. The chemical contributions, if included would not contribute to the homogenized mechanical response beyond their roles in delineation of the microstructure--information that is being provided to train the DNN and CNN representations. We also have confirmed that the DNN and CNN representations resolve the total free energy with accuracy that is fully equivalent to that shown in Fig. \ref{fig:psi-label-shift-dnn-cnn}c,f and \ref{fig:psi-label-shift-dnn-cnn-m-dns}c,f, respectively, for the base mechanical free energy, $\Psi^0_\text{mech}$.}

\subsection{Homogenized mechanical behavior of a single microstructure} \label{sec:sim-kbnn-one-microstructure}

\begin{figure}[t]
  \centering
  \subfloat[learning curve]{\includegraphics[height=50mm]{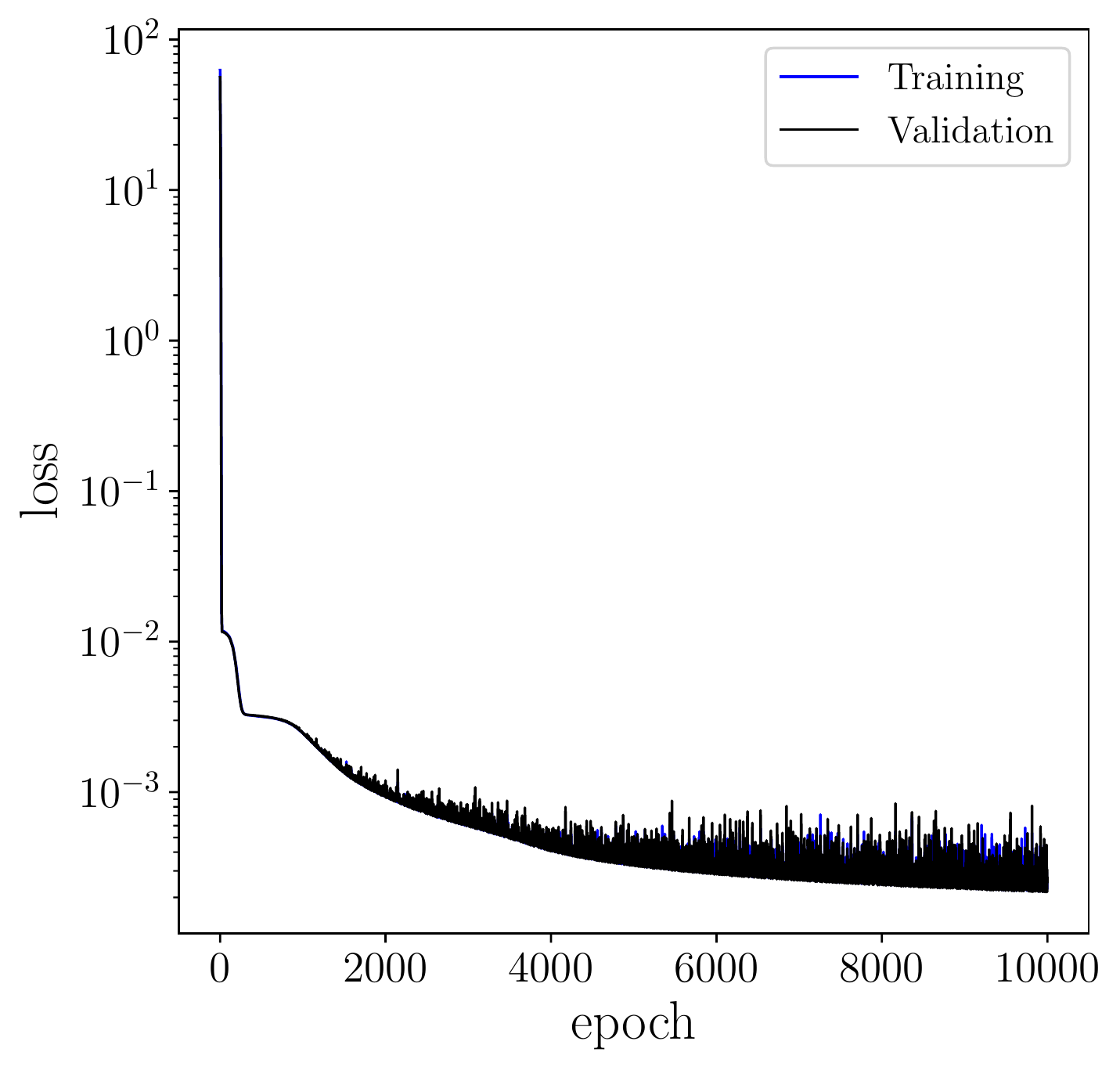}} \hfill
  \subfloat[test dataset prediction]{\includegraphics[height=50mm]{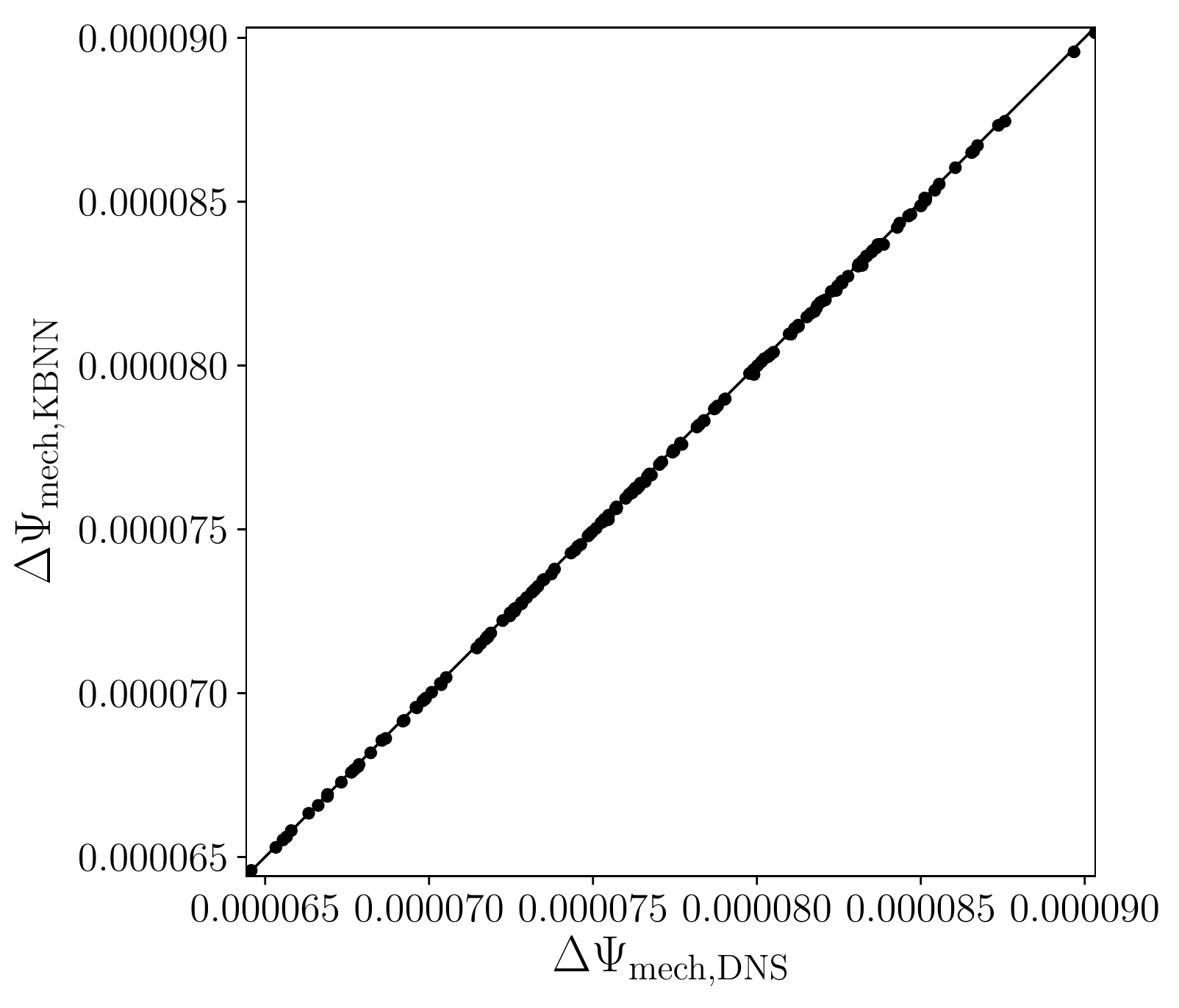}} \hfill
  \subfloat[P11]{\includegraphics[height=50mm]{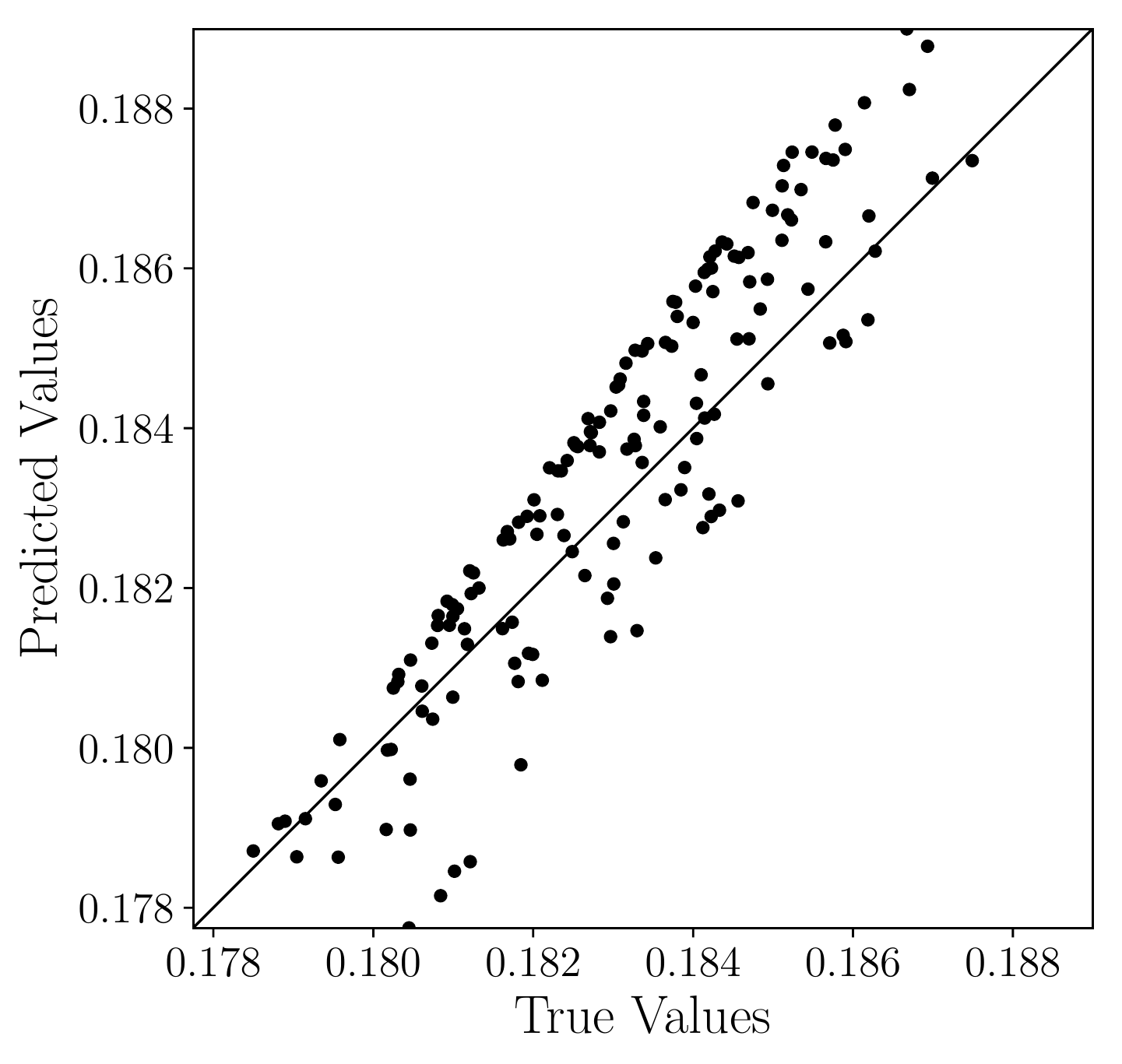}} \\
  \subfloat[P12]{\includegraphics[height=50mm]{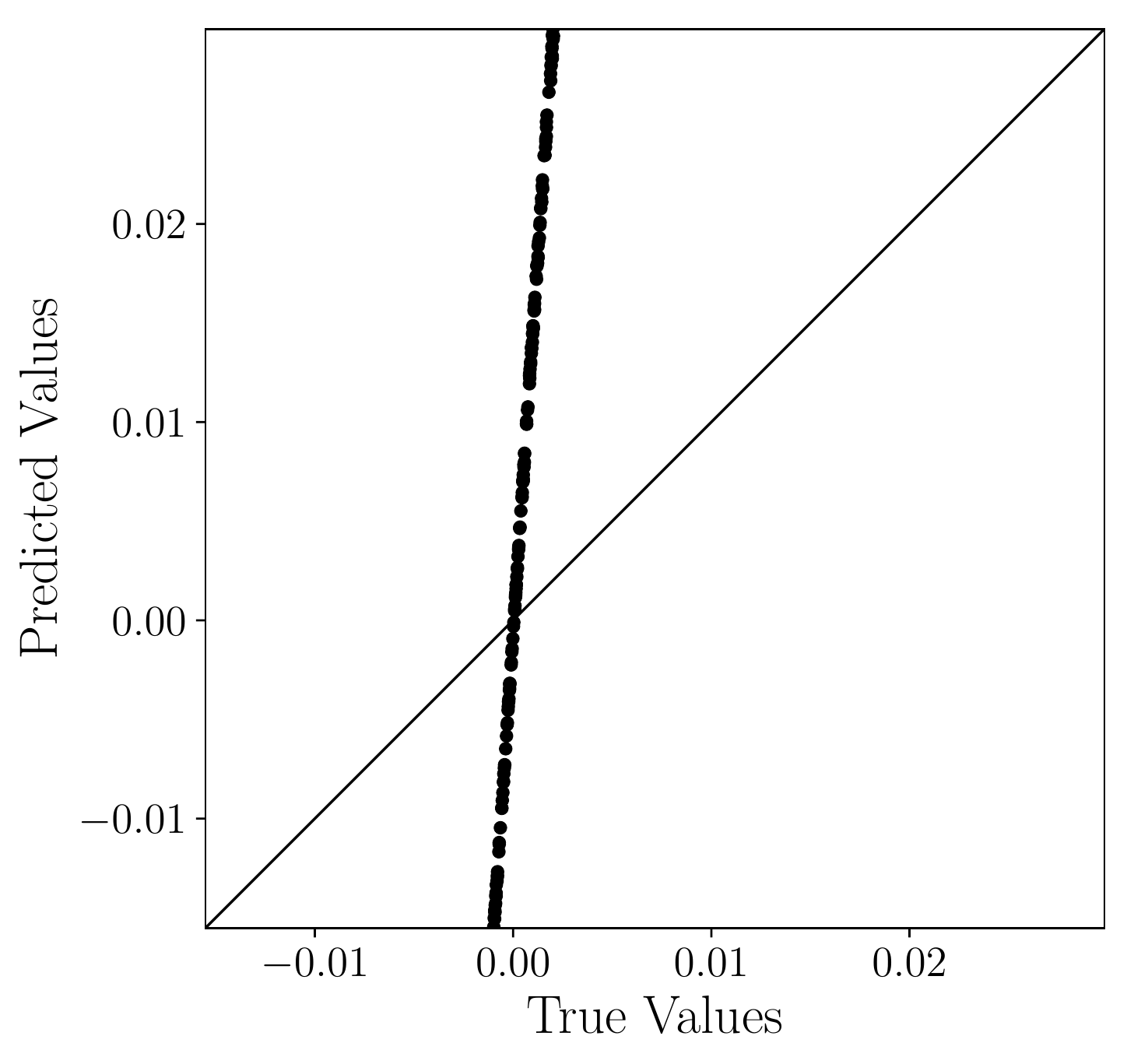}} \hfill
  \subfloat[P21]{\includegraphics[height=50mm]{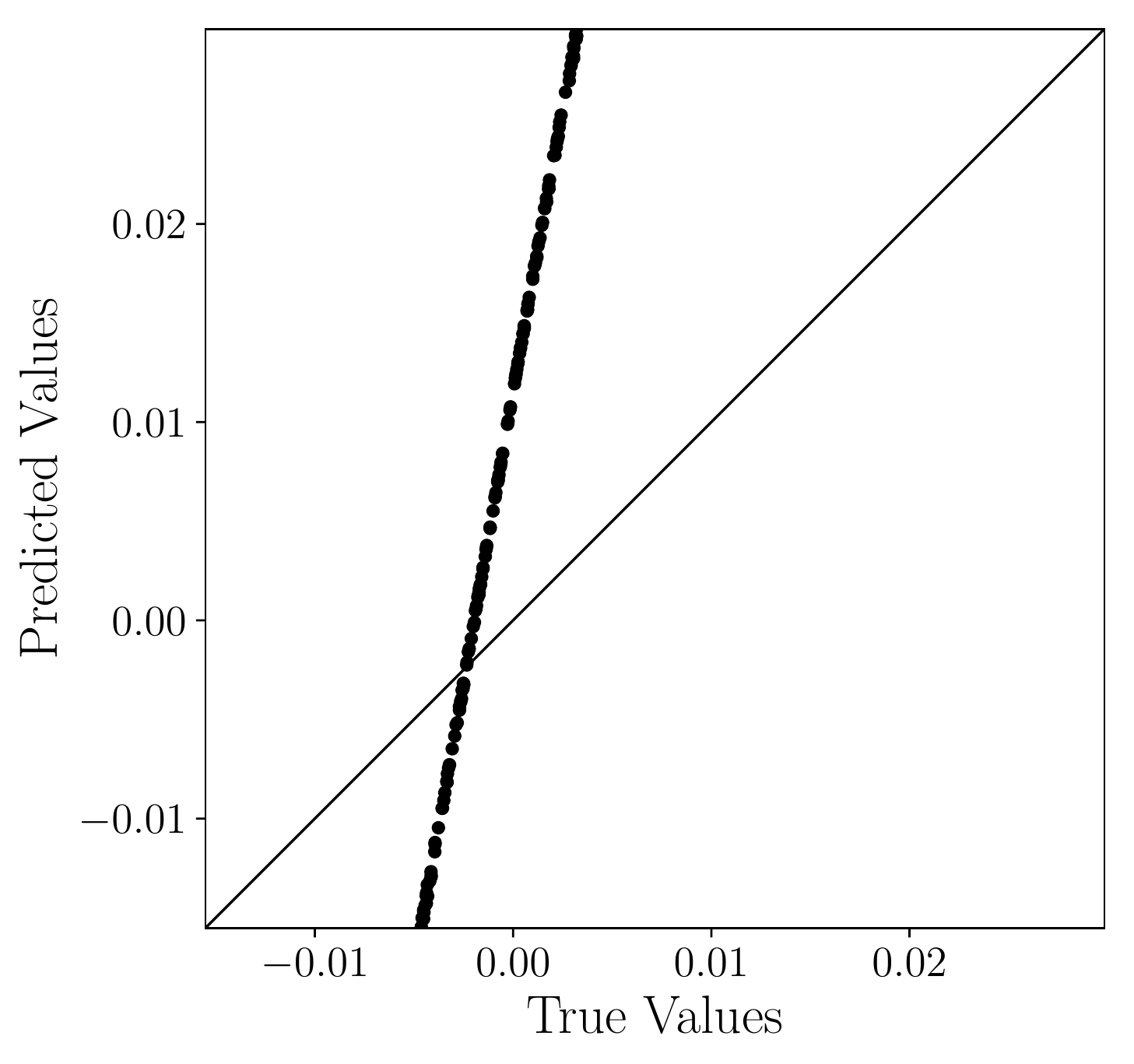}} \hfill
  \subfloat[P22]{\includegraphics[height=50mm]{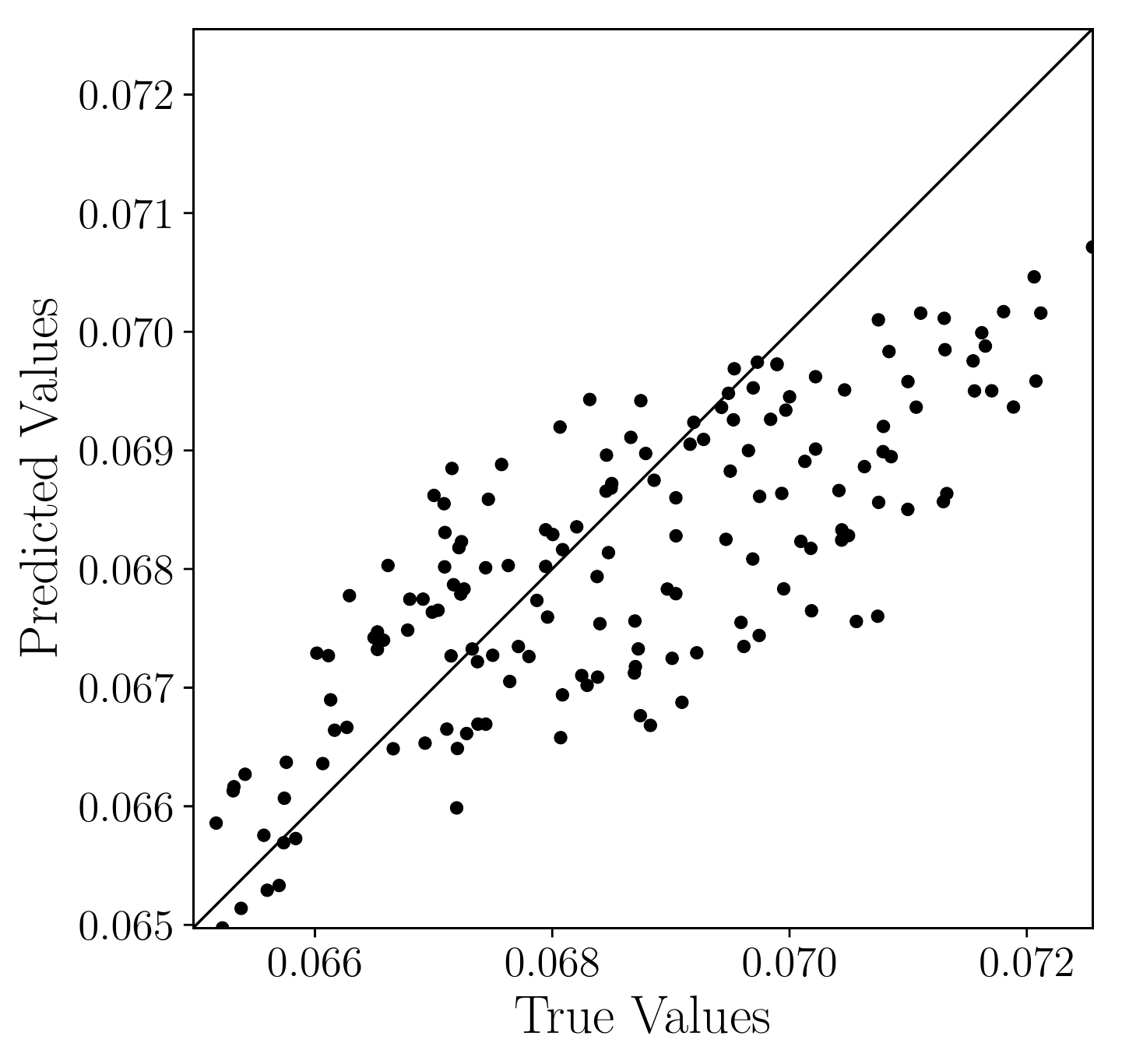}}
  \caption{DNN-based KBNN for one microstructure: 
  (a) learning curve; 
  (b) the KBNN predicted $\Delta \Psi_\text{mech,KBNN}$ vs the actual $\Delta \Psi_\text{mech,DNS}$;
  (c-f) the components of $\BP_\text{KBNN}$ vs $\BP_\text{DNS}$, where the KBNN learns a reasonable  derivative representation for the derived $P_{11}$ and $P_{22}$. The derivative representation of $P_{12}$ and $P_{21}$ is poor because these stress components are one order of magnitude smaller than $P_{11}$ and $P_{22}$ in the data. The inaccurate prediction of $P_{12}$ and $P_{21}$ leads to a wider axis range  for (d,e), resulting in less scattered plots than in (c,f).
}
  \label{fig:psi-800-dnn-kbnn}
\end{figure}

In this section, we discuss KBNNs constructed to study the homogenized mechanical behavior of a single microstructure (dataset $\text{D}_\text{III}$), with the ENNs being either pre-trained DNNs or CNNs.
The ENNs resolve the dominant {characteristics} present in the datasets to allow the KBNNs to capture the detailed {characteristics}.
This is achieved via a new MSE loss function of the form  
\begin{equation}
  \text{MSE} = \frac{1}{m} \sum_{i} \left( \mathbf{Y}- \mathbf{Z} \right)_i^2 
  \quad \text{with} \quad
  \mathbf{Y}= \Psi_\text{mech} - \Psi_\text{mech,NN}^0
  \label{eq:new-mse}
\end{equation}
where $\mathbf{Y}$ is the label, $\mathbf{Z}$ is the KBNN predicted value, $\Psi_\text{mech}$ is the DNS value of the elastic free energy after mechanical testing, and $\Psi_\text{mech,NN}^0$ is the ENN-predicted base elastic free energy of the microstructure before mechanical testing.
In \eref{eq:new-mse}, $\mathbf{Y}$ essentially represents the change of mechanical free energy, $\Delta \Psi_\text{mech}$, resulting from the mechanical testing. In this sense, the MNN resembles a discrepancy model. If the focus were only on capturing this difference (a small fluctuation) in the data resulting from the mechanical testing, the KBNN itself would be very similar to a discrepancy model trained on pre-computed differences in the data. However, the KBNN, as presented here, has the added advantage of describing the entire trend in the data--both the dominant characteristics via the ENN, and the fluctuation via the MNN. Thus, it is capable of multi-resolution representation.

\subsubsection{DNN-based KBNN}\label{sec:dnn-based-kbnn}

\begin{figure}[t]
  \centering
  \subfloat[learning curve]{\includegraphics[height=50mm]{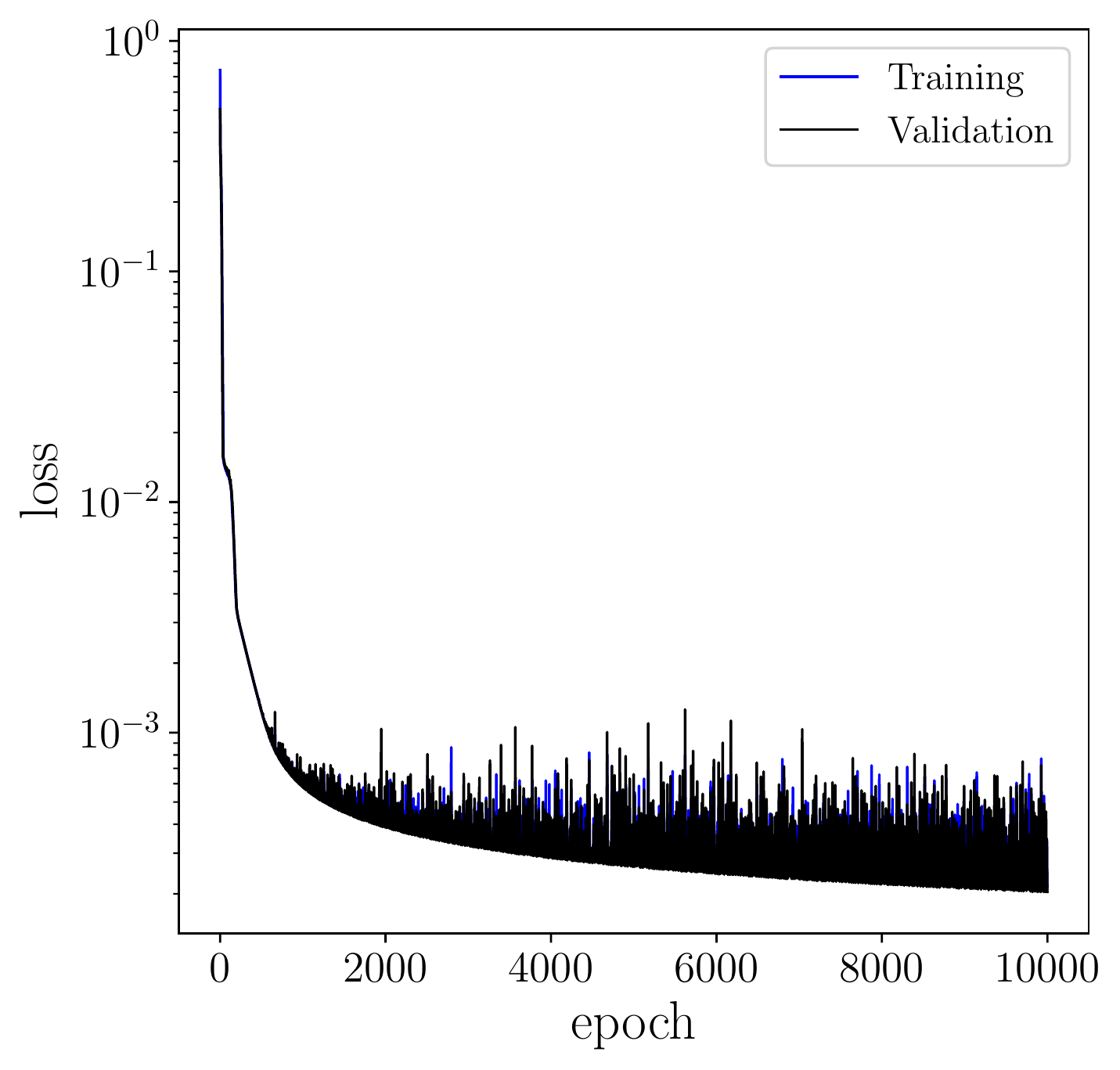}} \hfill
  \subfloat[test dataset prediction]{\includegraphics[height=50mm]{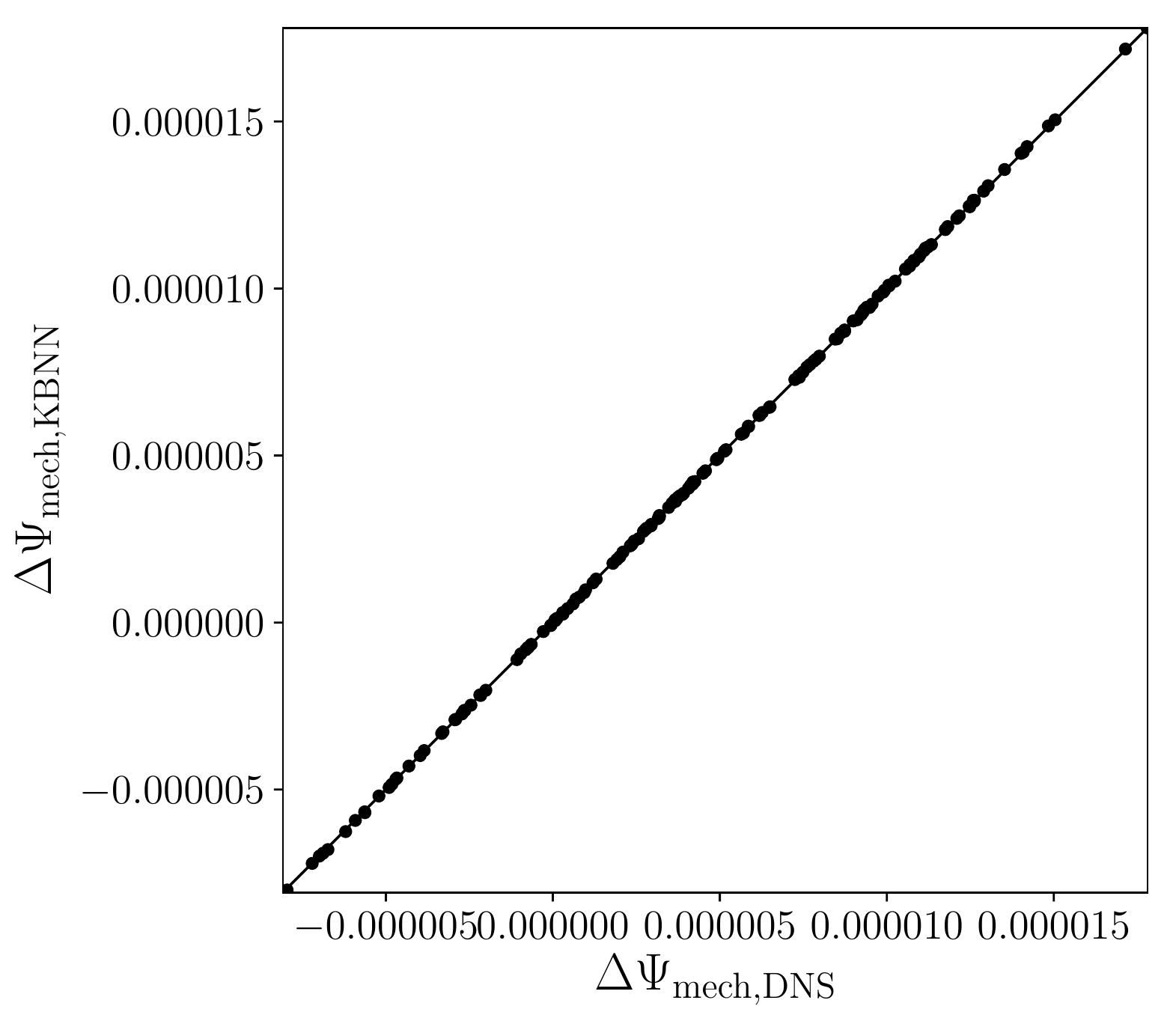}} \hfill
  \subfloat[P11]{\includegraphics[height=50mm]{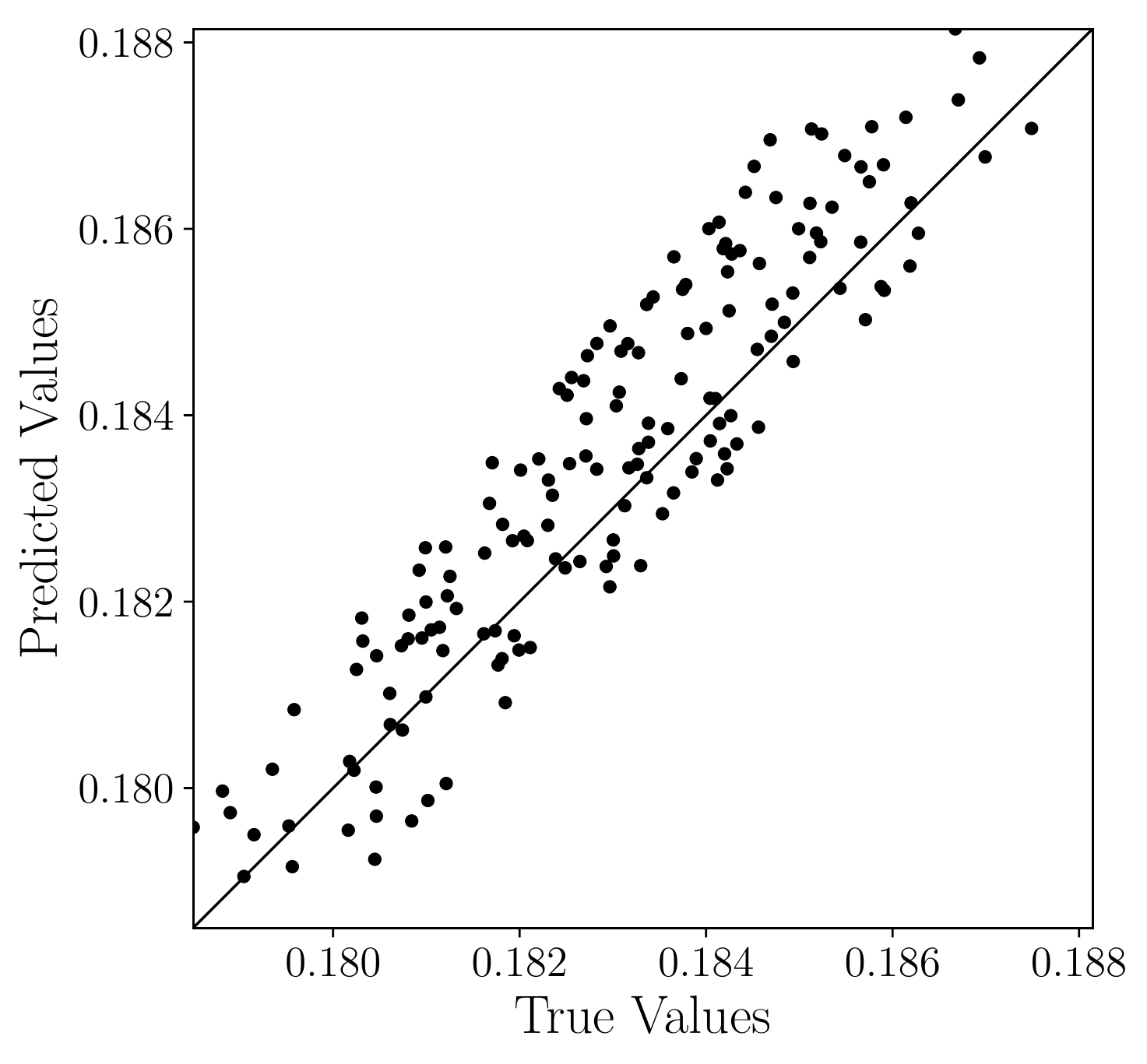}} \\
  \subfloat[P12]{\includegraphics[height=50mm]{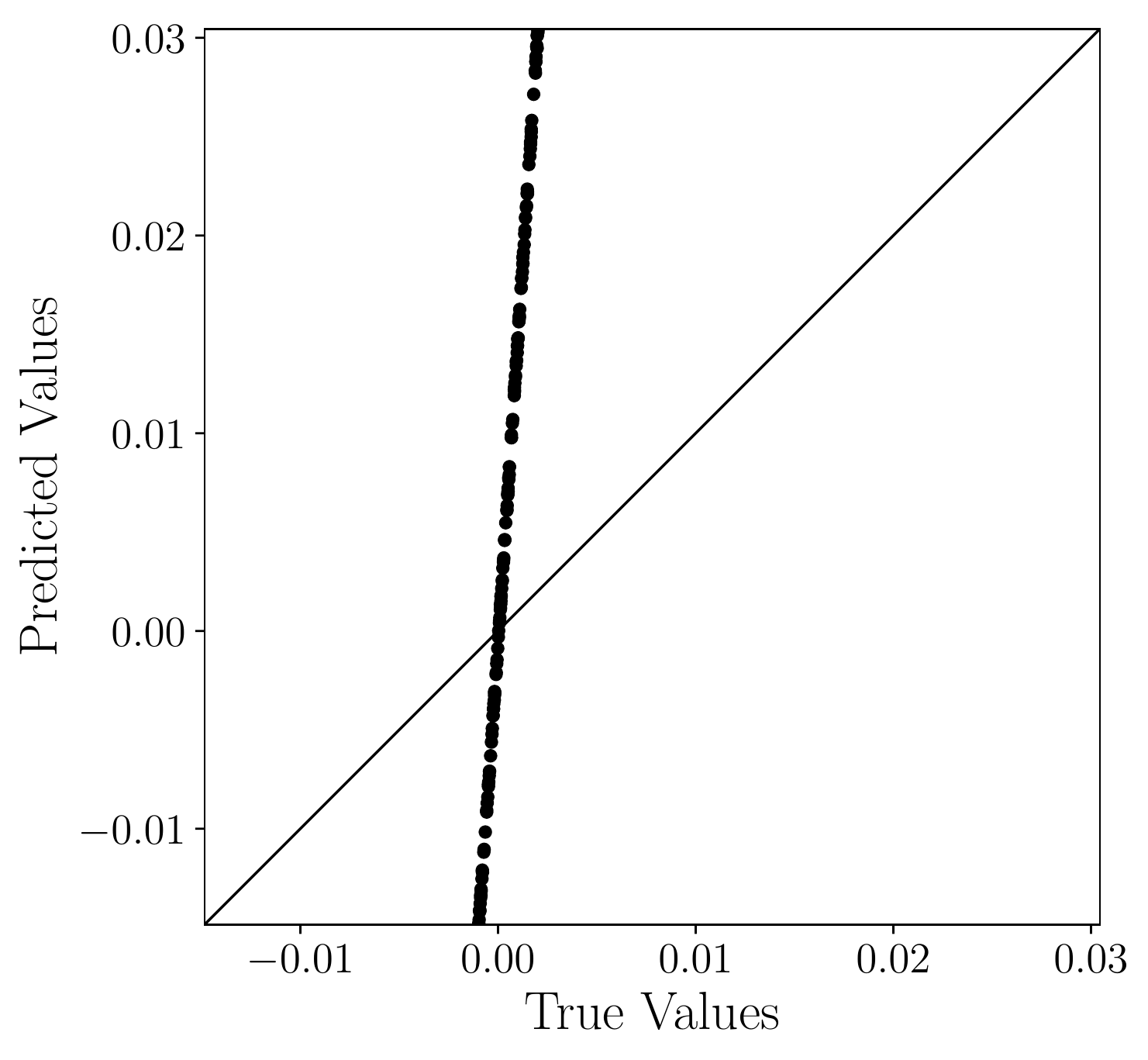}} \hfill
  \subfloat[P21]{\includegraphics[height=50mm]{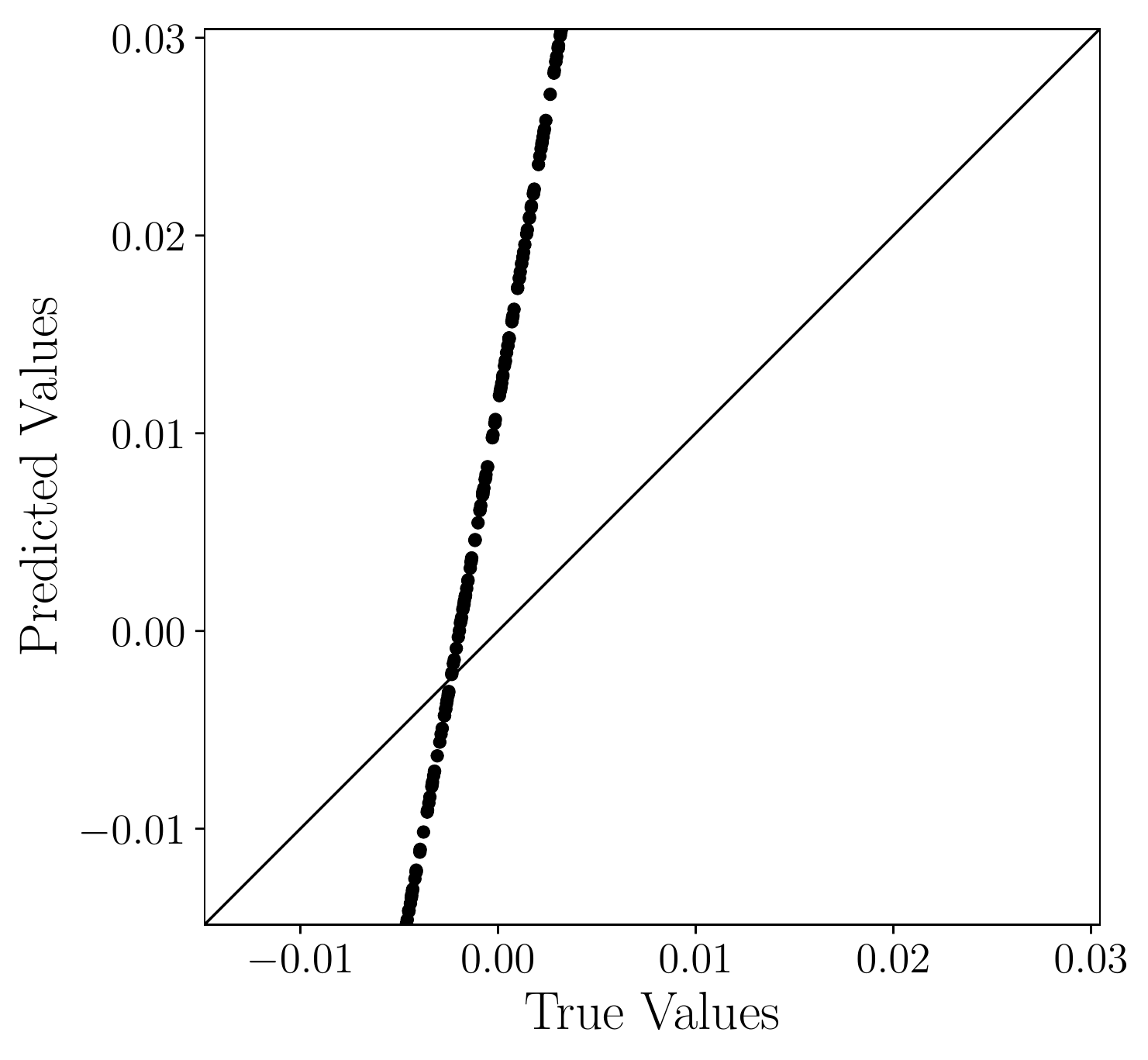}} \hfill
  \subfloat[P22]{\includegraphics[height=50mm]{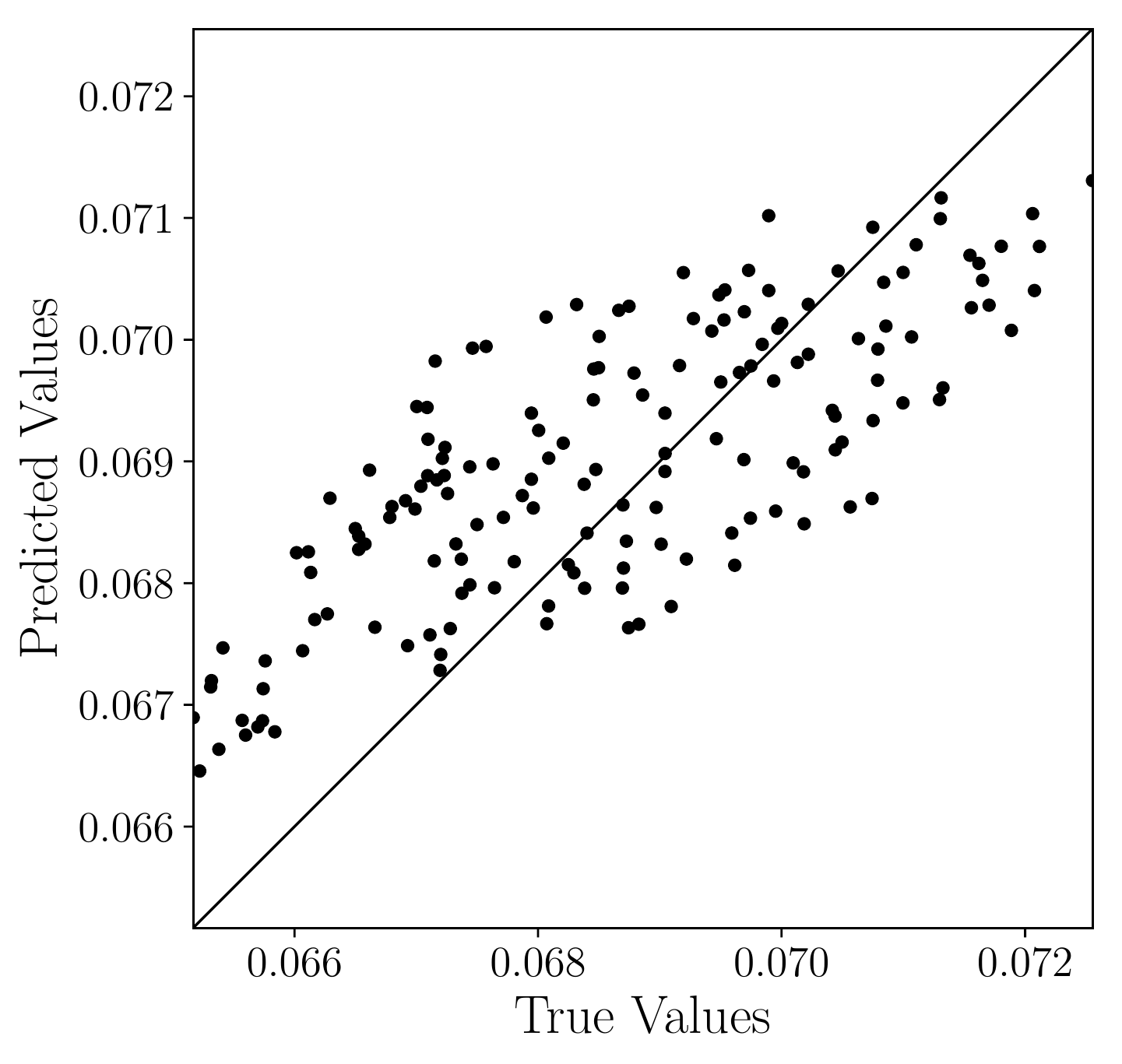}}
  \caption{CNN-based KBNN for one microstructure: 
  (a) learning curve; 
  (b) the KBNN predicted $\Delta \Psi_\text{mech,KBNN}$ vs the actual $\Delta \Psi_\text{mech,DNS}$;
  (c-f) the components of $\BP_\text{KBNN}$ vs $\BP_\text{DNS}$, where the KBNN learns an  derivative representation for $P_{11}$ over the DNN-based KBNN. The accuracy of this derivative representation is reasonable for $P_{11}$ and is approximately the same for $P_{22}$. The derivative representations of $P_{12}$ and $P_{21}$ are poor because these stress components are one order of magnitude smaller than $P_{11}$ and $P_{22}$ in the data.
}
  \label{fig:psi-800-cnn-kbnn}
\end{figure}

With the DNN in Section~\ref{sec:label-shift-dnn} in hand, we now build a KBNN model whose architecture has been presented in Fig.~\ref{fig:kbnn}, with $E_{11}$, $E_{12}$, $E_{22}$, $\phi_r^+$, $\phi_r^-$, $l_s^r$, $l^{r+}$, and $l^{r-}$ as features and $\Psi_\text{mech}$ as the label.
In this KBNN,  the embedded pre-trained DNN takes $\{\phi_r^+,~\phi_r^-,~l_s^r,~l^{r+},~l^{r-} \}$ to predict $\Psi_\text{mech,NN}^0$. 
The remaining features $\{E_{11},~E_{12},~E_{22} \}$ and the shifted label $\Delta \Psi_\text{mech} = \Psi_\text{mech} - \Psi_\text{mech,NN}^0$ are used to optimize the variables of the MNN.
The MNN is not exposed to the features $\{\phi_r^+,~\phi_r^-,~l_s^r,~l^{r+},~l^{r-} \}$, {and therefore, while trained against one microstructure it cannot represent the homogenized mechanical response of a different one. Such an extension to multiple microstructures is a refinement we undertake in Section \ref{sec:sim-kbnn-multi-DNS}.}
The optimal values of $N_\text{HL}$ and $N_\text{NPL}$ for the MNN are searched by following the procedures in Section~\ref{sec:hyperparameter}.
An $L^2$ kernel regularization with a factor of 0.001 is applied to the input layer to minimize the coefficients of less important features to reduce overfitting. 
The Softplus activation function is used.
An optimal MNN is obtained with $N_\text{HL}=2$,  $N_\text{NPL}=26$, and $V_\text{total} = 833$.
The KBNN is trained with the Adam optimizer for 10000 epochs with the exponentially decaying learning rate given in \eref{eq:lr-step} where $v_\text{decay} = 0.92$.
The learning curve for the KBNN is plotted in Fig.~\ref{fig:psi-800-dnn-kbnn}(a), where neither overfitting nor underfitting is observed.
Fig.~\ref{fig:psi-800-dnn-kbnn}(b) shows that the KBNN can capture the detailed {characteristics} of the data and predict $\Delta \Psi_\text{mech}$ with satisfactory accuracy. 
To further evaluate the model performance, we compute the KBNN-predicted first Piola-Kirchhoff stress
\begin{equation}
  \BP_\text{KBNN} = \BF^\text{avg}\BS_\text{KBNN}  
  \label{eq:p_kbnn}
\end{equation}
where $\BS_\text{KBNN}$ is the derivative of $\Delta \Psi_\text{mech, KBNN}$ with respect to the features $\BE$. We evaluate the difference between NN-predicted value and the one computed from DNS for $\BP$ instead of $\BS$ because $\BP$ is a quantity that can be directly computed based on surface traction from DNS. Labels are thus easier to generate for $\BP$. The comparison between $\BP_\text{KBNN}$ and $\BP_\text{DNS}$ are shown in Fig.~\ref{fig:psi-800-dnn-kbnn}(c-f). The KBNN shows good performance for the derivative representations $P_{11}$ and $P_{22}$, but is poor for $P_{12}$ and $P_{21}$ due to the fact that the data on these stress components are one order of magnitude smaller than those for $P_{11}$ and $P_{22}$ in the DNS.

\subsubsection{CNN-based KBNN} 

A CNN-based KBNN is built with $E_{11}$, $E_{12}$, $E_{22}$, and the image of the $e_2$ field solution as features and $\Psi_\text{mech}$ as the label.
In this KBNN, the embedded pre-trained CNN takes the image of the $e_2$ field solution of the base microstructure to predict $\Psi_\text{mech}^0$ {(recall the explanation in Section \ref{sec:label-shift-cnn} for using the $e_2$ field)}. 
The remaining features $\{E_{11},~E_{12},~E_{22} \}$ and the shifted label $\Delta \Psi_\text{mech} = \Psi_\text{mech} - \Psi_\text{mech}^0$ are used to optimize the variables of the MNN.
The  MNN trained and used here is identical to that in Section~\ref{sec:dnn-based-kbnn}.
Fig.~\ref{fig:psi-800-cnn-kbnn}(b) shows that the KBNN can resolve the detailed {characteristics} in the data and predict $\Delta \Psi_\text{mech}$ with satisfactory accuracy. 
The KBNN-predicted $\BP_\text{KBNN}$ in \eref{eq:p_kbnn} is compared with $\BP_\text{DNS}$ in Fig.~\ref{fig:psi-800-cnn-kbnn}(c-f), and represents a small improvement on the results in Fig.~\ref{fig:psi-800-dnn-kbnn}. Now, the CNN-based KBNN performs well at predicting $P_{11}$ and $P_{22}$ as derivative representations, but continues to perform poorly on $P_{12}$ and $P_{21}$ due to the order of magnitude difference in DNS data.

\begin{table}[h!]
  \centering
  \begin{tabular}{l | l | l}
    \hline
    Layers &   & Notes \\ \hline
    Input (1) & perturbed $e_2$ fields  & \\
    Conv2D & filters = 8 & kernel (3,3), stride (2,2), padding (2,2), ReLU \\
    MaxPooling2D & - & kernel (2,2), stride (1,1), padding (1,1)\\
    Conv2D & filters = 16 & kernel (3,3), stride (2,2), padding (2,2), ReLU \\
    MaxPooling2D & - & kernel (2,2), stride (1,1), padding (1,1)\\
    Conv2D & filters = 24 & kernel (3,3), stride (2,2), padding (2,2), ReLU  \\
    MaxPooling2D & - & kernel (2,2), stride (1,1), padding (1,1)\\
    Flatten & - & -\\
    Dense (*) & neurons = 8 & ReLU  \\
     Input (2) & $E_{11},~E_{12},~E_{22}$ & - \\
    Concatenate & Dense (*) + Input (2) & \\
    Dense & neurons = 48 & Softplus\\
    Dense & neurons = 48 & Softplus\\
    Dense & neurons = 48 & Softplus\\
    Output Dense Layer & label = 1 & Linear \\
  \end{tabular}
  \caption{Details of the MNN for predicting homogenized mechanical response of multiple microstructures sampled from different DNS.}
  \label{tab:cnn-enhanced-MNN}
\end{table}

\begin{figure}[h!]
    \centering
    \includegraphics[width=0.9\linewidth]{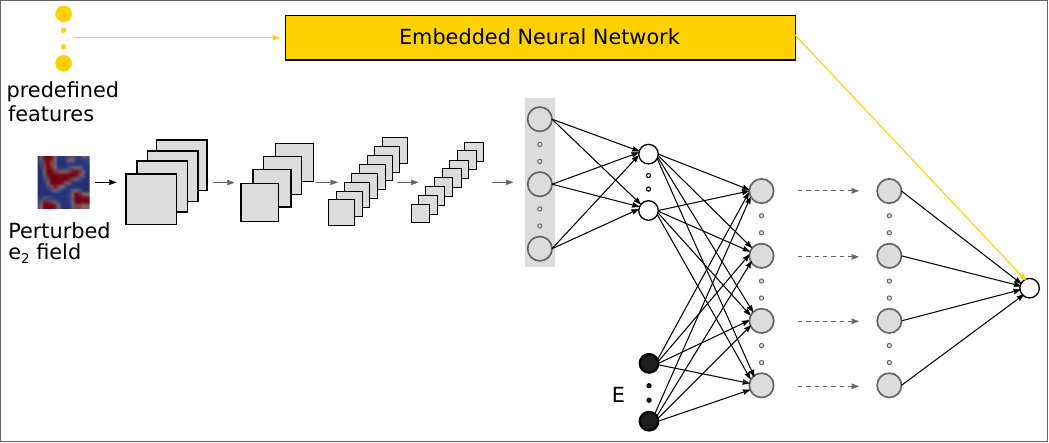}
    \caption{Illustration of the structure of CNN-enhanced KBNN. The ENN, which takes pre-defined features as inputs, is used to resolve the dominant characteristics. Alternatively, the ENN could take the microstructure images as inputs.  The MNN includes a CNN (shown horizontally across the middle), which takes the perturbed $e_2$ field information, and is used to identify the most relevant features for homogenized mechanical behavior prediction. The combination of the outputs from the CNN and the components of $\BE$ serves as the input for a fully connected DNN that resolves the detailed {characteristics} of the dataset.}
    \label{fig:cnn-enchanced-kbnn}
\end{figure}

\begin{figure}[h!]
  \centering
  \subfloat[learning curve]{\includegraphics[height=50mm]{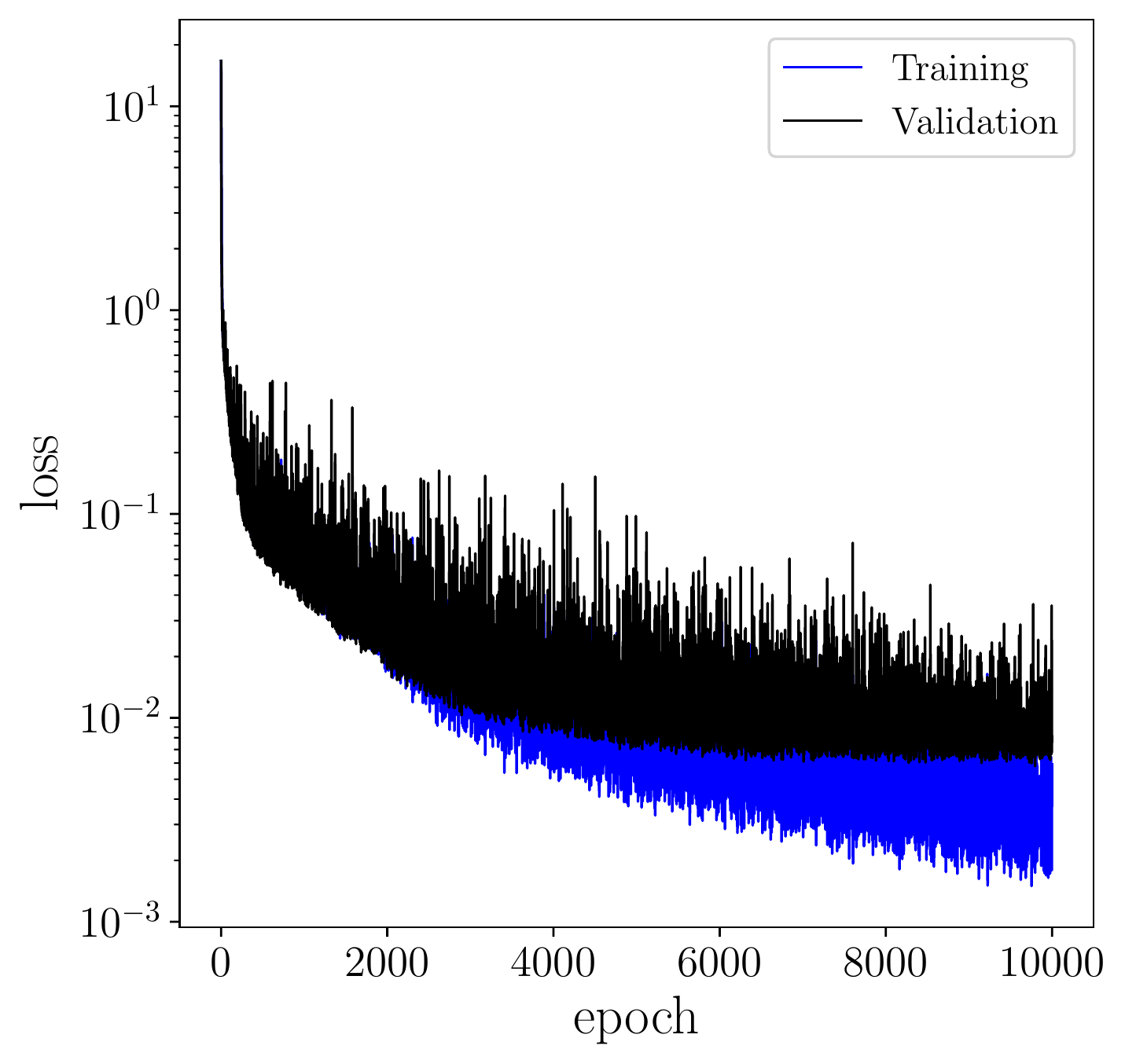}} \hfill
  \subfloat[test dataset  prediction]{\includegraphics[height=50mm]{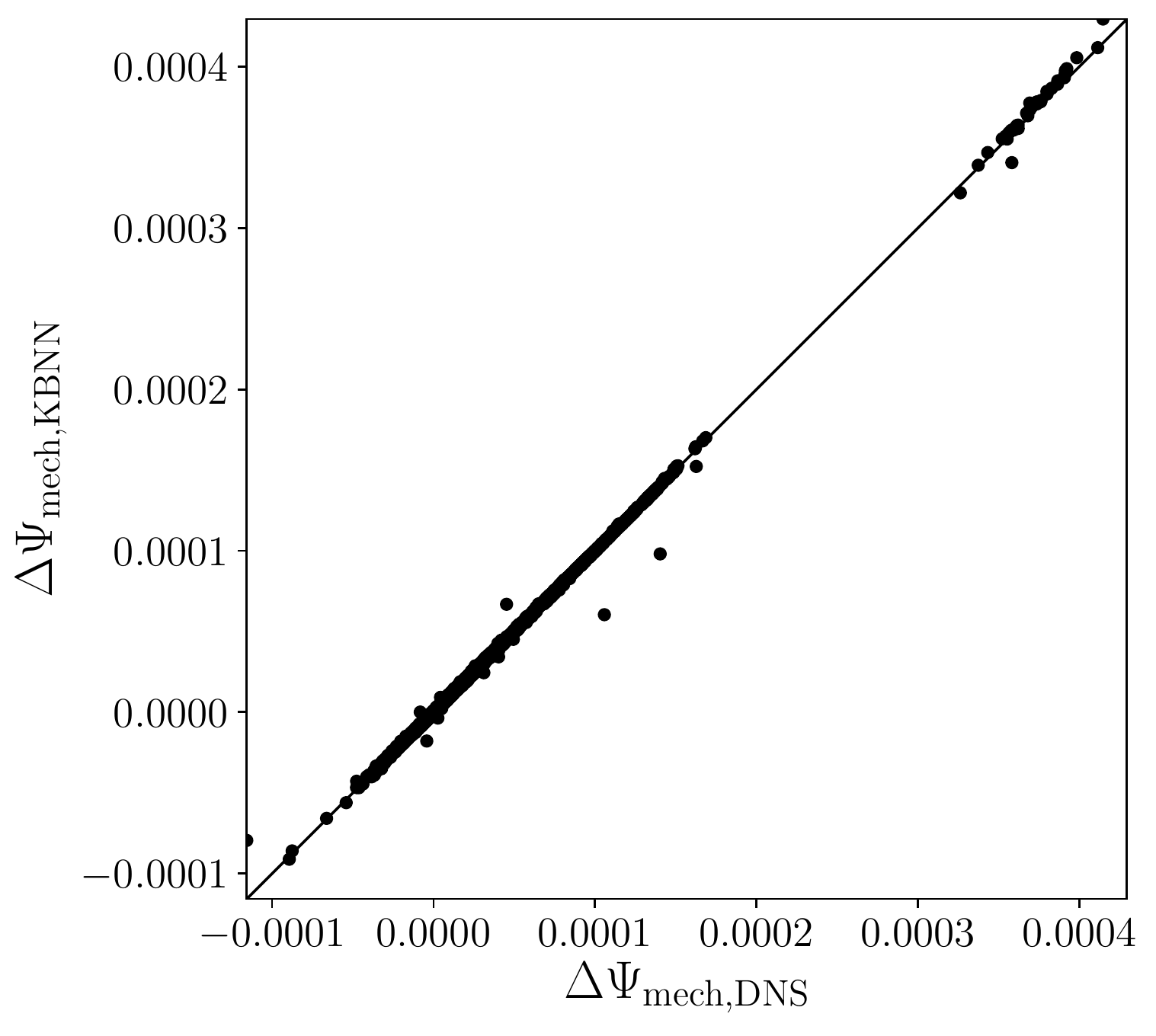}} \hfill
  \subfloat[P11]{\includegraphics[height=50mm]{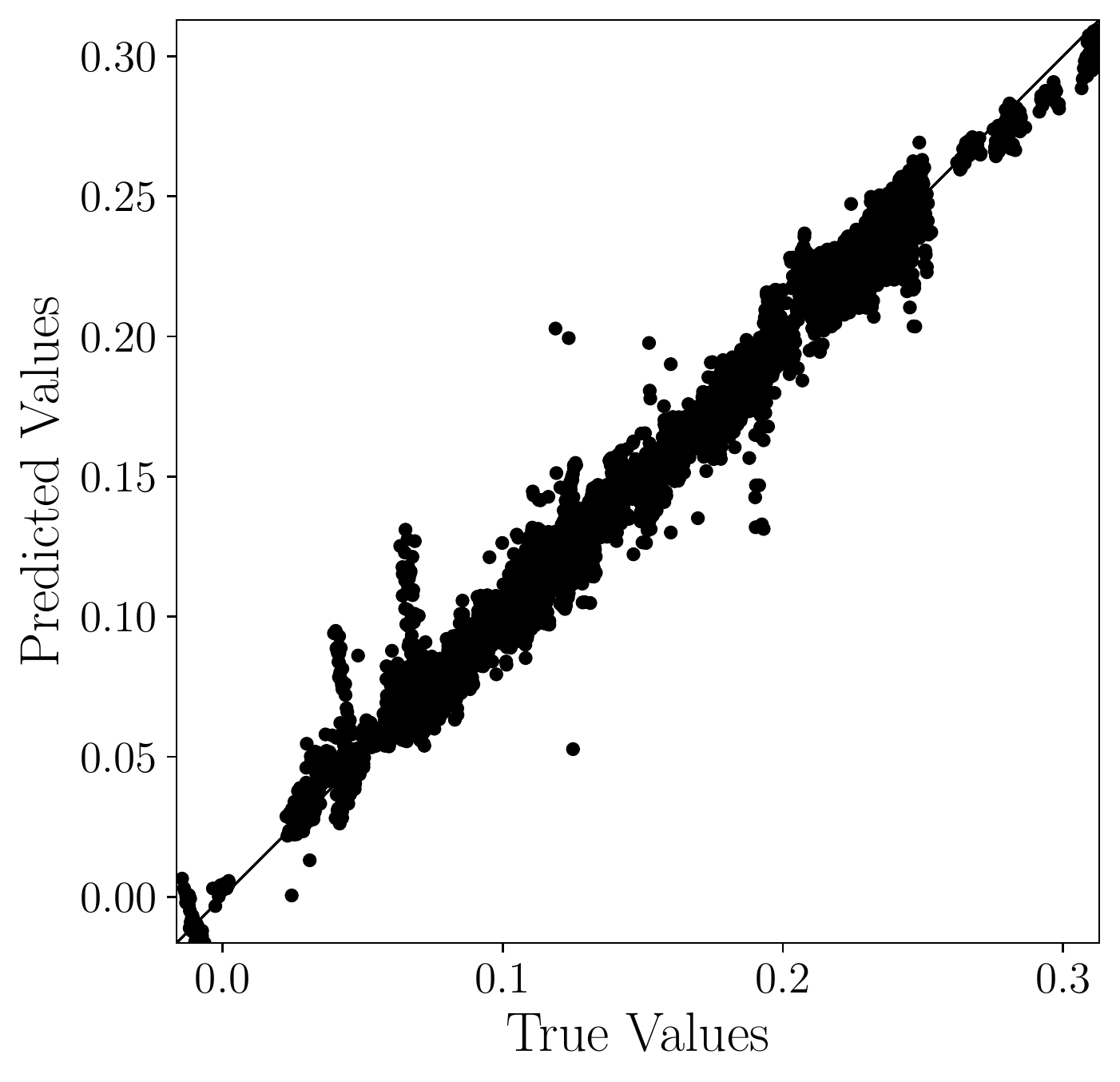}} \\
  \subfloat[P12]{\includegraphics[height=50mm]{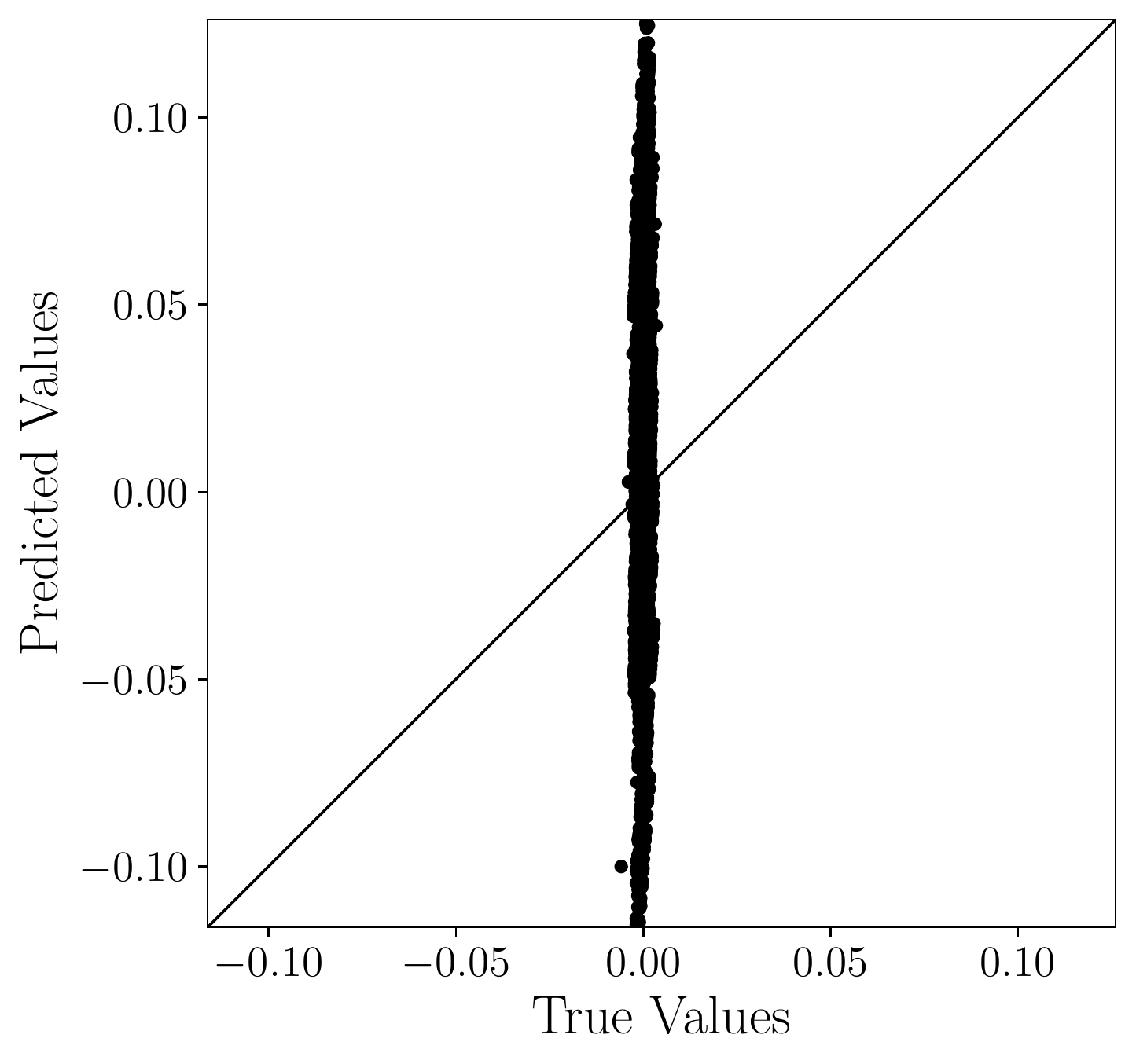}} \hfill
  \subfloat[P21]{\includegraphics[height=50mm]{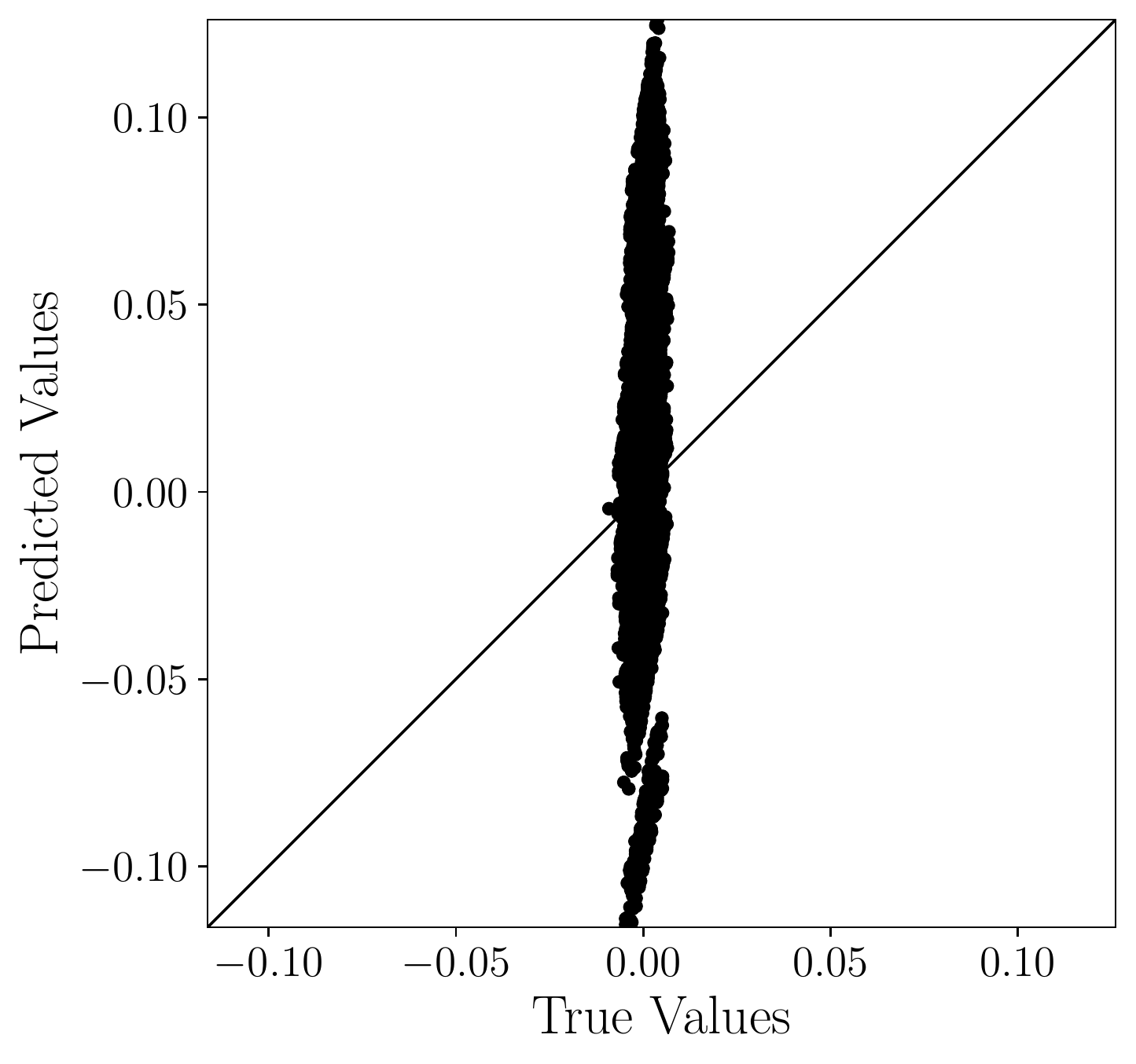}} \hfill
  \subfloat[P22]{\includegraphics[height=50mm]{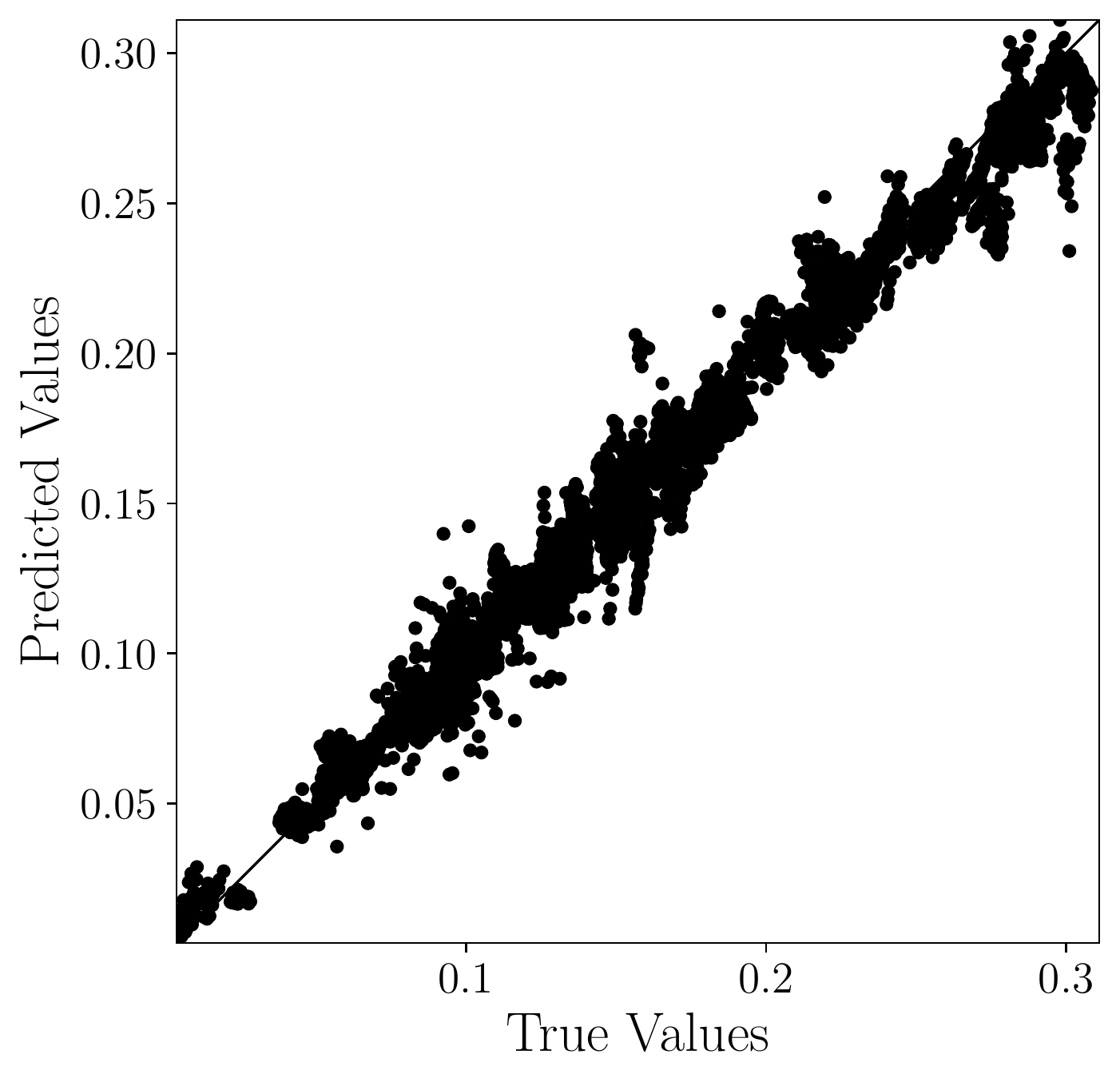}} 
  \caption{ { CNN enhanced KBNN for 180 microstructures from different DNS with the enhancing CNN being exposed to the entire perturbed $e_2$ field: }
  (a) learning curve; 
  (b) the KBNN predicted $\Delta \Psi_\text{mech,KBNN}$ \emph{versus} the actual $\Delta \Psi_\text{mech,DNS}$;
  (c-f) the components of $\BP_\text{KBNN}$ vs $\BP_\text{DNS}$, where the KBNN shows good derivative representations for $P_{11}$ and $P_{22}$, but not $P_{12}$ and $P_{21}$ because the DNS data for $P_{12}$ and  $P_{21}$ are one order of magnitude smaller than for $P_{11}$ and $P_{22}$.
  }
  \label{fig:kbnn-m-dns}
\end{figure}

\subsection{Homogenized mechanical behavior of microstructures from multiple DNS} \label{sec:sim-kbnn-multi-DNS}

Expanding beyond the studies for a single microstructure, KBNNs are constructed to predict the homogenized behavior of multiple microstructures from different DNSs (dataset $\text{D}_\text{IV}$).
 KBNNs similar to those used in Section~\ref{sec:sim-kbnn-one-microstructure} are investigated.
However, the MNN with $\{E_{11}$, $E_{12}$, $E_{22}\}$ as features is incapable of describing the homogenized mechanical behavior of different microstructures, as such a simple MNN is unexposed to the details of each microstructure as explained in Section \ref{sec:dnn-based-kbnn}.
Our studies also confirm that even the inclusion of pre-defined microstructure related features $\{\phi_r^+$, $\phi_r^-$, $l_s^r$, $l^{r+}$, $l^{r-}\}$ in the MNN shows insignificant improvement of the performance of KBNNs for multiple microstructures.

\subsubsection{CNN-enhanced KBNN} \label{sec:cnn-enhanced-kbnn}

Since the MNN with pre-defined features has insufficient expressivity to describe the homogenized mechanical response across microstructures, a CNN-enhanced KBNN structure, as shown in Fig.~\ref{fig:cnn-enchanced-kbnn}, is explored. {Now, the MNN takes both $\{E_{11}$, $E_{12}$, $E_{22}\}$ and the perturbed $e_2$ data as features, with the CNN enhancement being utilized to identify additional relevant features from the $e_2$ data for homogenized mechanical behavior prediction. The pre-trained DNN obtained in Section~\ref{sec:sim-base-free-energy-m-dns} is used as the ENN. The new KBNN takes $\Delta\Psi_\text{mech}$ as the label.} 
A manual hyperparameter tuning is performed.
The details of an MNN with satisfactory performance, which has a total variable number of 10257, are summarized in Table~\ref{tab:cnn-enhanced-MNN}. 
Our results, as shown in Fig.~\ref{fig:kbnn-m-dns}, confirm the effectiveness and good expressivity of the new KBNN structure, which can accurately predict the mechanical free energy on the test dataset.
Furthermore, the $P_{11}$ and $P_{22}$ components of $\BP_\text{KBNN}$, obtained based on \eref{eq:p_kbnn}, match well with respective components of $\BP_\text{DNS}$.

\subsubsection{CNN-enhanced KBNN with penalization}

\begin{figure}[h!]
  \centering
  \subfloat[learning curve]{\includegraphics[height=50mm]{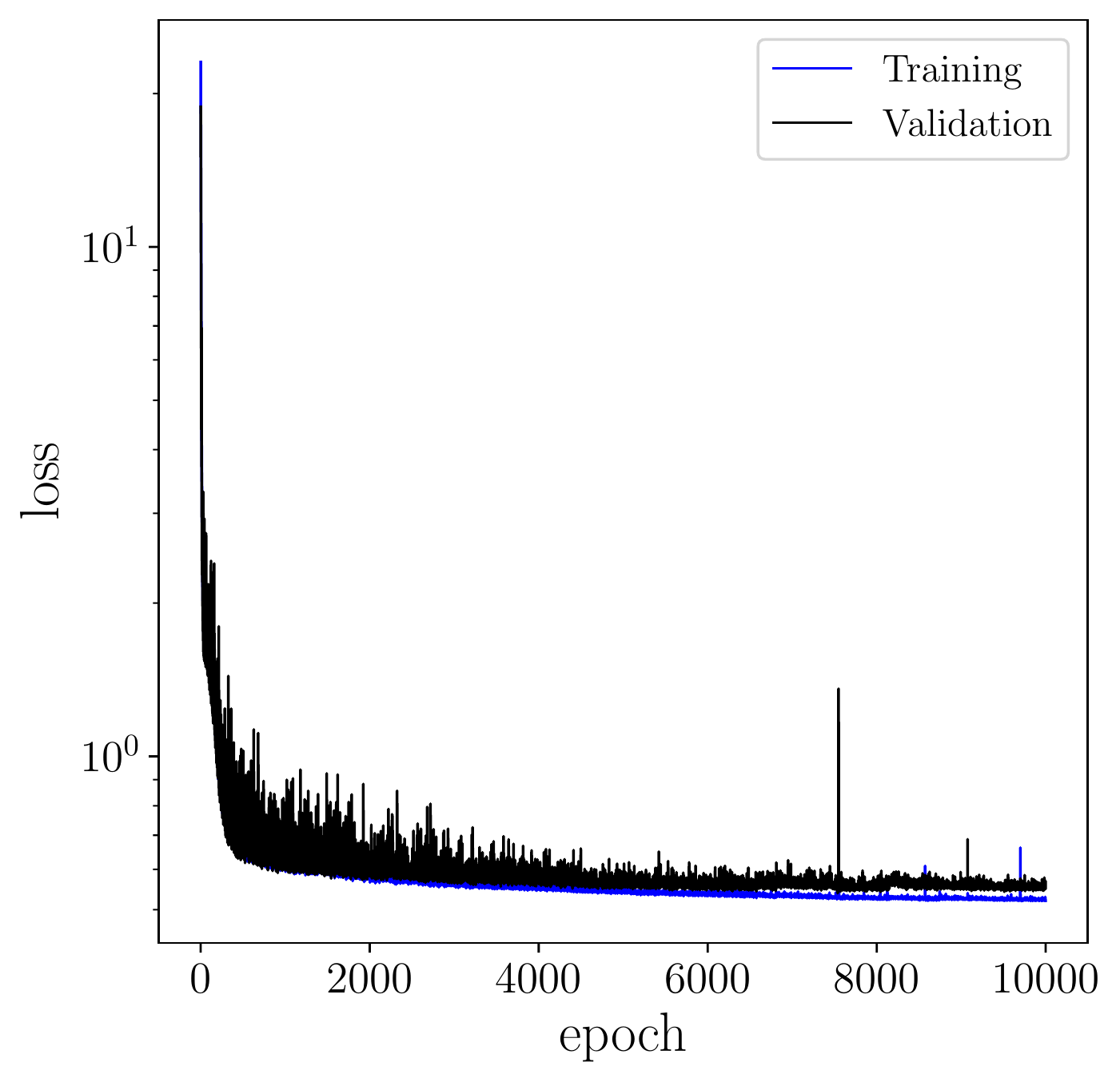}} \hfill
  \subfloat[test dataset prediction]{\includegraphics[height=50mm]{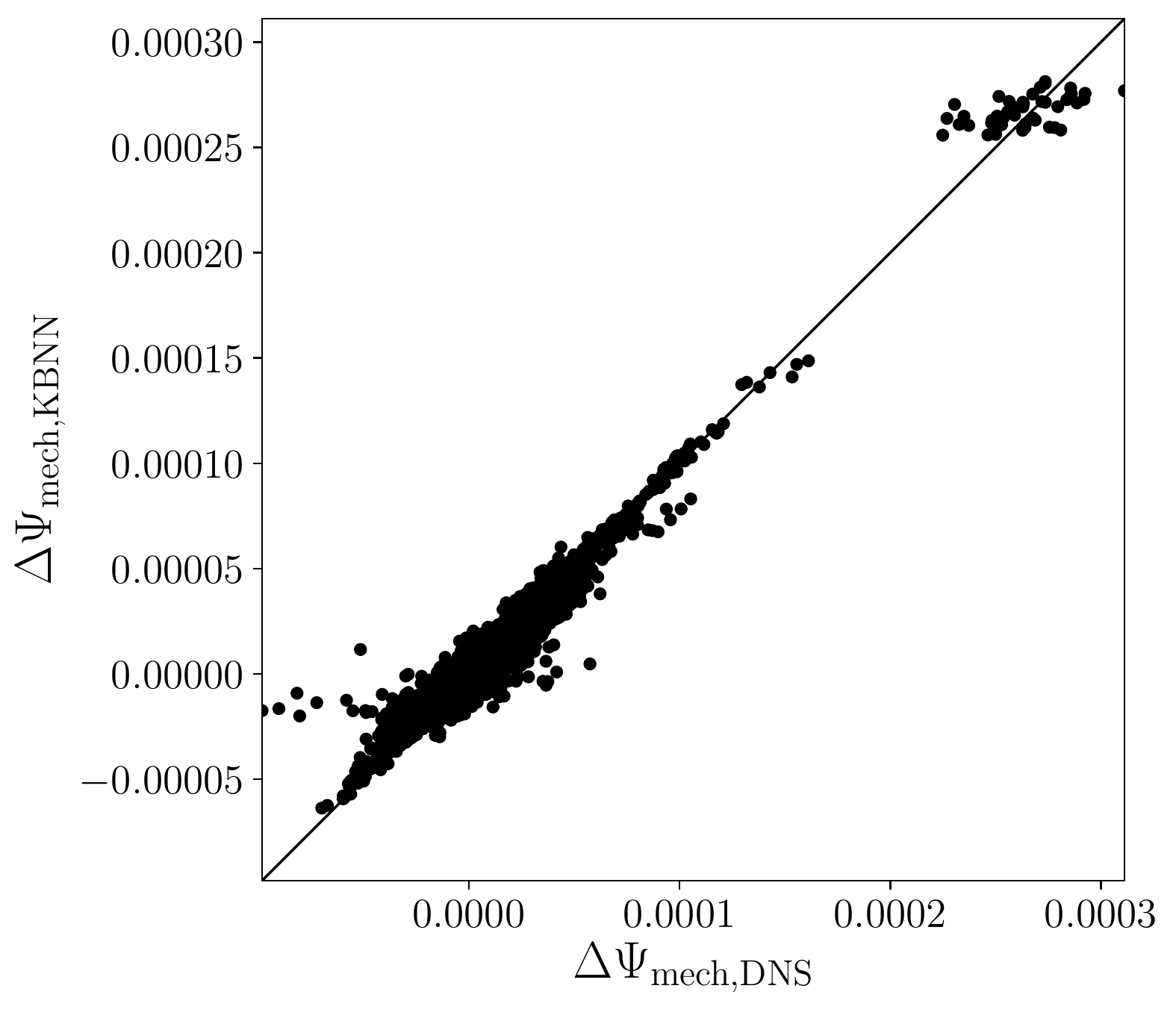}} \hfill
  \subfloat[P11]{\includegraphics[height=50mm]{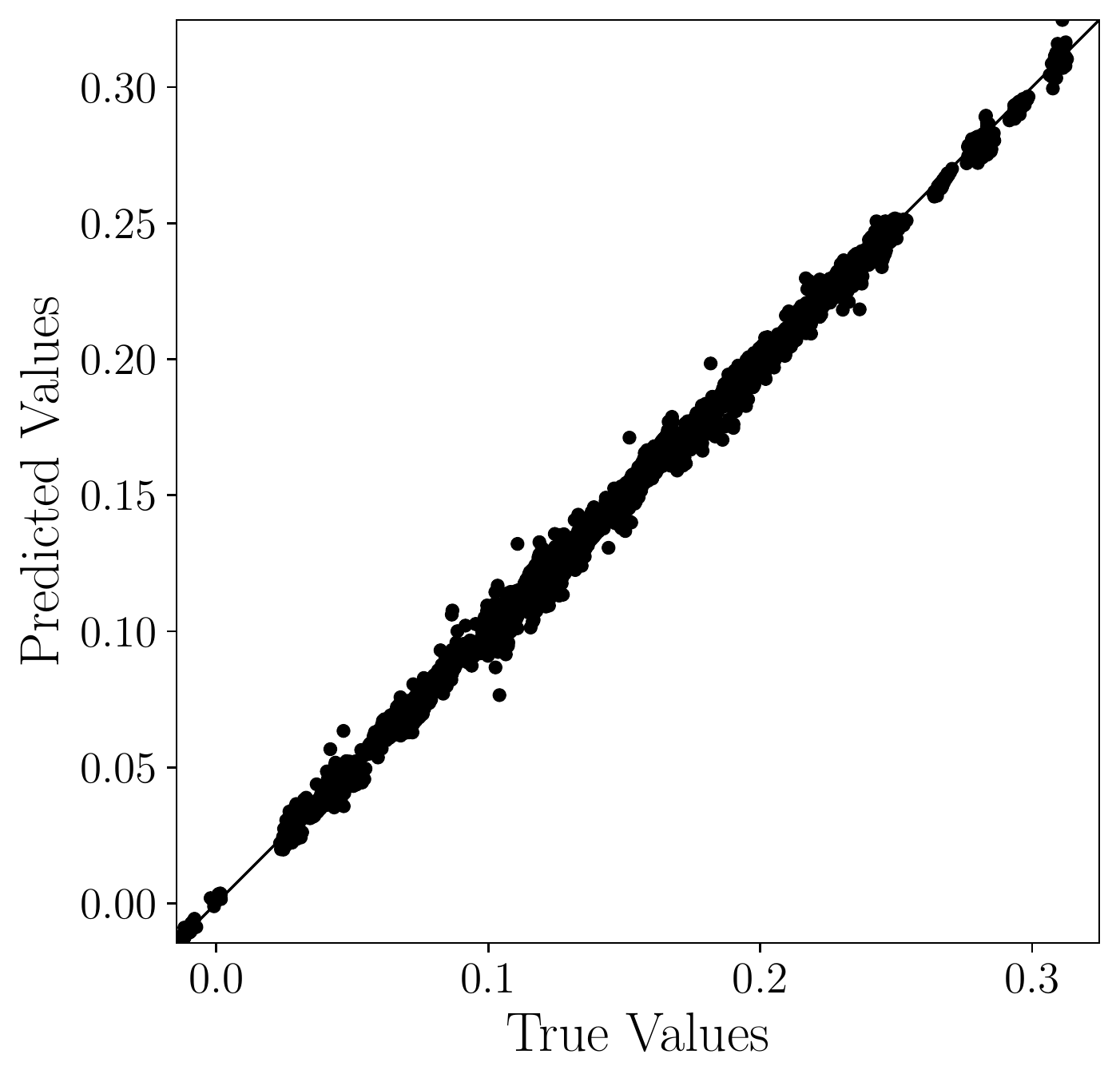}} \\
  \subfloat[P12]{\includegraphics[height=50mm]{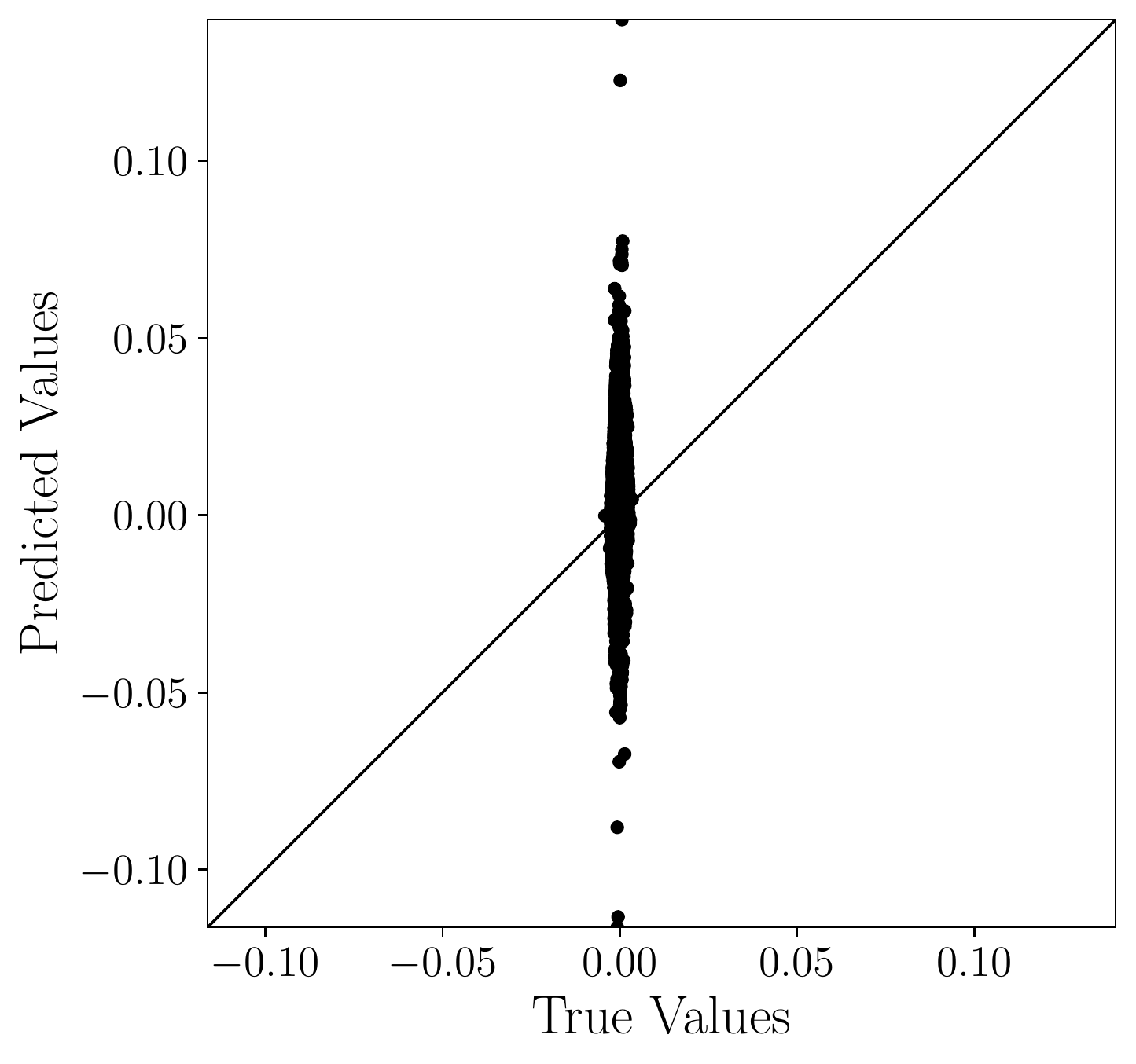}} \hfill
  \subfloat[P21]{\includegraphics[height=50mm]{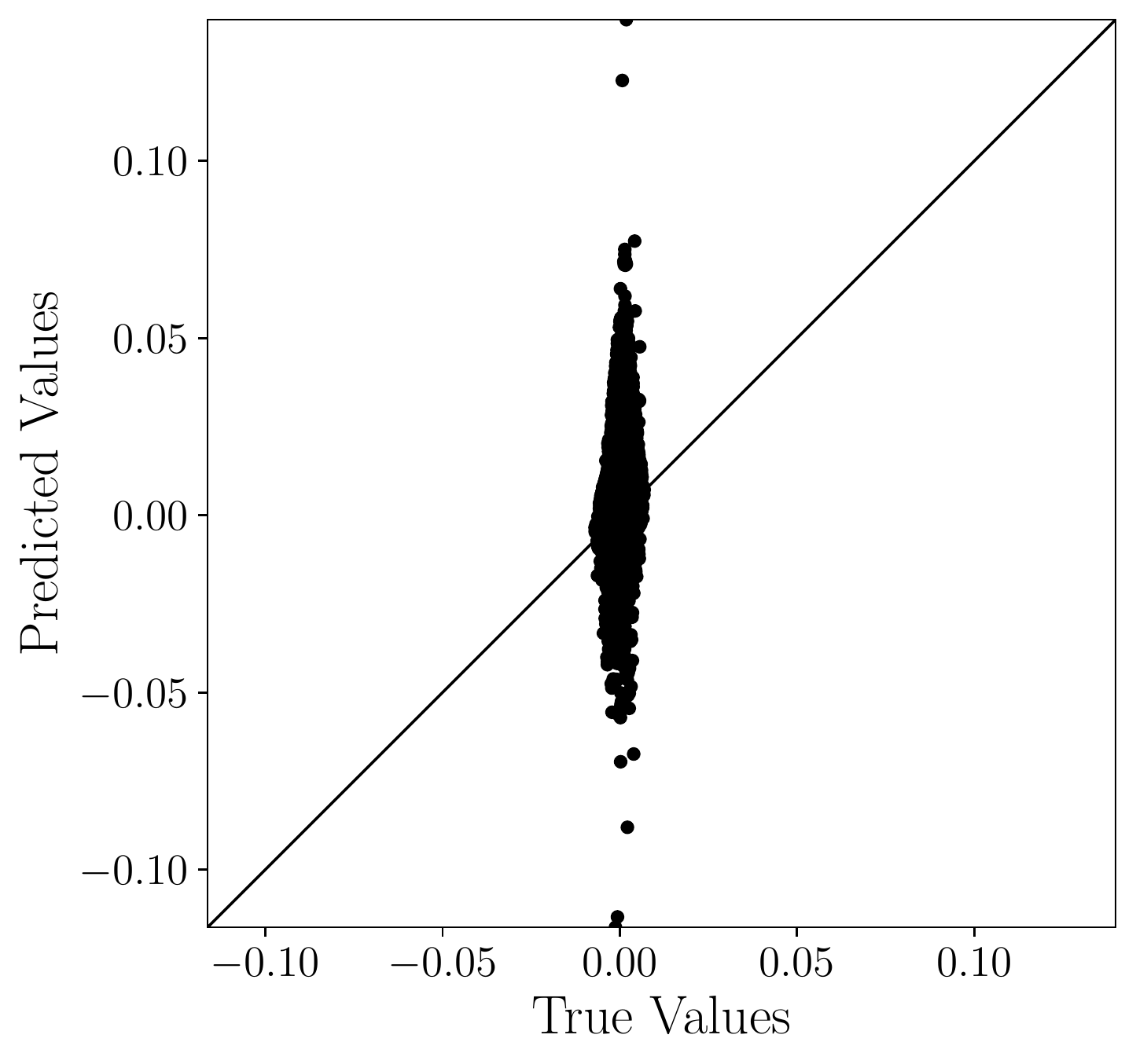}} \hfill
  \subfloat[P22]{\includegraphics[height=50mm]{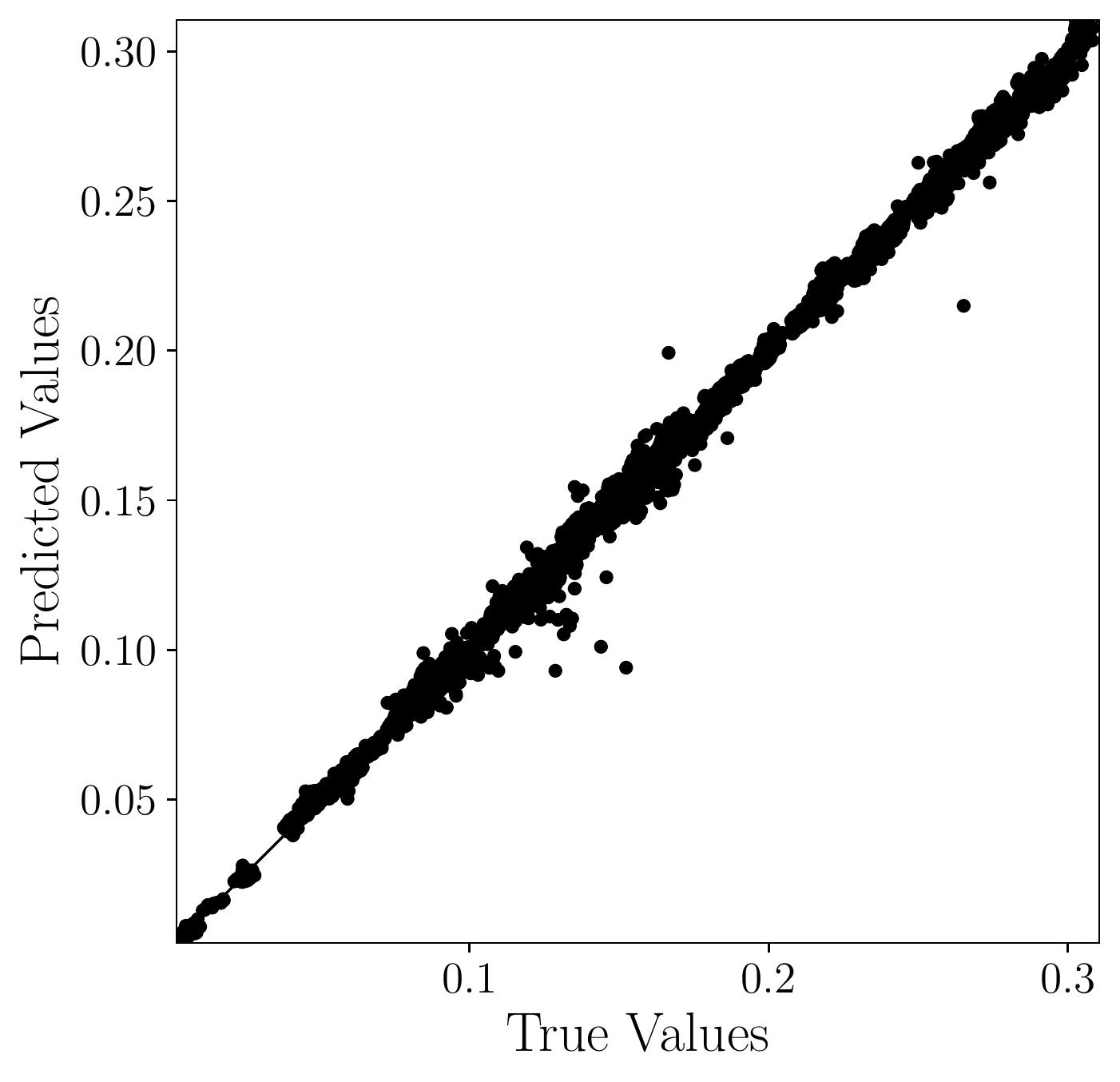}} 
  \caption{CNN enhanced KBNN for 180 microstructures from different DNS with the enhancing CNN being exposed to the original $e_2$ field:
  (a) learning curve; 
  (b) the KBNN predicted $\Delta \Psi_\text{mech,KBNN}$ \emph{versus} the actual $\Delta \Psi_\text{mech,DNS}$;
  (c-f) the components of $\BP_\text{KBNN}$ vs $\BP_\text{DNS}$, where the KBNN achieves improved derivative representations for $P_{11}$ and $P_{22}$, but not $P_{12}$ and $P_{21}$ because the DNS data for $P_{12}$ and  $P_{21}$ are one order of magnitude smaller than for $P_{11}$ and $P_{22}$.
  }
  \label{fig:kbnn-m-dns-penalize}
\end{figure}

The predictive capability of the CNN-enhanced KBNN is limited since it requires knowing the perturbed $e_2$ field, which often is not readily available. Furthermore, simply replacing the perturbed $e_2$ field with the original microstructure information (the original $e_2$ solution) results in an unsatisfied derivative representation of the free energy. To address such limitation, a penalized MSE loss function is used for the CNN-enhanced KBNN with the form 
\begin{equation}
  \text{MSE} = \frac{1}{m} \sum_{i} \left[ \left( \mathbf{Y}- \mathbf{Z} \right)_i^2
  + \beta \left\Vert \BP_\text{KBNN} -  \BP_\text{DNS} \right\Vert_i^2\right]
  \quad \text{with} \quad
  \mathbf{Y}= \Psi_\text{mech} - \Psi_\text{mech,NN}^0
  \label{eq:new-mse-penalization}
\end{equation}
where $\left\Vert \bullet \right\Vert$ is the Frobenius norm and $\beta$ is a penalization parameter with a chosen value of $0.01$ being used in this section. The penalization term in the new MSE serves as a physics-based guidance for the KBNN to find the proper derivative representation. This new MSE allows us to use 
the original $e_2$ solution\footnote{If the datasets used for training are collected from experiments, the original $e_2$ solution would correspond to the actual experimental image of a microstructure.} in the CNN-enhanced KBNN, which significantly improves the usability of the proposed approach in computational homogenization. The results of the CNN-enhanced KBNN using the original $e_2$ solution with the penalized MSE loss are presented in Fig. \ref{fig:kbnn-m-dns-penalize}. The NN representation of the mechanical free energy in Fig. \ref{fig:kbnn-m-dns-penalize}(b) is slight worse than that in Fig. \ref{fig:kbnn-m-dns}(b), which is expected, as the penalization term guides the NN toward improved representation of the stress components at the cost of accuracy in the energy. This is confirmed, and justified, by the outcomes in Figs. \ref{fig:kbnn-m-dns-penalize}(c,f), which shown more accurate presentation for $P_{11}$ and $P_{22}$ than those in Figs. \ref{fig:kbnn-m-dns}(c,f).

\begin{figure}[h!]
  \centering
  \subfloat[learning curve]{\includegraphics[height=50mm]{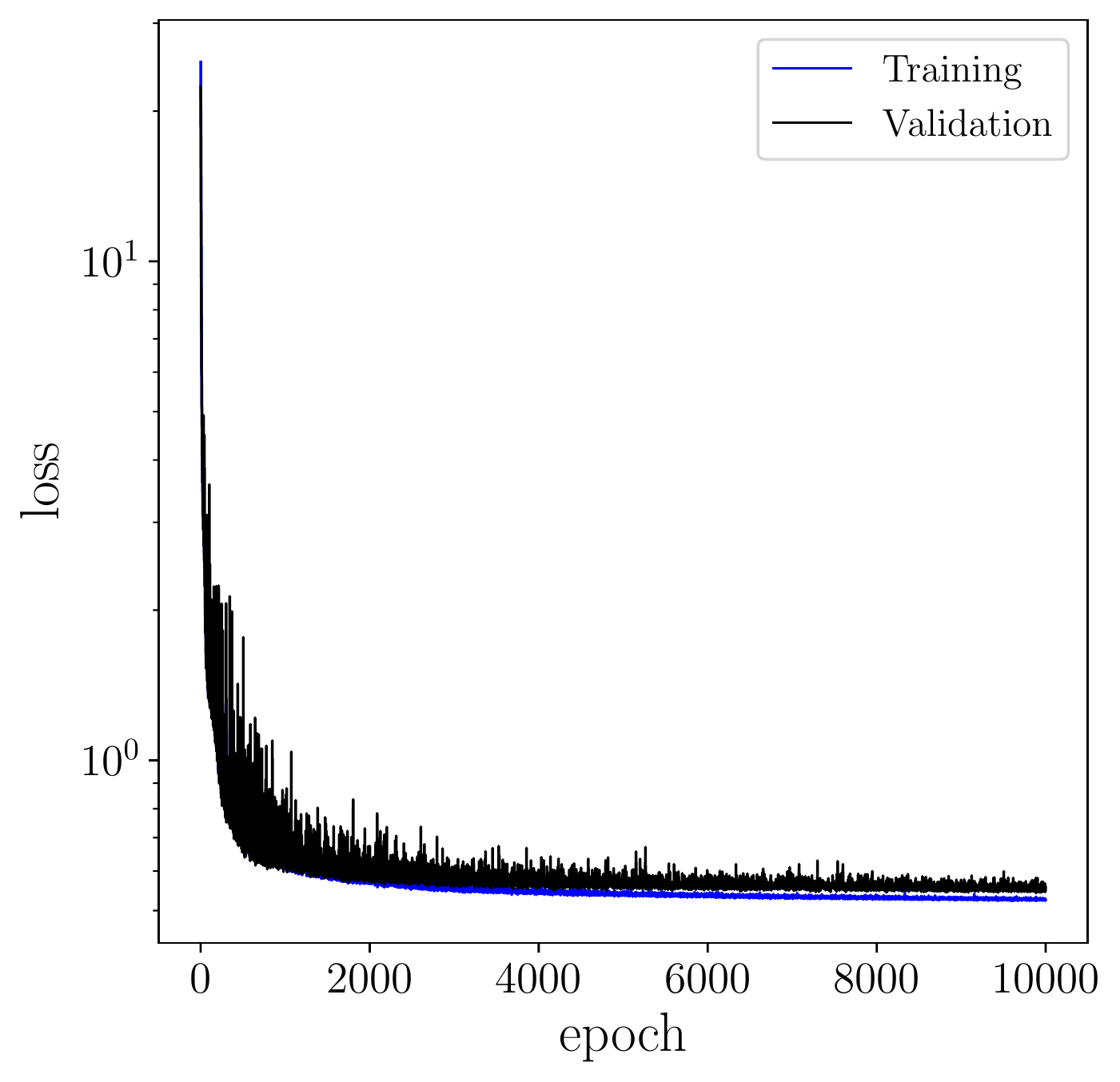}} \hfill
  \subfloat[test dataset prediction]{\includegraphics[height=50mm]{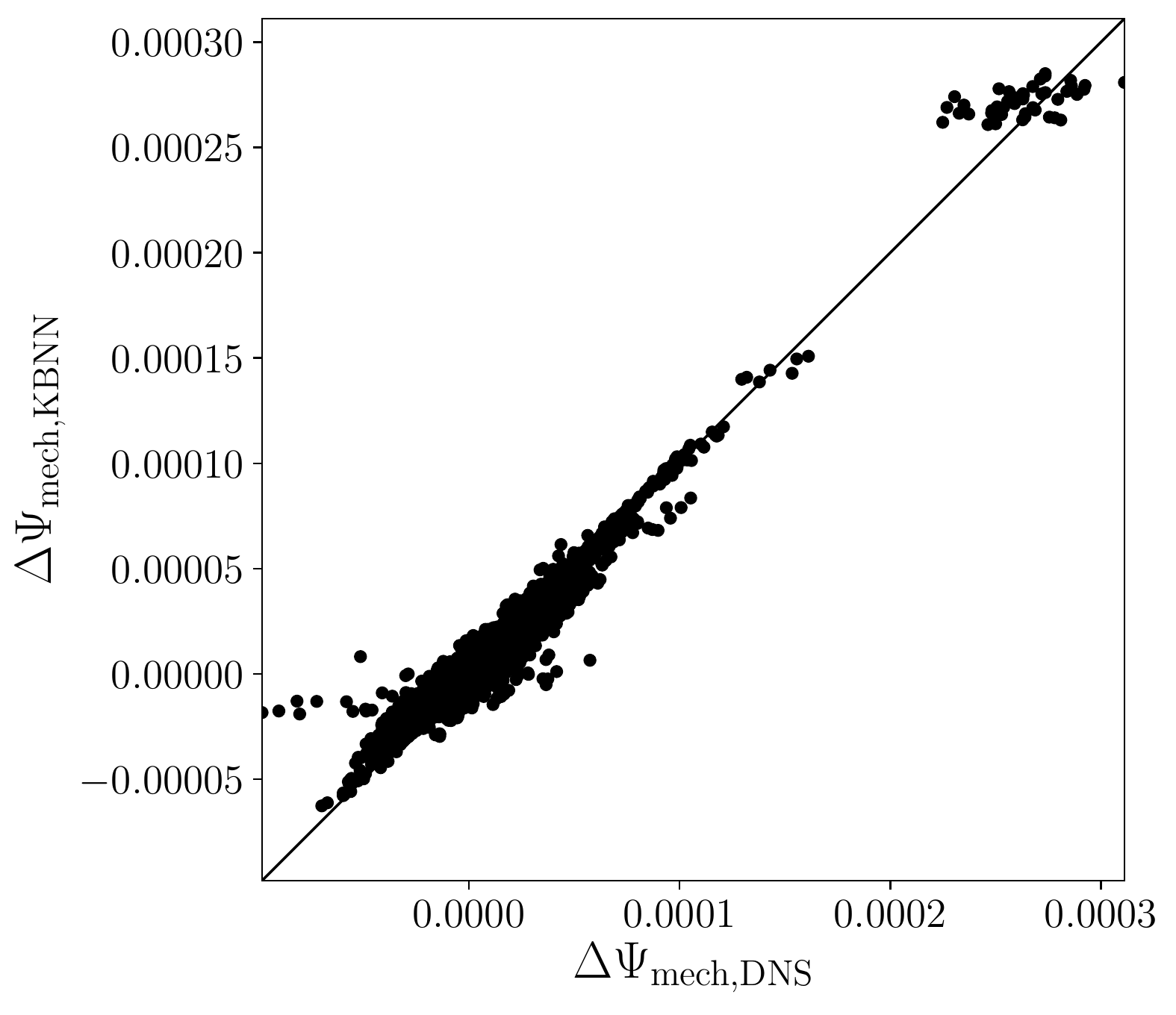}} \hfill
  \subfloat[P11]{\includegraphics[height=50mm]{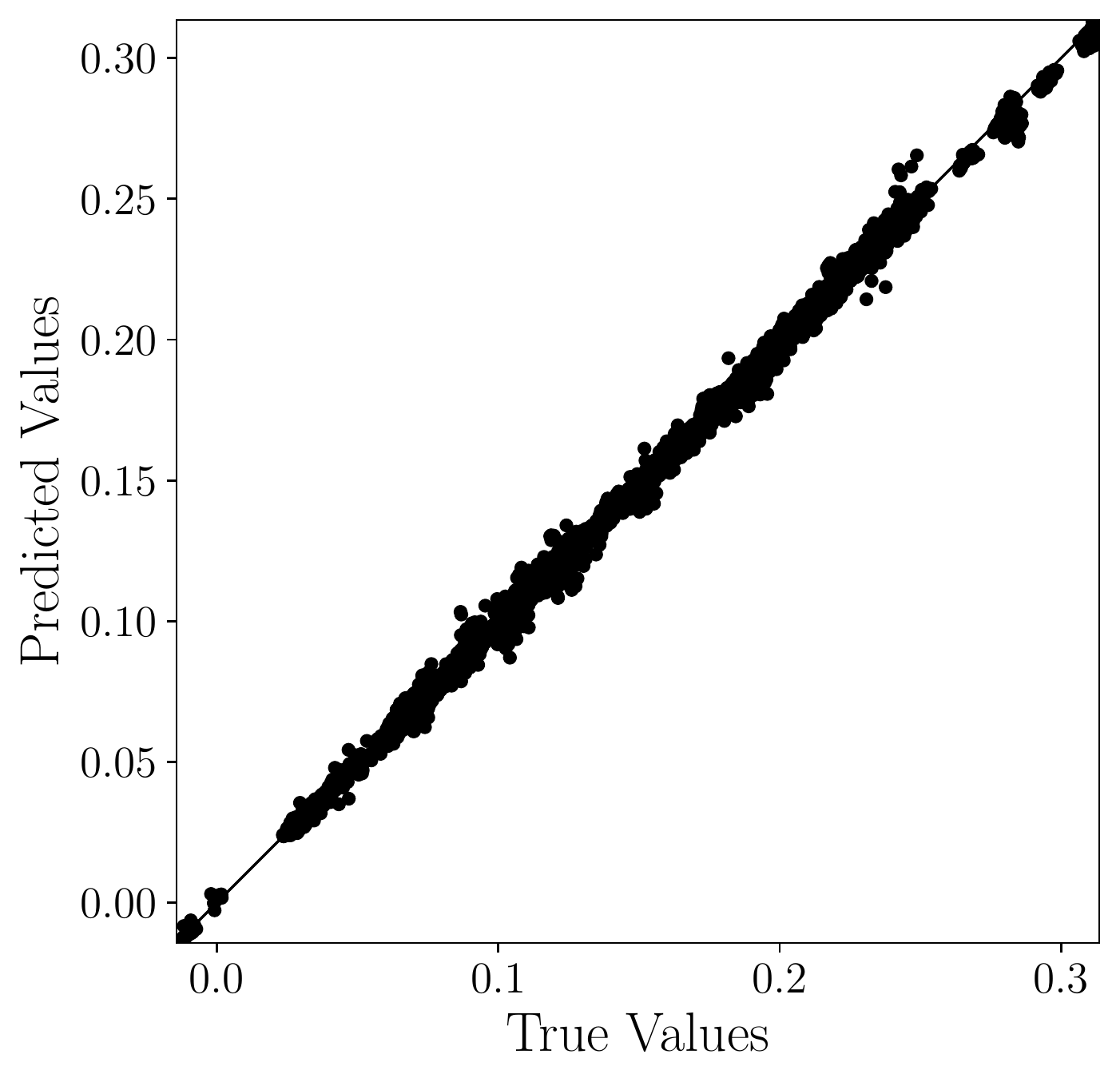}} \\
  \subfloat[P12]{\includegraphics[height=50mm]{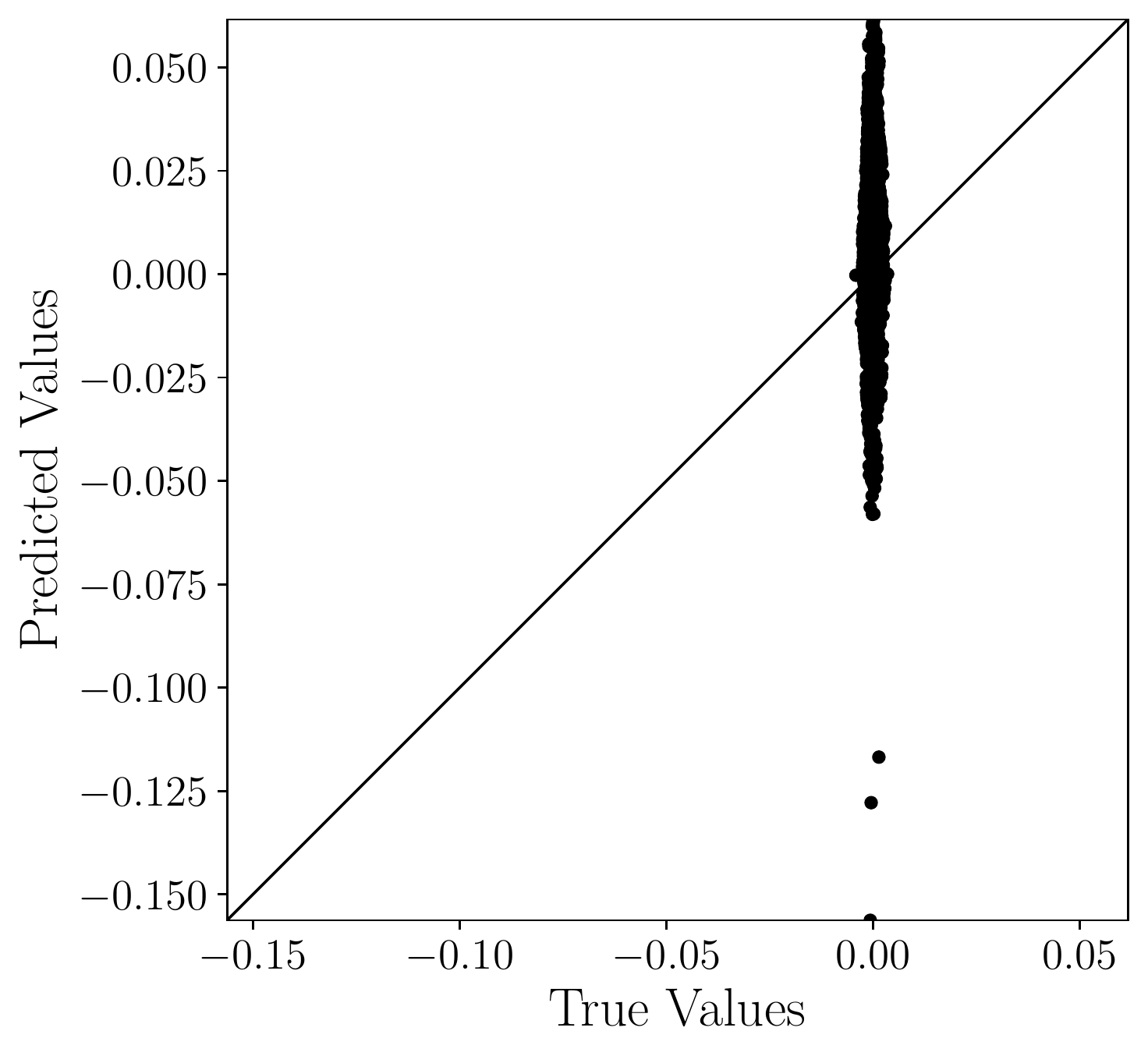}} \hfill
  \subfloat[P21]{\includegraphics[height=50mm]{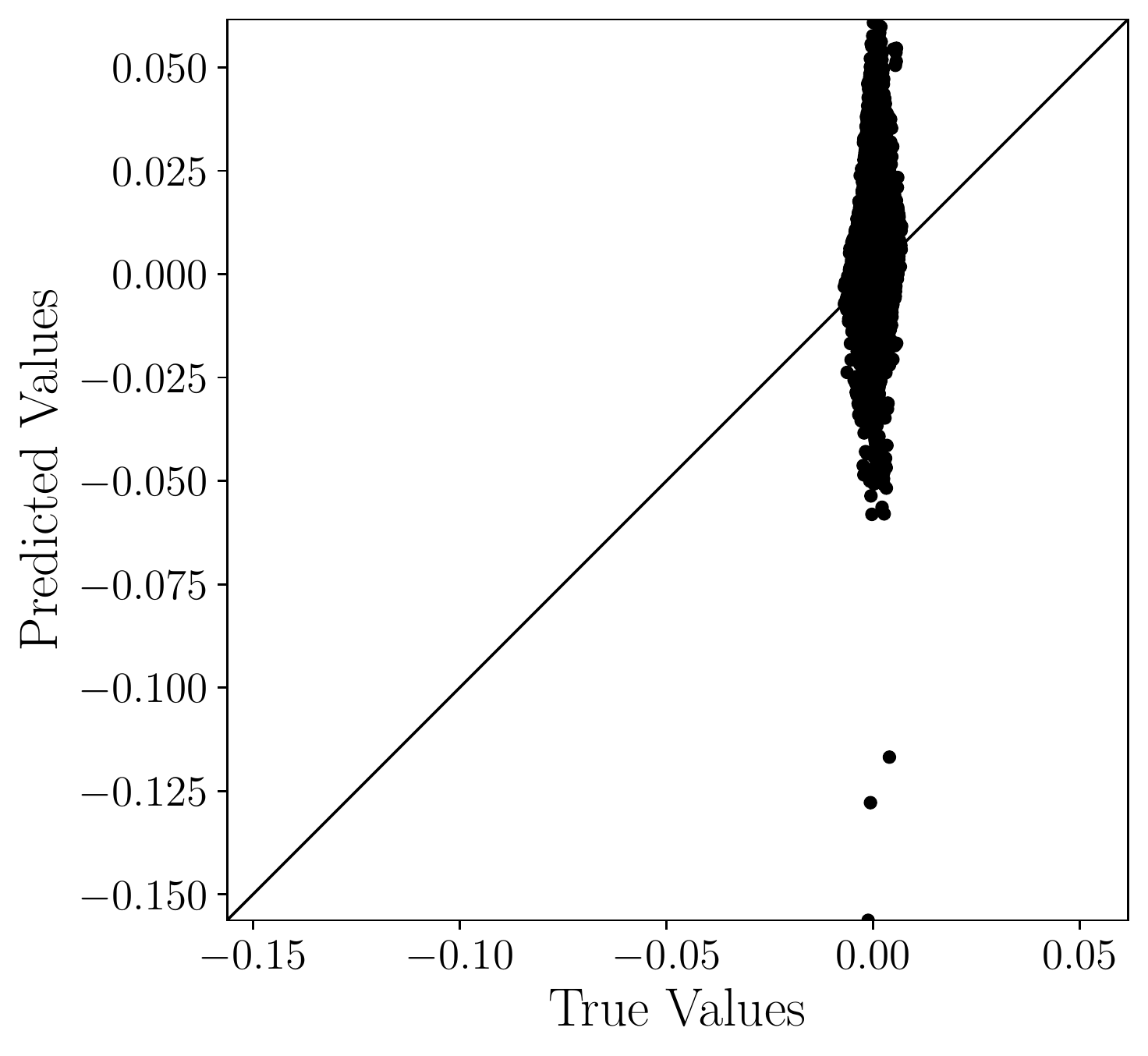}} \hfill
  \subfloat[P22]{\includegraphics[height=50mm]{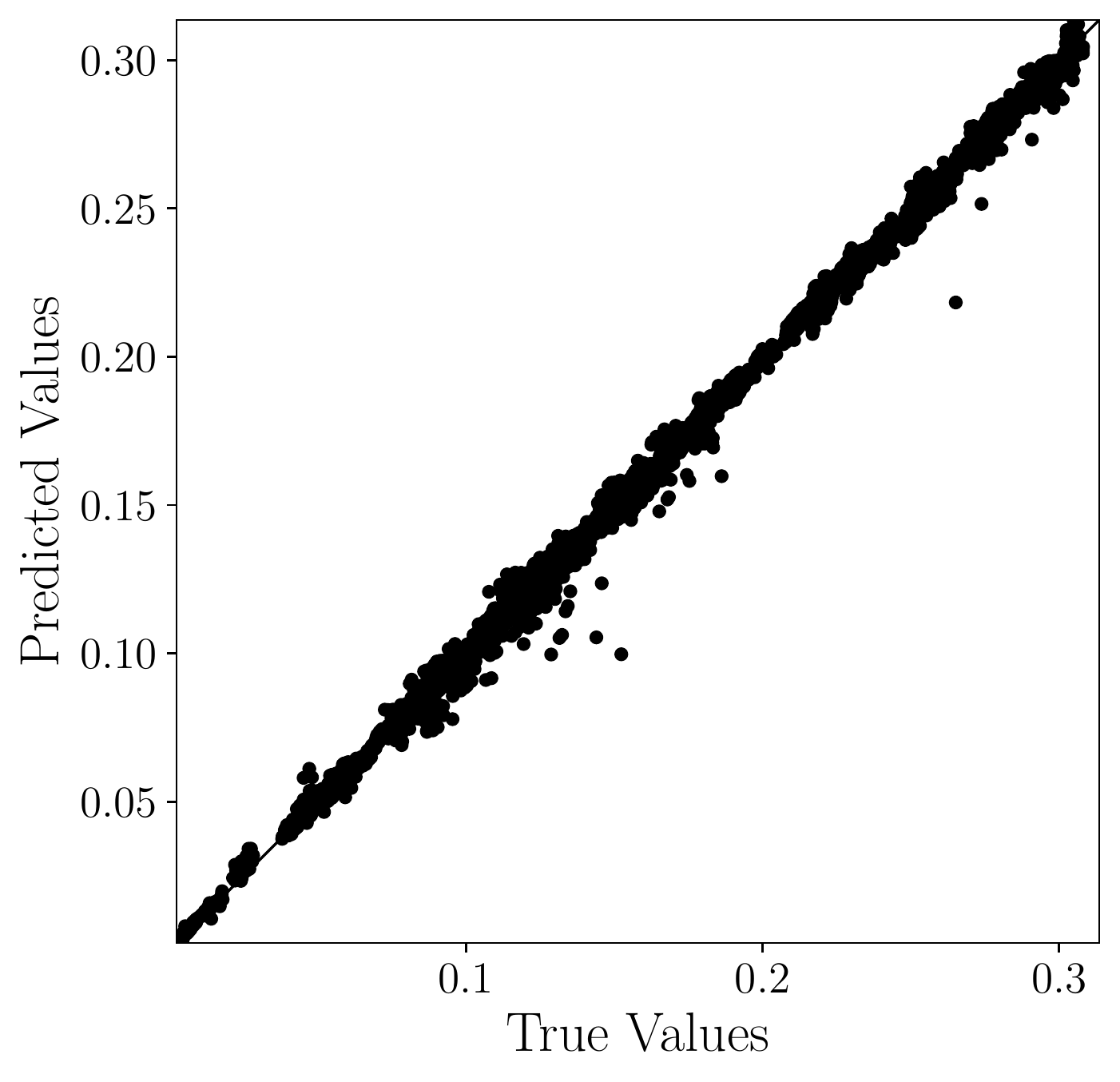}} 
  \caption{ 
    CNN enhanced KBNN for 180 microstructures from different DNS with the enhancing CNN being exposed to the original $e_2$ field only on the boundary:
  (a) learning curve; 
  (b) the KBNN predicted $\Delta \Psi_\text{mech,KBNN}$ \emph{versus} the actual $\Delta \Psi_\text{mech,DNS}$;
  (c-f) the components of $\BP_\text{KBNN}$ vs $\BP_\text{DNS}$, where the KBNN achieves good derivative representations for $P_{11}$ and $P_{22}$, but not $P_{12}$ and $P_{21}$ because the DNS data for $P_{12}$ and  $P_{21}$ are one order of magnitude smaller than for $P_{11}$ and $P_{22}$.
  }
  \label{fig:kbnn-m-dns-penalize-bc}
\end{figure}

We stress that the CNN that forms part of the MNN  in Fig.~\ref{fig:kbnn-m-dns-penalize} is exposed to the entire $e_2$ strain field over the domain. In the interest of minimal representations, we also explored this architecture's power of representation using a more parsimonious input.
We therefore retrained the CNN-enhanced KBNN with the same architecture as in Fig.~\ref{fig:kbnn-m-dns-penalize}, but with the CNN incorporated in the MNN only being exposed to the original $e_2$ data, on the microstructure boundaries. Interestingly, as shown in Fig.~\ref{fig:kbnn-m-dns-penalize-bc}, this KBNN could achieve equally good performance as Fig.~\ref{fig:kbnn-m-dns-penalize}, with the enhancing CNN being exposed to the original $e_2$ field only on the microstructure's boundary, and with $e_2$ values within the microstructural domain being hidden by setting to zero.
The use of the CNN within the MNN works to identify features in the original $e_2$ field over the microstructure, and to thus predict the homogenized response. However, the CNN is able to learn the most relevant features of the microstructure's detailed elastic response even from just the boundary data. Such a KBNN structure which performs well at learning the homogenized mechanical behavior of different microstructures demonstrates the advantage of utilizing CNNs in a multi-resolution learning framework for this instance of computational material physics applications, with heterogeneous microstructures.

\begin{table}
  \centering
  \begin{tabular}{l | l | l  }
    \hline
     Simulation & Description & Wall-Time   \\ \hline
     Base free energy for  & 1. DNN: Fig. \ref{fig:psi-label-shift-dnn-cnn}(a-c)  & GPU: 130 s  \\ 
     single DNS   & 2. CNN: Fig. \ref{fig:psi-label-shift-dnn-cnn}(d-f)  & GPU: 612 s \\  \hline
     Base free energy for & 1. DNN: Fig. \ref{fig:psi-label-shift-dnn-cnn-m-dns}(a-c) & GPU: 1621 s \\ 
     multiple DNSs    & 2. CNN: Fig. \ref{fig:psi-label-shift-dnn-cnn-m-dns}(d-f)  & GPU: 10057 s \\ \hline
     Mechanical behavior for  & 1. DNN-based KBNN: Fig. \ref{fig:psi-800-dnn-kbnn}   & GPU: 270 s\\
     single microstructure  & 2. CNN-based KBNN: Fig. \ref{fig:psi-800-cnn-kbnn}  & GPU: 260 s \\ \hline
     Mechanical behavior for  & 1. using perturbed $e_2$ solution: Fig. \ref{fig:kbnn-m-dns}  & GPU: 15168 s  \\
     multiple microstructures & 2. using original $e_2$ solution with penalization: Fig. \ref{fig:kbnn-m-dns-penalize} & GPU: 15702 s \\
     with CNN-enhanced KBNN    & 3. using boundary information of original $e_2$ solution  & GPU: 15051 s \\
    &  with penalization: Fig. \ref{fig:kbnn-m-dns-penalize-bc} &  \\ \hline
    Prediction of free energy and  & 1. CNN-enhanced KBNN with boundary information  & CPU: 0.38 ms \\
    homogenized stresses per  & of original $e_2$ solution &   \\
    microstructure per testing & 2. Physics-based DNS & CPU: $>$5.3 s \\
	\hline
  \end{tabular}
  \caption{Summary of the CPU/GPU-hrs for different numerical simulations. A single NVIDIA Kepler K80 GPU is used to train different NNs. An Intel i7-8750H CPU with 6 cores is used to predict free energy and homogenized stresses of microstructures with NNs and DNS, both of which are run under parallel mode. The wall-time for physics-based DNS can vary due to the convergence rate, which depends on the microstructure itself and the magnitude of the applied perturbed mechanical loading. }
  \label{tab:wall-time}
\end{table}

The required training time for each NN presented in this Section is summarized in Table \ref{tab:wall-time}. It is of interest to note that, for a well-trained CNN-enhanced KBNN with boundary information of the original $e_2$ solution, it can achieve a speed up of more than 10,000 times (5.3 s / 0.38 ms = 13,947) compared to the physics-based DNS for predicting free energy and homogenized stresses of each microstructure. Such a speed up would allow us to rapidly evaluate the macroscopic behavior of different microstructures for applications such as material optimization, discovery, and design.

\section{Conclusions} \label{sec:conclusion}

We have explored different NN architectures to represent the homogenized mechanical behavior of microstructures. For this purpose, we have generated synthetic data on microstructures  by mechanochemical spinodal decomposition, a coupled diffusional-martensitic phase transformation.
Our preliminary results show the promise of applying CNNs in computational material physics. 

{One motivation behind this study was to evaluate DNN- and CNN-based architectures for their expressivity in representing microstructural information. The DNN-based architectures were given input features based on the established understanding in materials physics that phase volume fractions, interface areas and effective strains must determine the homogenized elastic response. The CNN-based architectures, as is well understood, recognize these and other patterns from the images of microstructures. For microstructures resulting from the same or different initial conditions (and therefore following different trajectories in the latter case), DNN- and CNN-based architectures perform equally well at simply predicting the evolution of the free energy. However, when the homogenized elastic response also is of interest, the multi-resolution nature of the data must be accounted for. Here, with our introduction of multi-resolution learning via KBNNs, we used either DNNs or CNNs as the ENNs representing the dominant {characteristics} in the base free energy's evolution, while the MNN that represents the finely resolved elastic response remained a DNN. Notably, this architecture was applied only to  microstructures that evolved from the same initial condition. In this case, the architectures with CNN-based ENNs proved marginally superior to DNN-based ENNs. This already suggested that CNNs are able to discern more information in the microstructural patterns that determines their subsequent elastic response, than could be imparted via pre-selected features to DNNs.}

Even the above multi-resolution architecture, however, proved inadequate at predicting the elastic response across microstructures that evolve from different initial conditions. In this case, the MNN needed further enhancement that could only be met by a CNN that learned the detailed elastic response of the different microstructures that were being tested mechanically. DNN-based MNNs could not provide this expressivity. Furthermore, with a physics-based penalization in addition to the MSE, the enhancing CNNs seemed to be able to predict the elastic response within the microstructural domain by being exposed only to boundary data, and were capable of using this learned representation to accurately predict the homogenized elastic response via the stress representations obtained as derivatives on the CNN enhanced KBNN.

\begin{figure}[t!]
  \centering
  \subfloat[original microstructure]{\includegraphics[width=0.2\linewidth]{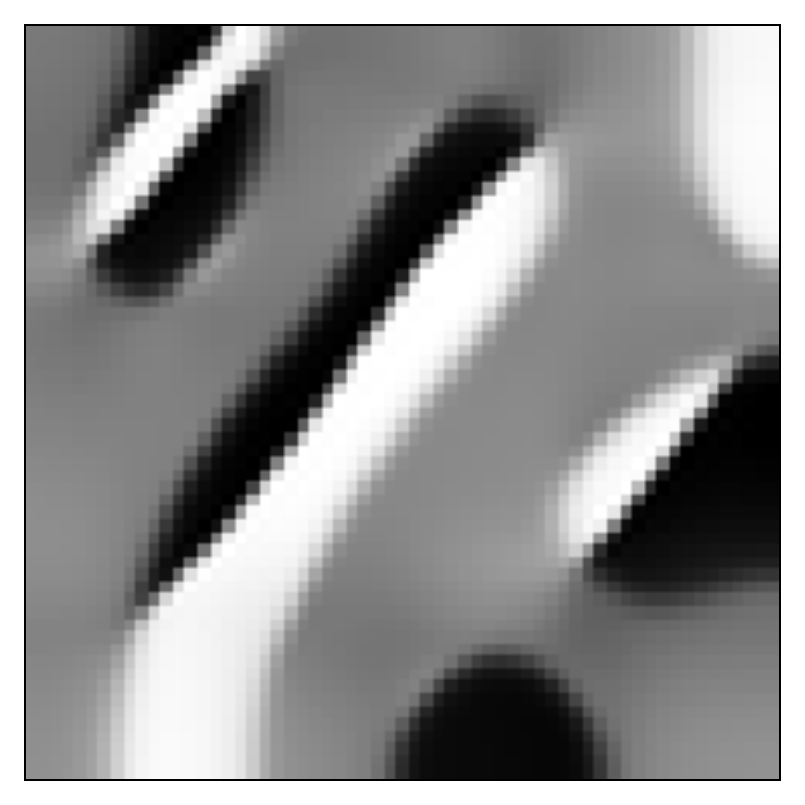}} 
  \subfloat[filter 1]{\includegraphics[width=0.2\linewidth]{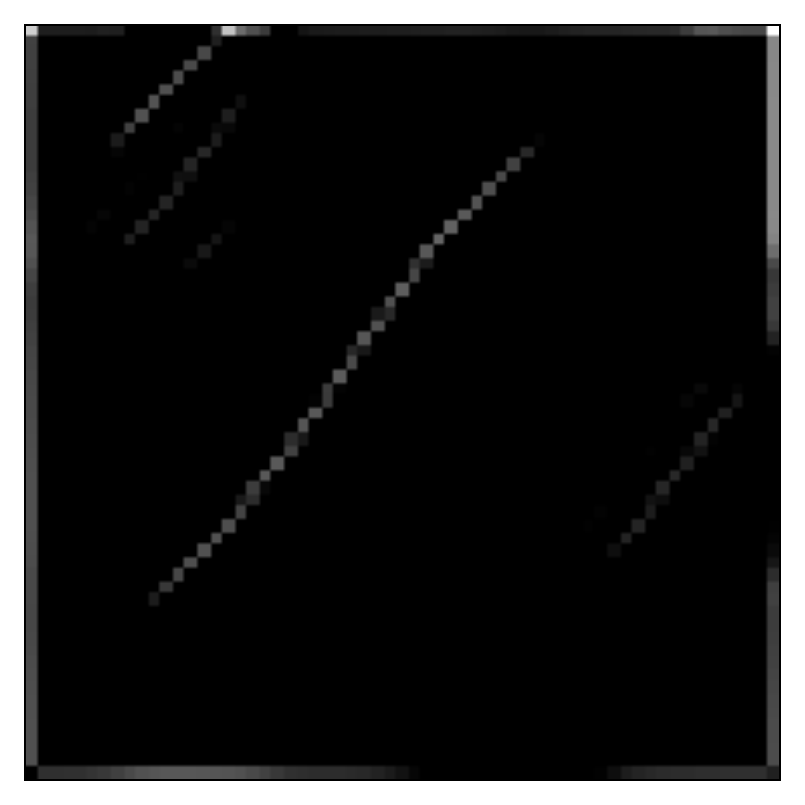}} 
  \subfloat[filter 2]{\includegraphics[width=0.2\linewidth]{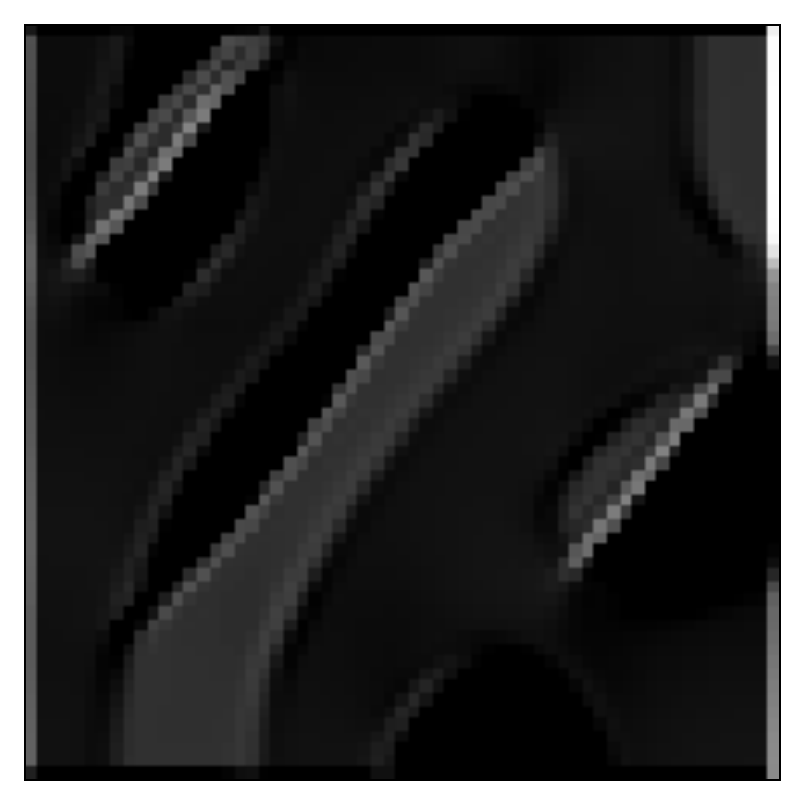}} 
  \subfloat[filter 3]{\includegraphics[width=0.2\linewidth]{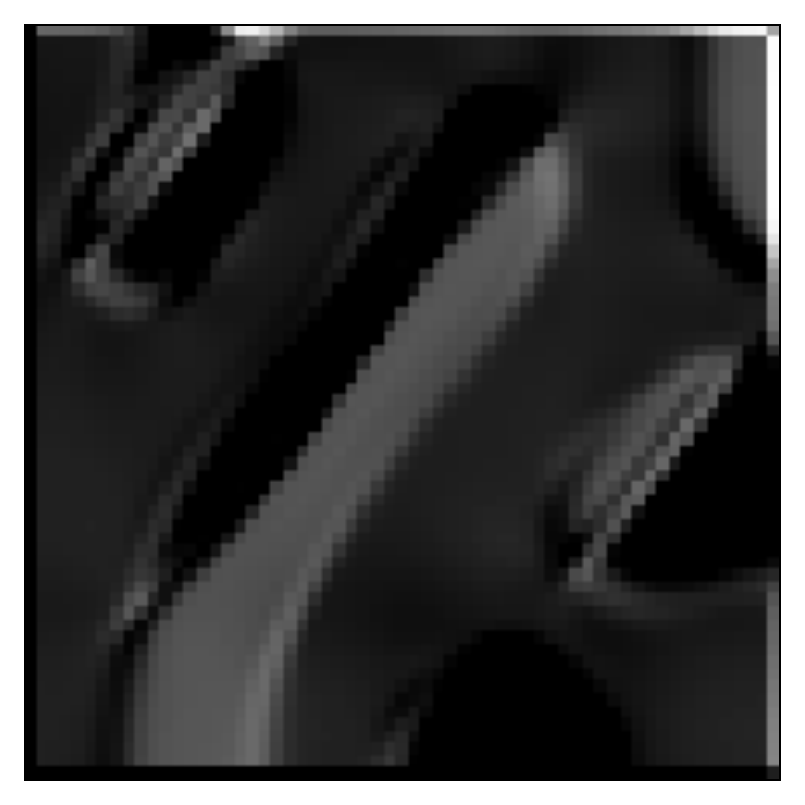}} 
  \subfloat[filter 4]{\includegraphics[width=0.2\linewidth]{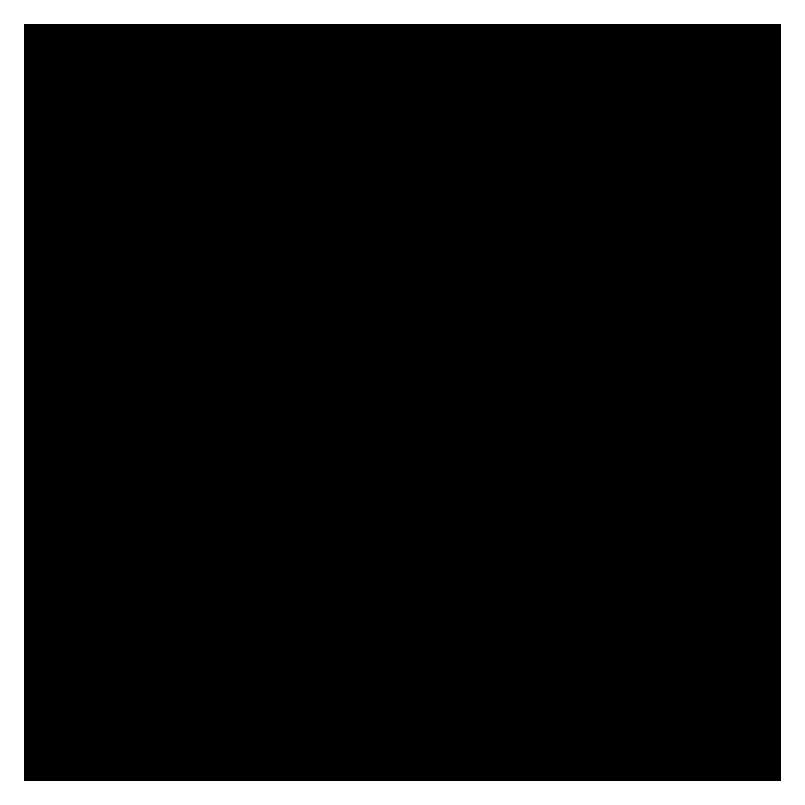}} \\
  \subfloat[filter 5]{\includegraphics[width=0.2\linewidth]{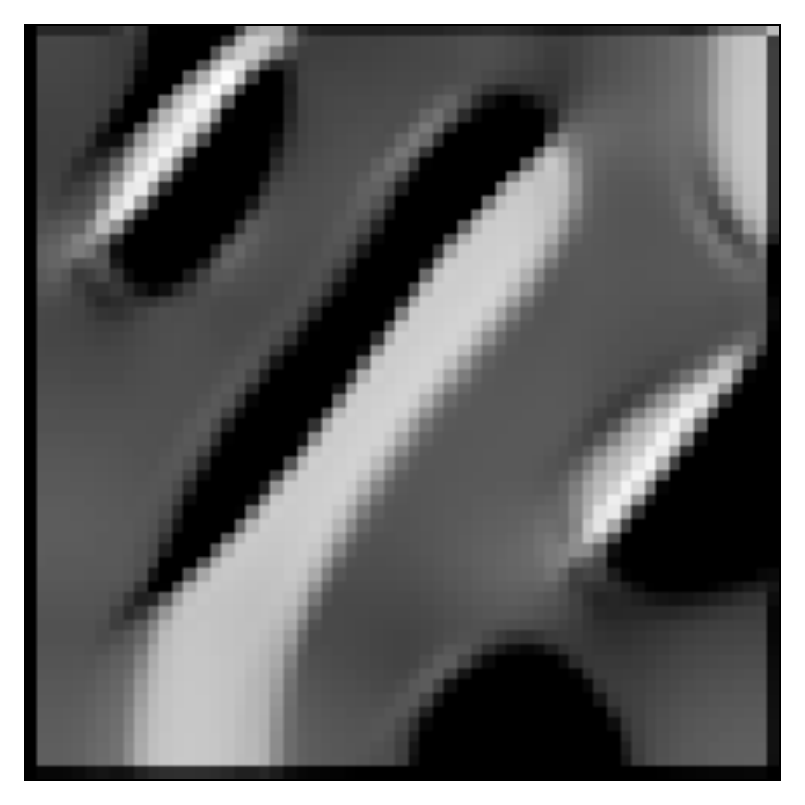}} 
  \subfloat[filter 6]{\includegraphics[width=0.2\linewidth]{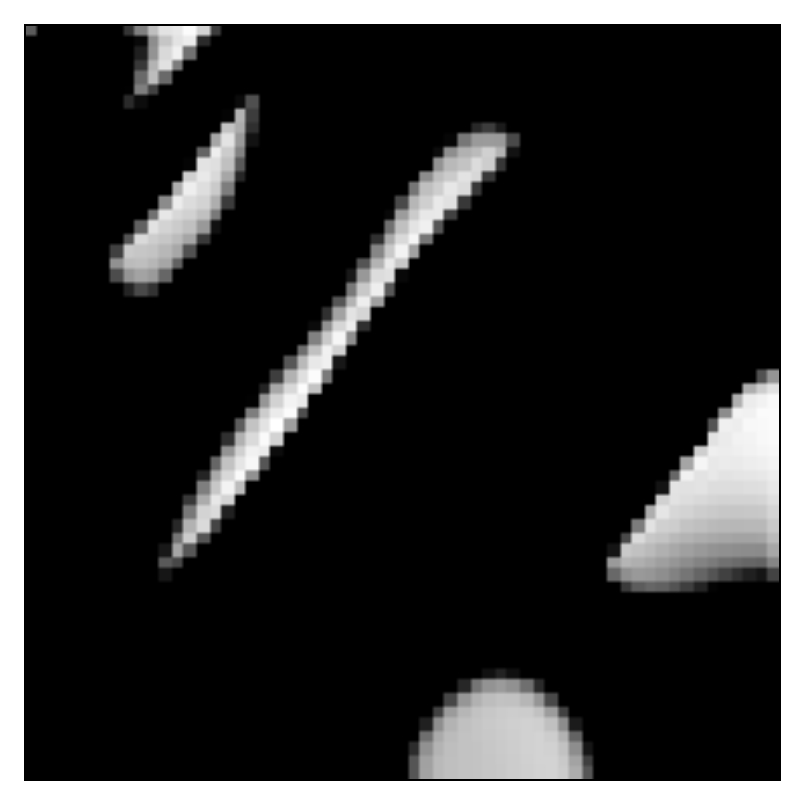}} 
  \subfloat[filter 7]{\includegraphics[width=0.2\linewidth]{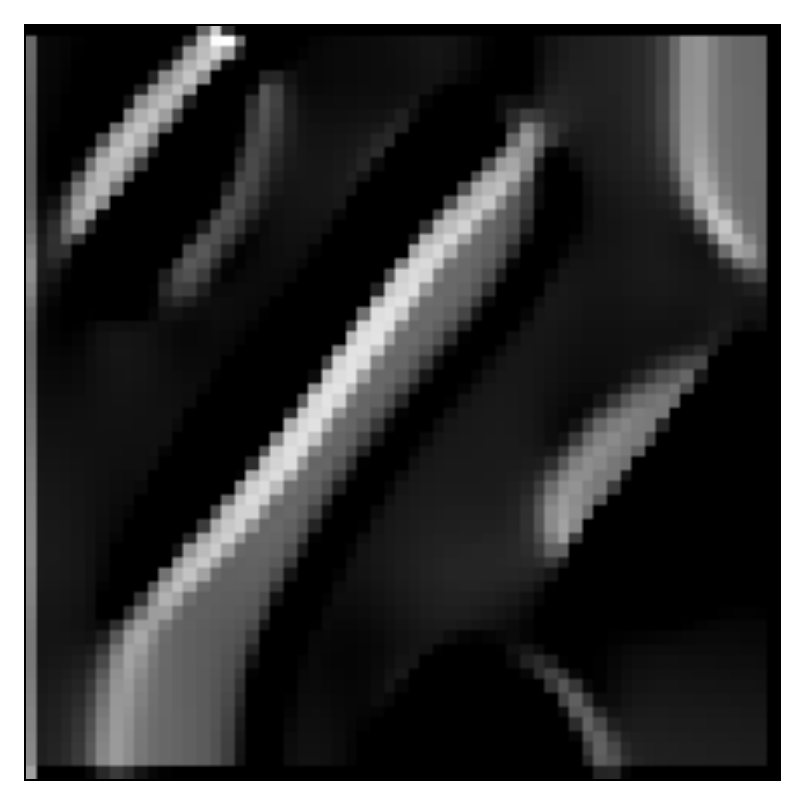}} 
  \subfloat[filter 8]{\includegraphics[width=0.2\linewidth]{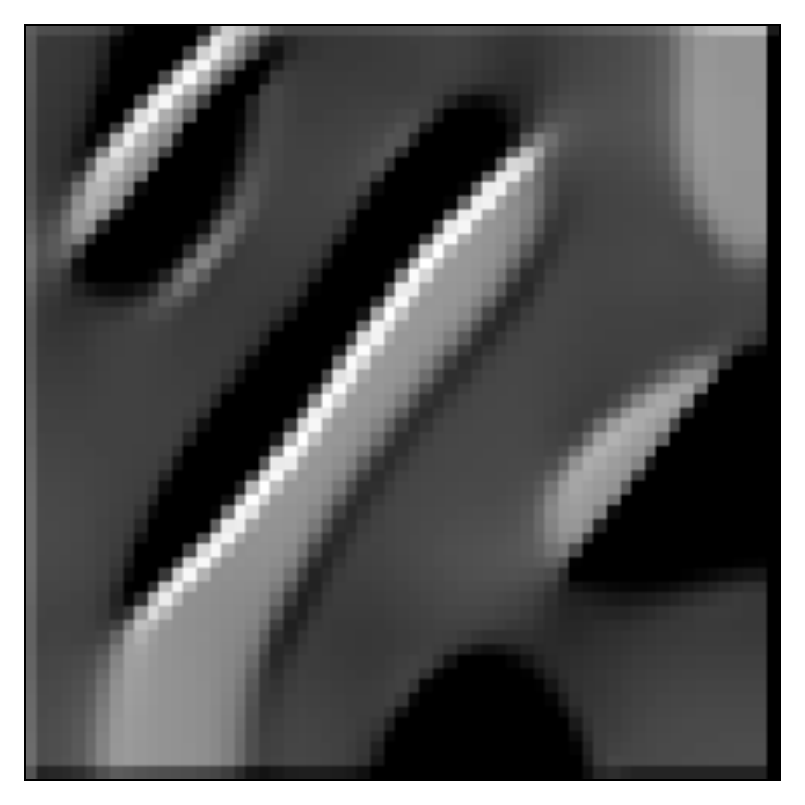}} 
  \subfloat[filter 9]{\includegraphics[width=0.2\linewidth]{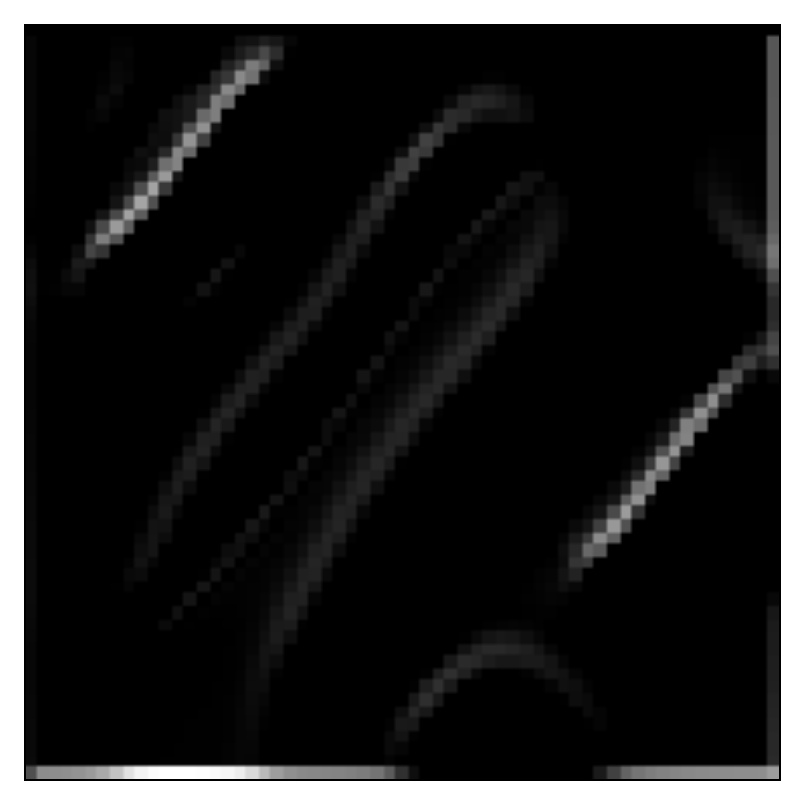}} 
  \caption{ Filter interpretation of the CNN, given in Table~\ref{tab:cnn-base-psi-m-dns}, for 17000 microstructures from 20 DNSs. (a) One randomly selected testing microstructure from $\text{D}_\text{II}$ that contains the 17000 microstructures. (b-j) Filter outputs of the first Conv2D layer after application of the activation function. Observe that the interfacial length between positive and negative rectangle phases is learned by filter 1, the volume fraction of both positive and negative rectangle phases is learned by filter 5, whereas the volume fraction of the negative volume fraction is learned by filter 6. The outputs of filters 2, 3, 7, 8, and 9 seem to contain mixed information of volume fraction, interfacial length, and/or other features that cannot be interpreted in a straightforward manner. The convolution operation of filter 4 to the original microstructure returns negative values, which are transformed to zero by the ReLU activation function, as shown in (e).}
  \label{fig:filter-m-dns}
\end{figure}

{This naturally led to questions on interpretability, and we sought to identify what features the CNNs were learning from the microstructures.} Our investigations toward infusing the better-performing CNN architectures with interpretability reveal that the convolutional layers isolate a greater number of microstructural features than those that we identified on the basis of domain knowledge: $\{\phi_r^+,~\phi_r^-,~l_s^r,~l^{r+},~l^{r-} \}$. {See Fig.}~\ref{fig:filter-m-dns}(b,f,g),
{ in which the phase volumes and interfaces appear as recognizable outputs from the convolutional layers.} {However, the convolutional layers clearly delineate additional features.}
While not presenting a set of features with the parsimony that the expert may postulate for the problem, it suggests that CNN architectures use redundancy to outperform DNNs, { as shown in Fig.~\ref{fig:filter-m-dns}(c,d,h,i,j)}, {where the output of the layers seem to recapitulate aspects of the features identified in Fig.~\ref{fig:filter-m-dns}(b,f,g)}. Interestingly, it also raises questions about the completeness of the feature set $\{\phi_r^+,~\phi_r^-,~l_s^r,~l^{r+},~l^{r-} \}$ that was imposed on the DNN model, suggesting that there are epistemic gaps in the experts' understanding of this problem.

Such findings are important for future studies on combining image data from experiments with multiphysics simulations.
Although this work focused on two-dimensional simulations, our results point to the CNN being more effective in three-dimensional studies.
Because three-dimensional data is more complex in its information content, our domain knowledge might harbor further inadequacies to identify the relevant features. The CNN, instead, could prove more effective at feature selection and dimensionality reduction.

\section*{Acknowledgements}

We gratefully acknowledge the support of Toyota Research Institute, Award \#849910: ``Computational framework for data-driven, predictive, multi-scale and multi-physics modeling of battery materials''.  
Computing resources were provided in part by the National Science Foundation, United States via grant 1531752 MRI: Acquisition of Conflux, A Novel Platform for Data-Driven Computational Physics (Tech. Monitor: Ed Walker). 
This work also used the Extreme Science and Engineering Discovery Environment (XSEDE) Comet at the San Diego Supercomputer Center and Stampede2 at The University of Texas at Austin's Texas Advanced Computing Center through allocation TG-MSS160003 and TG-DMR180072.

\bibliographystyle{unsrt}

\bibliography{lib.bib}
\end{document}